\documentclass[aps,twocolumn,pre,longbibliography,floatfix]{revtex4-2}

\usepackage{xcolor,hyperref}
\hypersetup{
   colorlinks,
   linkcolor={blue!50!black},
   citecolor={blue!50!black},
   urlcolor={blue!80!black}
}

\usepackage{epsfig, amsmath}
\usepackage{bm}
\usepackage{soul,ulem}
\usepackage{hyperref}
\usepackage{footmisc}
\usepackage{wrapfig}
\usepackage{comment}
\graphicspath{{figs/}{./}}
\definecolor{darkGreen}{RGB}{0,100,0}

\DeclareMathAlphabet{\mathitbf}{OML}{cmm}{b}{it}

\newcommand{\fig}[1]{Fig.~\ref{#1}}
\newcommand{\Fig}[1]{Figure~\ref{#1}}
\newcommand{\eq}[1]{Eq.~(\ref{#1})}
\newcommand{\Eq}[1]{Equation~(\ref{#1})}

\newcommand{\be}{\begin{equation}}
\newcommand{\ee}{\end{equation}}
\newcommand{\dv}{\mathitbf d}

\newcommand{\uv}{\mathitbf u}
\newcommand{\xv}{\mathitbf x}

\newcommand{\calBold}[1]{\mbox{\boldmath${\cal #1}$}}

\begin{document}

\title{Theories of the Glass Transition Based on Local  Excitations}     \date{\today}
\author{Massimo Pica Ciamarra}
\email{massimo@ntu.edu.sg}
\affiliation{Division of Physics and Applied Physics, School of Physical and
Mathematical Sciences, Nanyang Technological University, Singapore}
\affiliation{Consiglio Nazionale delle Ricerche, CNR-SPIN, Napoli, I-80126, Italy}
\author{Jeppe C. Dyre}
\email{dyre@ruc.dk}
\affiliation{Glass and Time, IMFUFA, Department of Science and Environment, Roskilde University, PO Box 260,
DK-4000 Roskilde, Denmark}
\author{Edan Lerner}
\email{e.lerner@uva.nl}
\affiliation{Institute for Theoretical Physics, University of Amsterdam, Science Park 904, 1098 XH Amsterdam, the Netherlands}
\author{Matthieu Wyart}
\email{matthieu.wyart@epfl.ch}
\affiliation{Department of Physics and Astronomy \& Institute of Physics, \\
Johns Hopkins University \& EPFL \\
  Baltimore, Maryland, USA \& Lausanne,  Switzerland   \\}

\begin{abstract}
The dramatic slowdown of dynamics in supercooled liquids approaching the glass transition remains one of the central unresolved problems in condensed matter physics. We review approaches that attribute this slowdown to growing thermodynamic or structural length scales and discuss their difficulties in accounting for recent numerical results. These limitations motivate the present review, which critically examines alternative theories in which the glassy slowdown is instead controlled by localized excitations and their elastic interactions. After reviewing key phenomenology with a focus on the fragility of liquids, dynamical heterogeneities, thermodynamics-dynamics correlation, and the effect of kinetic rules and swap algorithms, we compare elastic descriptions based on homogeneous and local heterogeneous elasticity to excitation-based theories incorporating nonlinear responses. Results are compiled to relate global and local elastic moduli, the Debye-Waller factor, and the density of excitations, leading to a quantitative theory testable in experiments. The thermal evolution of the excitation spectrum provides a parameter-free account of the activation energy, while their elastic interactions quantitatively reproduce dynamical heterogeneities via thermal avalanche processes. Synthesized together, these results lead to a framework where the evolution of the excitation spectrum, rather than the growth of a thermodynamic length scale, governs fragility in simple glass-forming liquids -- yet mean-field concepts of dynamical transitions remain central to describing excitations and building a real-space picture of relaxation.
\end{abstract}

\maketitle

\tableofcontents

\section{Scope}

Although glass making began over four thousand years ago \cite{sho07} and was already used for windows in Roman times \cite{fle99}, the physical mechanism that prevents glasses from flowing remains challenging to explain \cite{and95,deb01}. If a liquid is cooled sufficiently rapidly to avoid crystallization, its relaxation time $\tau$—the time scale below which it behaves as a solid—grows continuously from picoseconds at high temperatures to values comparable to the inverse cooling rate (typically minutes or more) at the glass transition temperature $T_{\rm g}$. Below this temperature the system falls out of equilibrium, forming a glass \cite{kau48,har76,ang85,bra85,ang95,deb96,edi96,dyr06,ber11,mck17,alb23} that inherits the liquid’s structural disorder and average isotropy, properties that are crucial, e.g., for optical applications. 

In so-called strong liquids such as silica ($\rm{SiO_2}$) \cite{ang85}, the relaxation time $\tau$ follows an Arrhenius law with an activation energy $E_a$ that is independent of the temperature $T$. By contrast, in fragile liquids by definition $E_a(T)$ increases as $T$ decreases, despite no obvious structural changes taking place.  Identifying the mechanisms that control $E_a(T)$ and cause its growth is a central problem in the theory of the glass transition.  A second key feature is that the dynamics become heterogeneous over a correlation length $\xi$ that increases upon cooling \cite{wee00,ber05,hur95,kob97,yam98,dal07,sil99,edi00,ber11,kar14}. 

\begin{figure}\begin{center}
      \includegraphics[width=7 cm]{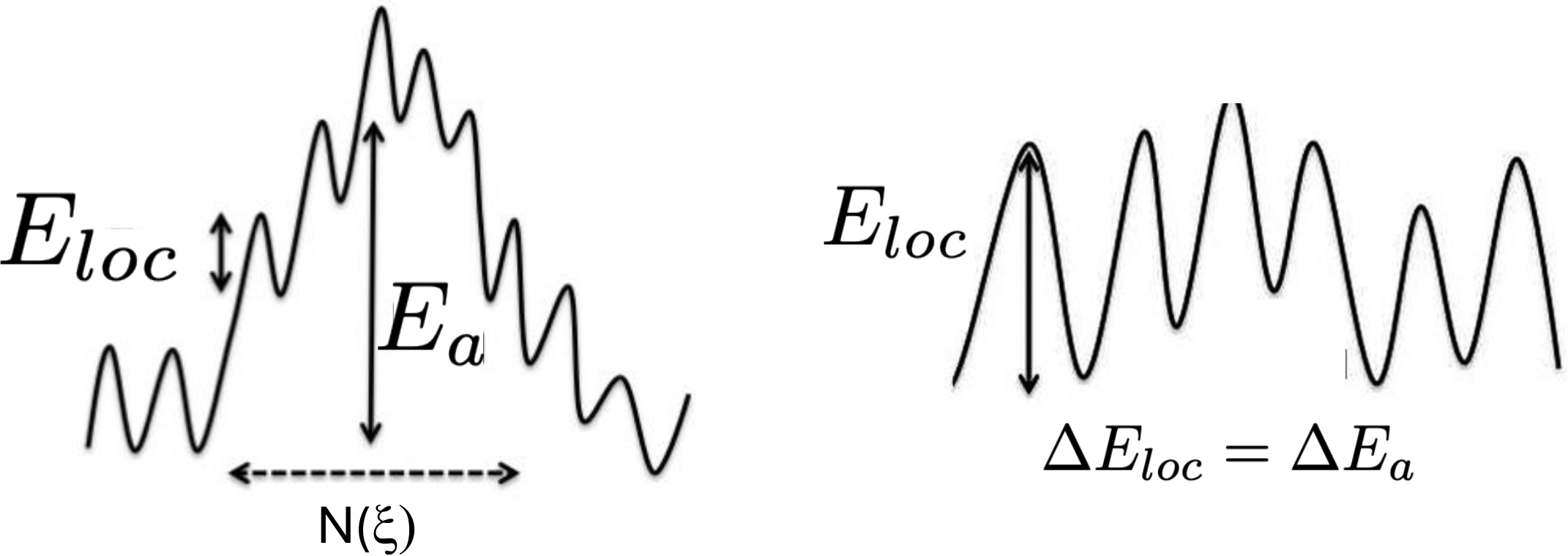}
\end{center}
        \caption{ Different scenarios for the glass transition. Left: In many popular views, the growth of a dynamical length scale $\xi$ when cooling, associated with an increasing number $N(\xi)$ of particles participating in structural relaxation, is responsible for the growth upon cooling of the activation energy $E_a$. Right: In alternative scenarios central to this review, local barriers tied to elementary rearrangements govern the change in activation energy $\Delta E_a$ under cooling. \label{sketch} }
\end{figure}

Although these features are readily observed in molecular dynamics simulations of model liquids where the trajectories of all particles can be tracked, very different theoretical frameworks have been proposed to explain them. One viewpoint assumes that the growth of a length scale \textit{drives} the increase of the activation energy upon cooling, as sketched in Fig.~\ref{sketch}, left.  This is the case for theories that posit the emergence of structural order upon cooling that must be broken for relaxation to occur, such as the Random First Order Transition (RFOT) theory \cite{kir89,lub07,bir23}. It also applies to certain kinetically constrained models \cite{fre84,rit03,Ton04,can10}, where defects facilitate motion with interactions that can generate fragile behavior \cite{Jac91,gar02,Gar03,gar07}.  An alternative view is that the two key observations—growing length scales and growing activation energies—are largely independent, and that $E_a(T)$ is set by local barriers associated with the elementary particle rearrangements, as sketched in Fig.~\ref{sketch}, right.  The oldest example of this is the free-volume model \cite{doo51,coh59,gre81}, which assumes that relaxation depends on the excess volume (beyond the sum of molecular volumes).  More modern versions of this ``local-barrier'' perspective are reviewed below.

Although the hypothesis that a growing length scale drives the slowdown of fragile liquids has long been dominant, several recent observations indicate that it does not apply, e.g., to polydisperse liquids \cite{nin17,ber23} that can be very deeply supercooled in computer simulations and display the hallmarks of the glass transition:  
(i) direct measurements of the distribution of the activation energies of local rearrangements -- termed ``excitations'' in this review -- show that their thermal evolution directly governs the change of activation energy \cite{cia24,jic25};  
(ii) both the time scale and the length scale of structural relaxation can be tuned at will by modifying kinetic rules in simulations while thermodynamic and structural properties remain unchanged \cite{wya17,ber19b,gav24}; theories ignoring kinetic rules therefore cannot be quantitatively predictive;  
(iii) mesoscopic models \cite{Bul94,reh10,gui22,oza23} and theoretical analyses \cite{tah23,deg25} show that elastic interactions among local rearrangements \cite{Lemaitre14} naturally generate dynamical heterogeneities, even when activation energies are controlled by local barriers.  
This framework can quantitatively reproduce dynamical observations in molecular dynamics simulations \cite{gui22,gav24}. After reviewing the current state of the field, our goal is to present different frameworks that seek to describe local barriers. 

The theoretical frameworks that attribute the slowing down of dynamics to the liquid's elastic properties are inspired by a seminal paper from 1969 by Goldstein~\cite{gol69}. Introducing the potential-energy landscape picture, Goldstein proposed the existence of an onset temperature $T_0$, such that for $T > T_0$ the system explores regions near saddles of the (free) energy landscape, while for $T < T_0$ the dynamics becomes activated and relaxation proceeds through barrier-crossing events. In real space, Goldstein proposed that these events (excitations) are in some sense ``local,'' in that in the rearrangement process leading from one minimum to a nearby one, most atomic coordinates change very little, and only those in a small region of the substance change by appreciable amounts~\cite{gol69}. Below we review a sequence of theoretical frameworks building on these ideas that have progressively relaxed simplifying assumptions about elasticity: from descriptions in which relaxation barriers are associated with the homogeneous, linear elastic response, through approaches that relate excitation energies to local but still linear elastic properties, to a final framework based on the full spectrum of excitations—an intrinsically heterogeneous and nonlinear description:

\begin{itemize}
\item {\it Elastic models}, which exist in different closely related versions \cite{hal87,buc92,dyr96,sta02,bor04,dyr04,dyr06,lar08,mei20,zho20,mei21}, in the simplest cases make the simplifying assumption that (i) the geometry of rearrangements is temperature-independent and the same throughout the liquid; 
(ii) On short time scales, the liquid is described well as an isotropic linear elastic material.  These models lead to the prediction that the activation energy growth under cooling is primarily governed by the thermal evolution of either the high-frequency shear modulus \cite{dyr96,dyr06} or the closely related Debye–Waller factor \cite{hal87,buc92,sta02,bor04,dyr04}. The corresponding strong correlations  are indeed observed in many experiments \cite{hec15a}. 
\item {\it Local elastic model} \cite{kap21} refines this picture by relating the magnitude of local barriers to an average elastic dipolar response. Spatial fluctuations are still neglected, but some temperature dependence of the geometry of rearrangements is accounted for. Linear elasticity is still considered. This approach yields  predictions for fragility that can be tested in simulations.
\item {\it Excitation-based theory} goes beyond linear elasticity and average response, by considering instead the full spectrum of local barriers or excitations. Their energy turns out to display a broad  spectrum. The growth of activation energy is attributed to a shift of this spectrum upon cooling \cite{cia24}.  This shift, as well as the evolution of the excitation architecture  under cooling \cite{jic25}, are associated to the presence of a dynamical transition \cite{gol69,kir89,bir23} -- thus building a  bridge between mean-field theories of the glass transition and the excitations of the glass.
\end{itemize}

This review is organized as follows.  In Sec.~\ref{S2} we set the stage by summarizing key facts of the glass transition, including classic observations—fragility, dynamical heterogeneities, and correlations between dynamics and thermodynamics—which are reviewed in more detail elsewhere \cite{har76,ang85,bra85,ang95,deb96,edi96,mck17,alb23}.  We also discuss recent numerical findings showing that local kinetic rules, such as swap moves in polydisperse systems, can dramatically accelerate or slow down the dynamics.  In Sec.~\ref{S2bis} we discuss the historical difficulty of measuring precisely the activation energy, and present a novel method to do so. In this review this measure plays a key role to test theories beyond mere correlations, so as to stringently assess their success.  In Sec.~\ref{S3} we briefly summarize theories where a growing length scale controls the activation energy; much more detailed reviews of this perspective already exist \cite{Got92,lub07,ber11,rit03}. We focus on updating their pros and cons in light of recent results, see also Ref. \onlinecite{bou24} for (partially) alternative views.  Sec.~\ref{sec:global} and Sec.~\ref{sec:local} review global and local elastic models of the glass transition, respectively, whereas Sec.~\ref{S4} reviews an excitation-based theory of that phenomenon. An empirical procedure is introduced to relate that theory to the Debye-Waller factor accessible in experiments. Sec.~\ref{comparison} compares quantitatively these theories: their predictions are correlated, but only the excitation-based theory quantitatively captures the evolution of the activation energy under cooling. Sec.~\ref{S7} reviews a scaling theory for how the architecture of excitations depends on both temperature and energy. Sec.~\ref{S8} first discusses how elastic interactions among excitations can quantitatively reproduce dynamical heterogeneities, establishing a link between the glass transition and avalanche-type response in disordered materials.  It then considers the link between dynamical heterogeneities in liquids and those observed during creep phenomena that occur when disordered systems (such as frictional interfaces,  amorphous solids or crumpled sheets) are loaded and age due to thermal activation. Sec.~\ref{S9} reviews an excitation-based theory for the dependence of fragility on some properties of the glass considered, including the presence of soft elastic modes in their vibrational spectrum (the so-called ``Boson peak'') or the valence for network glasses. Finally, Sec.~\ref{S12} summarizes our results and  Sec.~\ref{S11} discusses the open questions that remain and provides an outlook.

\section{Key Phenomenology}
\label{S2}
\begin{figure}\begin{center}
      \includegraphics[width=7 cm]{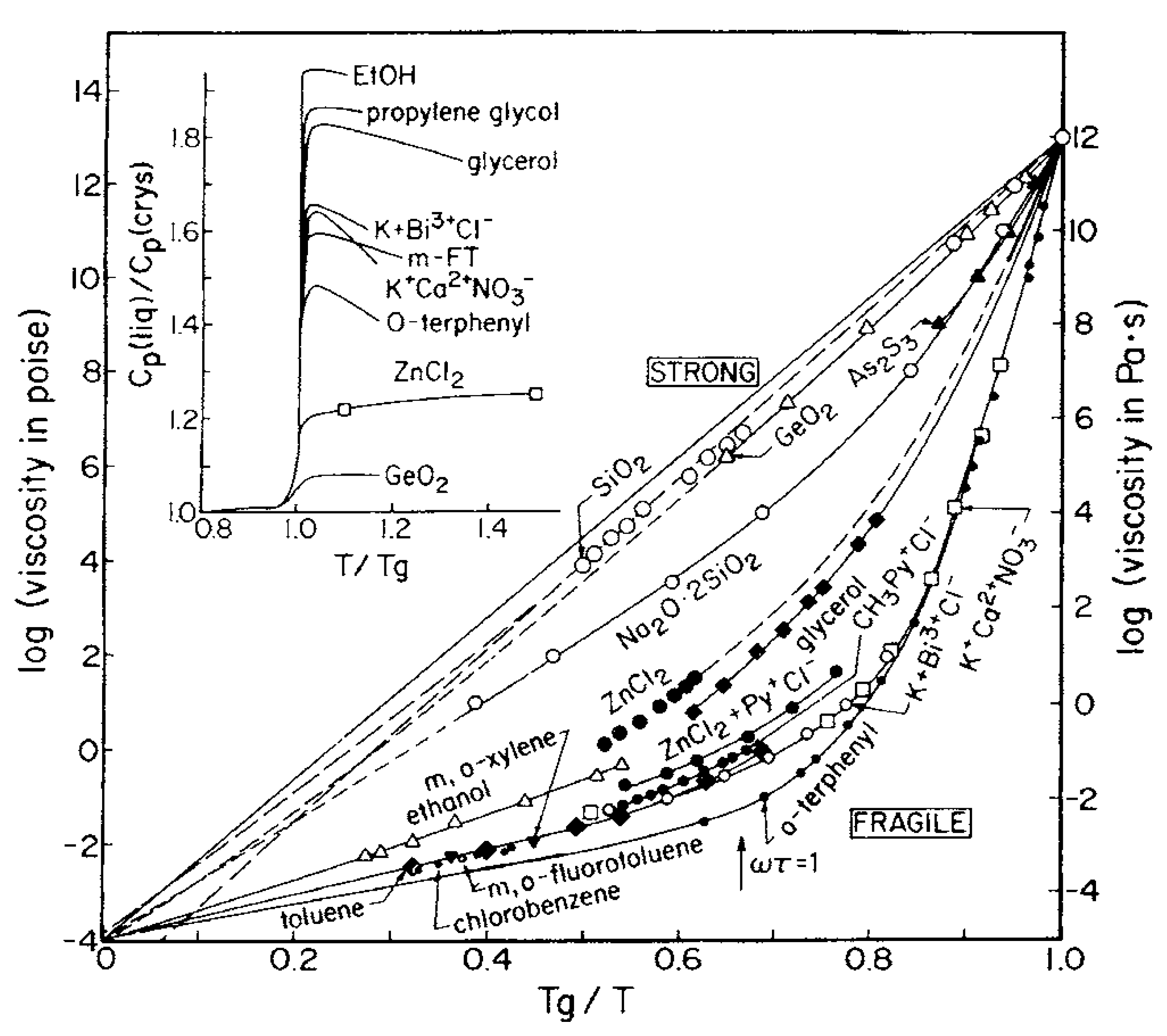}
\end{center}
        \caption{The classical Angell plot showing the super Arrhenius temperature dependence of the viscosity of several glass-forming liquids (reproduced from Ref. \onlinecite{ang85}). The x-axis gives the inverse temperature normalized to unity at the glass transition temperature $T_{\rm g}$, the y-axis gives the logarithm (base 10) of the viscosity. As is customary, $T_{\rm g}$ is here defined as the temperature at which the equilibrium (metastable) liquid has viscosity $10^{12}$ Pa$\cdot$s.  Reprinted with permission. \label{fig:angell} }
\end{figure}

This section summarizes briefly the experimental facts a satisfactory theory of the glass transition should account for.

\subsection{The fragility of liquids}

{\bf Definitions:} If relaxation is governed by jump over barriers, we expect the relaxation time  to be activated with: 
\be
\label{eq:act_en}
\tau(T)=t_0\exp\left(\frac{E_a(T)}{k_BT}\right)\,.
\ee
Here $k_B$ is the Boltzmann constant and $t_0$ is a time scale discussed in the next section. A system is termed Arrhenius if $E_a$ does not depend on temperature. This rarely happens for glass-forming liquids (pure silica is almost Arrhenius); in almost all cases $E_a(T)$ increases upon cooling. This is the famous non-Arrhenius characteristic termed \emph{super-Arrhenius}. There appears to be no glass-forming liquids for which $E_a(T)$ decreases upon cooling. Why are some glass-forming liquids strongly non-Arrhenius? This old simple question remains one of the most important ones of the field.

Experimental data for $\tau(T)$ determined, e.g., as the inverse dielectric or mechanical loss peak frequency \cite{har76} are often reported in terms of the so-called fragility $m$ defined as the slope at $T_{\rm g}$ of the data plotted in the Angell plot (\fig{fig:angell}):

\be\label{eq:frag_def}
m\equiv \left(\frac{d\log_{10}\tau(T)}{T_{\rm g}\,d (T^{-1})}\right)_{T=T_{\rm g}}
=\frac{1}{\ln(10)}\left|\frac{d\ln\tau(T)}{d\ln T}\right|_{T=T_{\rm g}}\,.
\ee
Taking the characteristic relaxation time at $T_{\rm g}$ to be 1000 seconds and $t_0$ to be 0.1 picosecond, an Arrhenius $\tau(T)$ corresponds to $m=16$. Typical values of $m$ are between 40 and 100, however, with some polymers reaching as high as 200, testifying to the general super-Arrhenius trend. A generic quantification of this is the ``temperature index'', $I$, defined \cite{hec08} as minus the logarithmic derivative of $E_a(T)$,

\be\label{eq:ti}
I(T)\equiv -\frac{d\ln E_a(T)}{d\ln T}\,.
\ee
If for instance $I=4$, then a 1 \% decrease of temperature leads to a 4\% increase of the activation energy. It is straightforward to show that $m=16\left(I(T_{\rm g})+1\right)$, the Arrhenius case of which is $I=0$ and $m=16$, while e.g. $I=4$ corresponds to $m=80$ \footnote{There is some leeway in the number 16 here. Thus, the classical \fig{fig:angell} instead operates with 17 because this plot uses a smaller high-temperature limiting viscosity than the 0.1 mPa s we prefer. Likewise, there is leeway in defining the glass transition temperature, which is nowadays often defined by a relaxation time of 100 s rather than the above. None of the reasoning detailed below is affected by such differences, however, which are minor on the logarithmic scale.}. Note that the (standard) isobaric fragility is usually significantly larger than the isochoric fragility \cite{nis06,alb22}.

{\bf Correlations between fragility and properties of the glass:} A number of elastic and vibrational observables measured in the glass \emph{correlate} with the kinetic fragility $m$ of their ancestral liquid. First, the glass Poisson ratio $\nu$ (or equivalently the ratio of bulk to shear moduli) shows a positive correlation with fragility across many inorganic and polymeric glasses: materials with larger $\nu$ tend to be more fragile, consistent with the notion that a small shear modulus $G$ relative to $K$ fosters rapid super-Arrhenius growth of relaxation times upon cooling \cite{nov05pre}. Second, the magnitude of the boson peak \cite{Buc84,Bal10,Schir07,deg14}---often characterized by the normalized amplitude $A_{\rm BP}$ of $Z(\omega)\equiv D(\omega)/D_D(\omega)$ (where $D(\omega)$ is the vibrational density of state and $D_D(\omega)$ the Debye prediction of that quantity in a continuous medium) at the peak frequency -- is 
\emph{anticorrelated} with fragility: strong liquids display a pronounced excess of soft modes (large $A_{\rm BP}$), whereas fragile liquids have a weaker excess \cite{nov05pre,ngai97jcp,tan08nm}. This trend holds in both molecular and covalent network glasses, yet hydrogen-bonded liquids and systems with strong directional interactions can show significant deviations \cite{nov05pre,ngai97jcp}. Third, chalcogenides (e.g.\ Ge--As--Se) provide a compositionally tunable test-bed in which the mean coordination (average valence $r$) controls rigidity: beyond some valence threshold, covalent bonds alone can maintain the elastic stability of the material  \cite{phillips79jncs,thorpe85ssc}. In these systems the liquid is empirically strongest and the specific-heat jump at $T_g$, $\Delta C_p$, is smallest near the rigidity threshold $r\simeq r_c$, while moving away on either side increases both the fragility and $\Delta C_p$ \cite{tatsu90prl,bohmer91prb,kam91prb}. Note that these observations also hold for silica, a network of tetrahedra that is marginally connected and strong \cite{Micoulaut2022}. These observations collectively suggest that features of \emph{linear} elasticity and vibrational spectra, though defined locally in the landscape, encode robust information about the activated dynamics and thereby about the liquid's fragility. In contrast to other aspects of the glass transition, theoretical frameworks to explain these findings are scarce.

\subsection{Dynamical heterogeneities}
An important characteristic of glass-forming liquids, which distinguishes them from ordinary low-viscosity liquids above the melting temperature $T_m$, is spatial heterogeneity of their dynamics.  At any given time, some regions of the liquid display intense activity while others remain almost frozen \cite{edi00,ric02}.  On the time scale of the structural relaxation $\tau$, the active regions migrate through the system, such that on long time scales the liquid appears homogeneous. Dynamic heterogeneity was firmly established experimentally in the 1990s, through four-dimensional NMR experiments that demonstrated the persistence of inactive regions \cite{sch91a}, via optical dynamic hole burning \cite{cic96}, and via solvation dynamics measurements \cite{wen00}, and later quantified through four-point correlation functions and the associated susceptibility $\chi_4$ \cite{Berthier2005,lac03}. 

Dynamical heterogeneities are associated with structure, as demonstrated by simulation studies of \emph{dynamic propensity}, introduced by Harrowell and Widmer-Cooper~\cite{isoconfigurational_prl_2004}. They showed that the spatial distribution of particle mobility in supercooled liquids is reproducible from the initial configuration, and that short-time fluctuations (e.g., Debye–Waller factors) can predict long-time dynamics~\cite{wid06}. These findings provide direct evidence that dynamic heterogeneity is structurally encoded.

\begin{figure*}[hbt!]
\centering
\includegraphics[width=\linewidth]{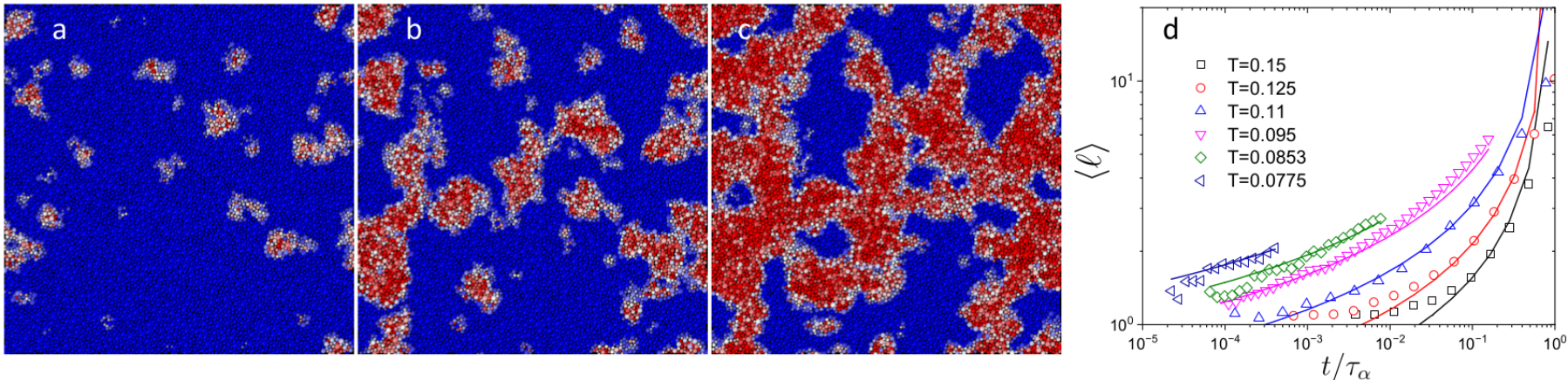}
\caption{ \textbf{a--c}: Rearranging regions (in red) at successive times in a molecular dynamics (MD) simulation of a supercooled liquid \cite{gui22}. Reproduced from: arXiv:2103.01569. 
\textbf{d}: From \cite{tah23}. Comparison between MD results (symbols) for the coarsening length $\ell_{\rm c}(t,T)$ in Ref.~\cite{sca22} and the theoretical prediction based on interacting excitations (curves) of Eq.~(\ref{eq:ellc}) presented in Sec.~\ref{S8} below. }
\label{figglass1}
\end{figure*}

Since then increasingly refined computer simulations have confirmed dynamic heterogeneity as a striking feature of both two- and three-dimensional viscous liquids.  As the glass transition temperature $T_{\rm g}$ is approached, the dynamics become heterogeneous over a correlation length $\xi$ that grows upon cooling \cite{wee00,lac03,ber05,wid06,hur95,kob97,yam98,dal07,kar14}.  At fixed temperature, spatial maps showing the regions that have already relaxed reveal a coarsening process characterized by a length scale $\ell_{\rm c}(t,T)$ \cite{cha10,hed09,gui22}, illustrated in Fig.~\ref{figglass1}a--c.  A central question is how to predict the growth of $\ell_{\rm c}(t,T)$, quantified in Fig.~\ref{figglass1}d.  This length increases slowly with time and saturates at its maximal value $\xi$, which is reached on the relaxation time scale $\tau$.

\subsection{Thermodynamics-dynamics correlation}

\begin{figure}[hbt!]
\includegraphics[width=0.6\linewidth]{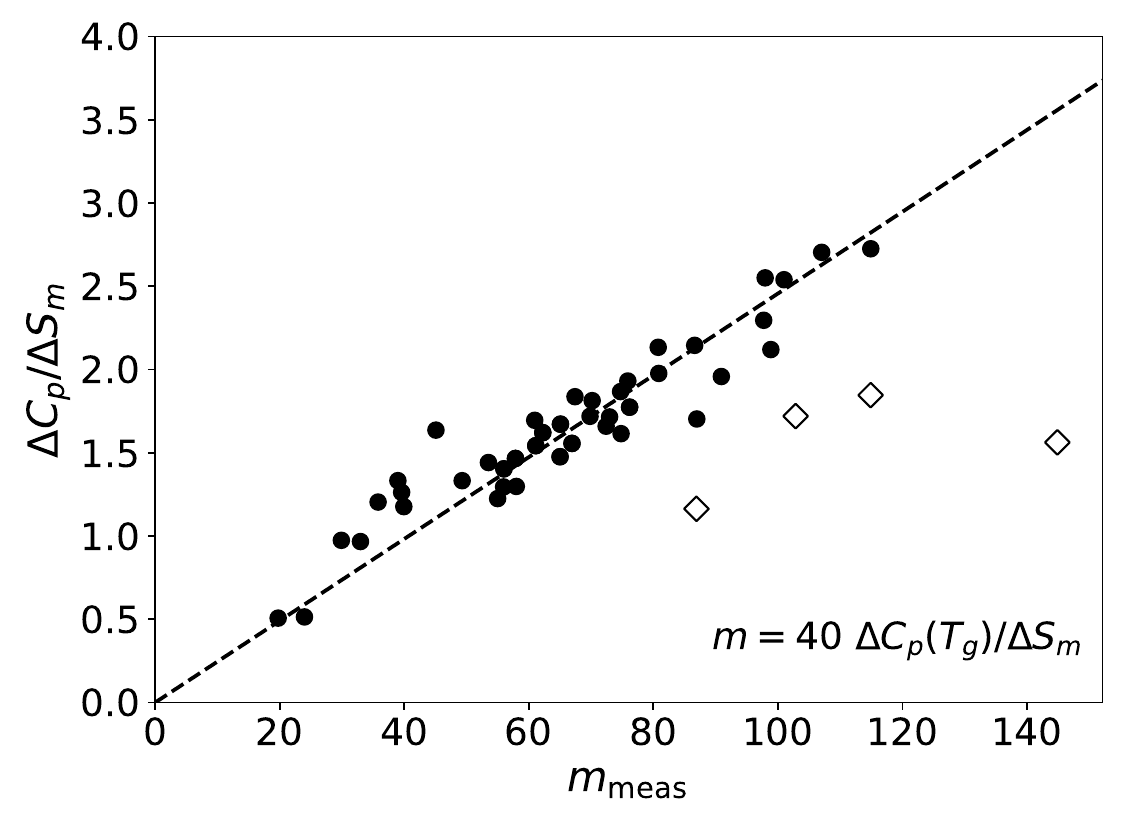}
\caption{Correlation between the fragility (\eq{eq:frag_def}) and the specific-heat jump at $T_{\rm g}$ relative to the melting entropy, $\Delta C_p/\Delta S_m$, for 53 non-polymeric glass-forming liquids (redrawn from Ref. \onlinecite{wan06}). The dashed line marks the empirical relation $m=40\Delta C_p/\Delta S_m$. The open symbols are selenium, toluene, triphenylphosphite (TPP), and decalin (decahydronaphthalene); many mono-alcohols and polymers also deviate from the line (data not shown).}
\label{figthermo}
\end{figure}

Another important observation is the correlation between dynamics and thermodynamics \cite{ang95,boh98a,mar01,alb22,nov22,alb23,mei24,loi25}, which was noted in a qualitative sense already by Kauzmann in his famous 1948 review \cite{kau48}. The correlation is illustrated in the inset of Angell's famous fragility plot (\fig{fig:angell}) and since then confirmed in many experiments \cite{nov22}. All glass-forming liquids show an apparent heat-capacity jump at the glass transition -- i.e, a sensible drop over a finite temperature interval--, reflecting kinetic arrest: the system falls out of equilibrium and remains trapped in a metastable basin rather than continuing to explore lower-energy states. In relative terms, the jump is larger the more fragile the liquid is. \Fig{figthermo} presents data for several nonpolymeric liquids, illustrating the correlation in a plot for which the specific-heat jump is measured relative to the entropy of fusion.

Note that, despite the convincing data of \fig{figthermo}, some systems do not follow the correlation marked by the dashed line. Thus mono-alcohols generally have only moderate fragility but a quite large specific heat jump \cite{boh14a} while, on the other hand, many polymers have fragility above 100 and only moderate specific-heat jumps \cite{nov22}.

\subsection{Effect of kinetic rules and swap algorithms}

Many modern numerical models of liquids are polydisperse, allowing one to efficiently use the `swap' algorithm where pairs of particles are exchanged \cite{Glandt84,gutierrez2015static,nin17,bri18} to reach thermal equilibrium on a range of temperatures comparable to laboratory experiments. Such models are currently receiving considerable attention \cite{nin17, ber23, gui22, sca22, ber19,simon2026molecular}, as they are fairly easy to simulate and capture the hallmarks of the glass transition.

The fact that swap algorithms can accelerate the dynamics by many orders of magnitude is highly informative about the nature of relaxation near the glass transition (at least in polydisperse systems). In particular, it provides a stringent test for theories in which dynamical slowing down is controlled by growing thermodynamic order (discussed below). In such theories, any dynamics based on local kinetic rules \footnote{Although swapping two particles appears at first sight as a non-local move, swap dynamics is equivalent to a purely local dynamics in which particles can effectively breathe by adapting their radius according to a chemical potential~\cite{Glandt84,bri18}.} should lead to comparable relaxation times~\cite{wya17}, up to pre-asymptotic effects that these approaches do not aim to describe.

\begin{figure}\begin{center}
      \includegraphics[width=7 cm]{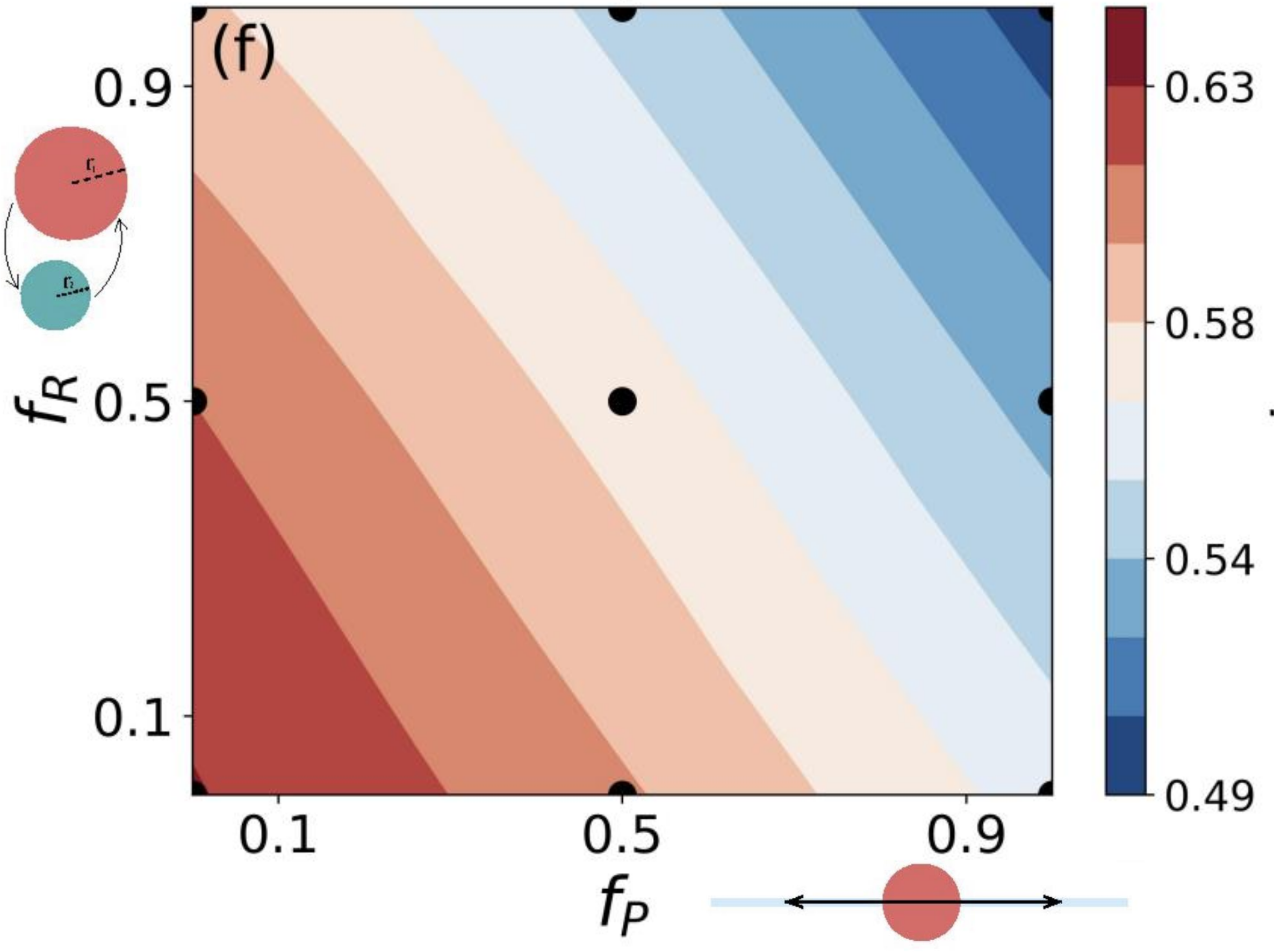}
\end{center}
        \caption{ {Kinetic rules greatly affect the glass transition:} the glass transition packing fraction of a poly-disperse hard sphere system is indicated in color as a function of both the fraction $f_P$ of particles whose motions are restricted on arbitrary planes and the fraction $f_R$ of particles not allowed to swap. All these different kinetic rules preserve thermal equilibrium and lead to identical static properties. Yet, gigantic difference in the time (and also length, see Section \ref{S8}) scales characterizing relaxation appears, which cannot be explained within theories based on thermodynamic properties alone. The normal kinetic rule is just one point in this diagram, $f_P=0,f_R=1$, which these theories do not distinguish from other points. From \cite{gav24}, with permission.  \label{swap} }
\end{figure}

This expectation is, however, strongly violated. As illustrated in Fig.~\ref{swap}, the location of the glass transition can be shifted dramatically by controlling the fraction of particles allowed to swap~\cite{gop22,gav24}, or by introducing kinetic constraints. From this perspective, standard dynamics corresponds to a single point in a broader space of admissible local kinetic rules. This point exhibits an intermediate degree of dynamical slowing, whereas nearby choices of rules can lead to much faster or much slower relaxation. Theories that do not distinguish this special point from the rest of this space therefore lack predictive power for the actual dynamics of glass-forming systems.

\section{Measuring the activation energy}
\label{S2bis}
\subsection{Status of the field}
A challenge in the field of the glass transition is that one of its central quantities, the activation energy, is hard to measure. Indeed, according to the standard rate theory, the relaxation time follows
\begin{equation}
\tau = \tau_{\rm vib} \exp \left(\frac{E_a - T S_a}{k_B T}\right)
       = t_0 \exp\left( \frac{E_a}{k_B T} \right),
       \label{entro}
\end{equation}
where $\tau_{vib}$ is a microscopic, vibrational time scale of order of picoseconds and $S_a$ is the barrier entropy that is largely unknown. It contains vibrational terms associated with the curvature (i.e., the Hessian spectrum) of saddles, as well as a term associated with the number of paths connecting one meta-stable state to another specific one (see further discussion in Sec.\ref{S4}.E). Since $t_0=t_{\rm vib} \exp(-S_a/k_B)$, $t_0$ itself is not known, and $E_a$ cannot be directly obtained from a measurement of $\tau$.

In the absence of knowledge of $t_0$,  the activation energy can only be known up to a term linear in temperature -- corresponding to a considerable indeterminacy, that hinders tests of theories. Any linear temperature dependence of the activation energy can be absorbed into a different estimate of the microscopic time~\cite{dyr95a,Struik1997}. Indeed,
\begin{equation}
\tau = t_0 \, e^{B/k_B} \, \exp\left(\frac{E_a(T) - B T}{k_B T}\right)
\end{equation}
does not depend on $B$. 

In the same spirit, the so-called apparent activation energy (the tangent slope in a plot of the logarithm of the relaxation time versus inverse temperature) ~\cite{Coslovich2018}
\begin{equation}
E_a^{\mathrm{app}}(T) = -T^2 \, \frac{d\ln \tau}{dT} = E_a - T \, \frac{dE_a}{dT},
\end{equation}
grossly overestimates the true activation barrier \cite{dyr95a}. Moreover, this quantity does not depend on $T$ if $E_a$ is an affine function of temperature, which corresponds to straight lines in Angell’s fragility plot. Such straight lines often occur near the glass transition and are sometimes misinterpreted as signaling ``fragile-to-strong behavior''. As we will now illustrate, the activation energy can still increase upon cooling in these situations, in an affine fashion.

\subsection{Fast heating or quenching to measure $E_a$ and $t_0$}
\begin{figure}[t!]
\centering
\includegraphics[width=0.8\linewidth]{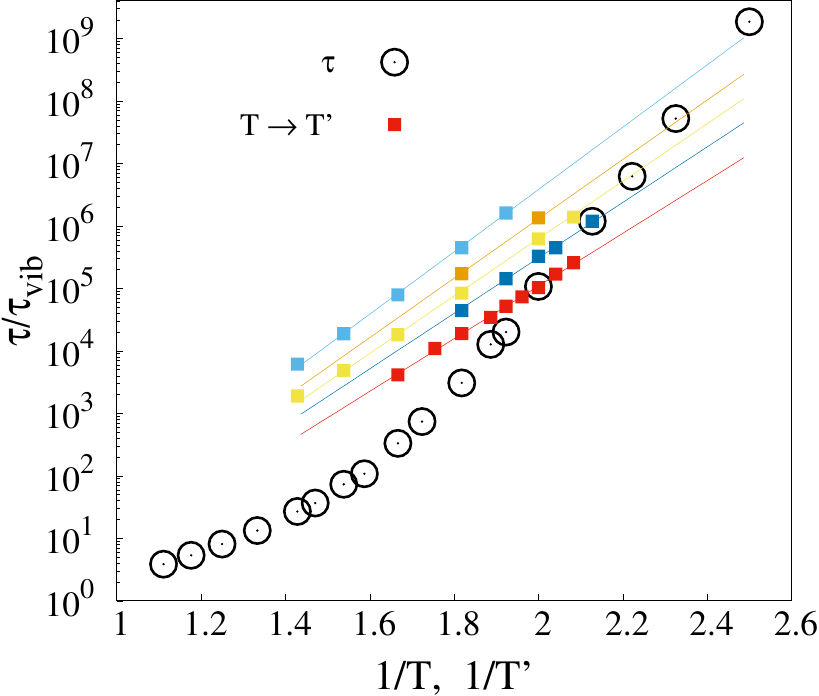}
\caption{
Inverse temperature dependence of the relaxation time $\tau$, normalized by the vibrational timescale, for the model considered in Ref.~\cite{cia24} (circles). 
Squares illustrate the $T'$ dependence of the relaxation time $\tau(T \to T')$ during a reheating procedure as $T'$ is varied, with the colors reflecting the initial $T$ values; for example, red squares correspond to $1/T=2$. Lines are fit to $\tau(T \to T') = t_0 e^{E_a(T)/T'}$, from which $t_0(T)$ and $E_a(T)$ can be extracted. These fits reveal that $t_0(T)$ is temperature-independent, within errors, while $E_a(T)$ increases with cooling. 
}
\label{fig:heating}
\end{figure}

An approach to estimating $t_0$ and $E_a(T)$ through a reheating procedure has recently been proposed in Ref.~\cite{cia24}. 
This method posits that, following an instantaneous temperature change $T \!\to\! T'$, the relaxation time conforms to an Arrhenius relation:
\begin{equation}\label{tau_act_en}
\tau(T \to T') = t_0 \exp\!\left(\frac{E_a(T)}{T'}\right)
\end{equation}
from which both $t_0$ and $E_a(T)$ can be extracted by varying $T'$. In an Angell plot, varying $T'$ leads to a line of slope $E_a$ and offset $t_0$, as exemplified in Fig.~\ref{fig:heating}. The underlying assumption that $E_a$ does not significantly evolve during the reheating-induced relaxation is expected to hold when $T'$ is not substantially larger than $T$. For larger temperature jumps, relaxation may proceed via the nucleation of a hot-liquid region invading the system~\cite{Berthier2020,Mehri2022}. This assumption can be verified directly by checking whether $\tau(T \!\to\! T')$ exhibits an Arrhenius dependence on $1/T'$. For the model system considered in Ref.~\cite{cia24}, a polydisperse systems of particles interacting via inverse-power law potentials (poly-IPL10), this approximation holds, as illustrated by the lines obtained in Fig.~\ref{fig:heating}. 

In the poly-IPL10 model, we find that $t_0(T) \simeq 2.4\cdot10^{-3} t_{\rm vib}$ does not vary noticeably with temperature.  Here, $t_{\rm vib}$ is the vibrational time, operatively defined as the time the self-scattering function at a wavevector corresponding to the first peak of the structure factor reaches its plateau value at the critical temperature. From this measurement, we get from Eq.\ref{entro} that $\frac{S_a}{k_b} = -\log\left(\frac{t_0}{t_{\rm vib}}\right) \simeq 6$.

\begin{figure}[t!]
\centering
\includegraphics[width=0.8\linewidth]{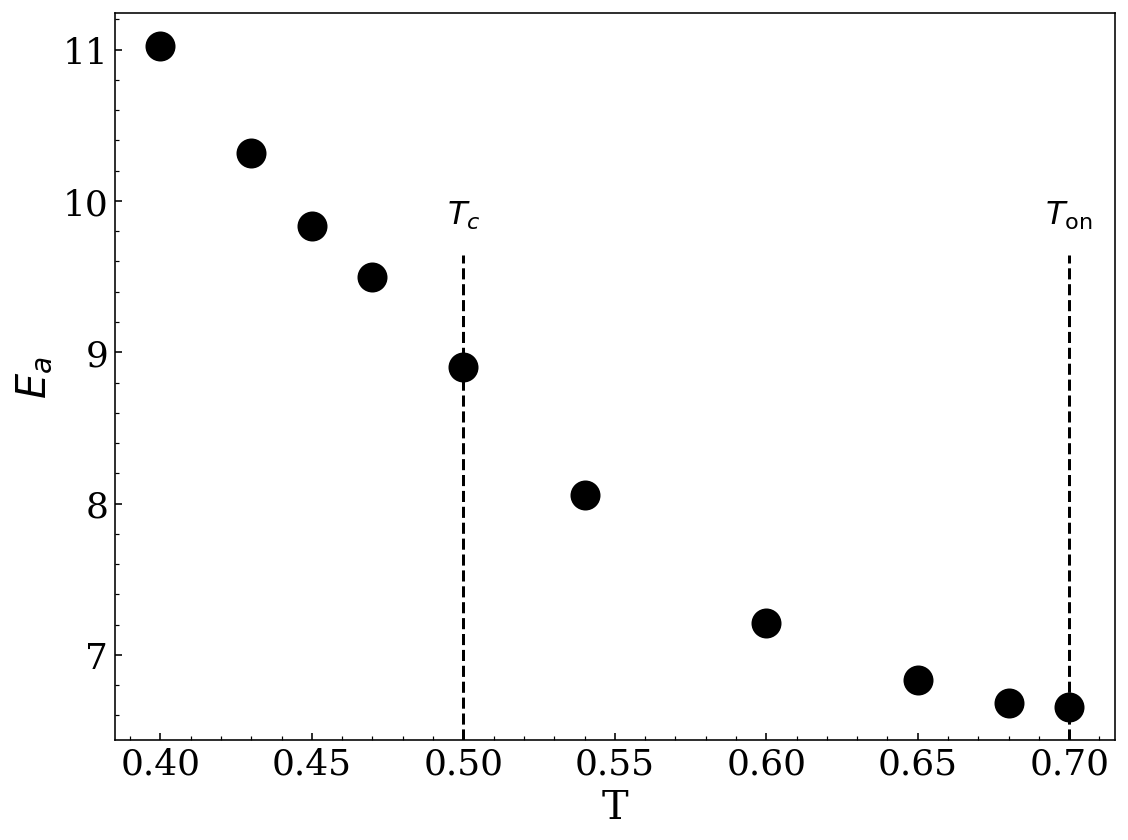}
\caption{
\label{fig:Ea_measurement}
Temperature dependence of the activation energy $E_a = T \log(\tau/t_0)$ (\eq{eq:act_en}) for the model considered in Ref.~\cite{cia24}. This measurement relies on that of $t_0$, which has been obtained via the re-heating procedure detailed in Fig.~\ref{fig:heating}. As a reference, using traditional methods in this liquid, the mode-coupling temperature $T_c$ is approximately $0.53$, while the onset temperature is $0.7$.} 
\end{figure}

Finally, the estimation of $t_0$ allows one to compute the activation energy from $E_a = T \log(\tau/t_0)$. Figure~\ref{fig:Ea_measurement} shows that $E_a(T)$ increases upon cooling, as expected for fragile liquids. Note that the dependence of activation energy at low $T$ is mostly affine. Thus, it cannot be properly estimated without knowledge of $t_0$, and is not captured by the apparent activation energy.

{\bf Experimental implementation}
As will be made clear in this review, measuring $t_0$ (instead of fitting it as most theories do) is key to make stringent test of theories, and this is very much missing in the current state of the field. A practical limitation for implementing this protocol experimentally is the competing requirement of achieving a sufficiently fast temperature step while maintaining adequate measurement signal. The relevant time scale for a ``sudden'' temperature change is the structural relaxation time $\tau_\alpha(T)$, which near the glass transition is of order $10^2$--$10^4$\,s; thus, a temperature jump occurring within $0.1$--$1$\,s is effectively instantaneous for the dynamics we aim to probe. Such time scales have been comfortably achieved in several set-ups where re-heating can occur in a few tens of milliseconds \cite{henot2024crossing, hec19}. Measuring $t_0$ so as to extract the activation energy at all temperatures from the relaxation time scale thus appears achievable.

Note that although the assumption of \eq{tau_act_en} that there is no change of activation energy on time scales much shorter than the main relaxation time is expected to hold in constant-volume experiments,  in constant-pressure experiments the volume will instantaneously increase \cite{hec19}.  This will change the activation energy by some $\Delta E_a=\partial E_a/\partial V|^{Glass}_T \Delta V$, where the index $Glass$ indicates that the derivative is computed in the glass and thus out-of-equilibrium, and $\Delta V$ is the volume change. In a set-up where the pressure can be controlled and changed and the sample volume measured, $\partial E_a/\partial V|^{Glass}_T$ can be obtained by changing the pressure to some $P'$ and measuring the instantaneous relaxation time and volume change, since $\tau(P')/\tau(P)=\exp(\Delta E_a/k_B T)$. From such measures, the change of activation energy can be removed, so that the absolute activation energy and $t_0$ can be extracted.

\section{Fragility and growing length scales}

We consider here theories seeking to connect fragility and dynamical heterogeneities, and for simplicity focus the discussion on these two observables. Needless to say, some of these theories make additional predictions, such as the shape of linear-response spectra, aging,  the Boson peak, the properties of glass-forming liquids under high pressure or confinement, etc., that are discussed, e.g., in the reviews referred to in the Introduction (Sec. I).

\label{S3}
\subsection{Adam-Gibbs and entropy crisis}
In the Adam-Gibbs model from 1965 \cite{ada65}, the starting point is that the decrease of entropy upon cooling necessarily implies fewer relevant states. A transition must obviously involve at least two states. As the number of states decreases upon cooling, the states differ more and more, in which case it makes good sense to assume that the energy barrier between the states increases. Assuming \textit{ad hoc} that the activation energy is proportional to the minimal volume corresponding to two states, the Adam-Gibbs model results in the prediction \cite{ada65}

\be\label{eq:Adam_Gibbs}
\tau(T)\sim\,\exp\left(\frac{A}{TS_c(T)}\right)\,.
\ee
where $A$ is a constant, and $S_c$ is the configurational (inherent) entropy per particle. The fact that the latter decreases as $T$ decreases leads to super-Arrhenius behavior. An ``entropy crisis" occurs if $S_c$ vanishes, leading to a divergence of the relaxation time scale. 

\subsection{Random First Order Transition} 
While the classical Adam--Gibbs model makes several \textit{ad hoc} assumptions \cite{dyr09}, the more recent random first-order transition (RFOT) theory provided a firmer theoretical basis for the central idea that entropy controls relaxation \cite{lub07,wol12,bir23}. RFOT is a broad framework containing many ingredients, some of which have been proven correct in infinite dimensions \cite{parisibog}. Later in this review we argue that certain RFOT ideas are central for predicting the architecture of local barriers in two or three dimensions. In this subsection, however, we focus on the RFOT description of activated dynamics near the glass transition.

Within RFOT, the configurational entropy vanishes at a temperature $T_K$, the so-called Kauzmann temperature \cite{kau48}. If equilibrium were attainable, the system would reach an ``ideal glass'' with sub-extensive entropy at $T_K$. As the entropy decreases, configurations become increasingly correlated over a growing static length scale $\xi$. This length is obtained by balancing the configurational entropy on scale $\xi$, i.e.\,$\propto S_c \xi^d$, with a surface tension term proportional to $Y(T)\xi^{d-1}$ (see \cite{lub07} for generalizations), where $Y(T)$ denotes a surface tension. In this picture, the glass is a mosaic of inherent states tiled on the scale $\xi$. Assuming that the activation energy grows as a power of $\xi$ leads to

\begin{equation}\label{eq:RFOT}
\tau(T)=t_0 \exp\!\left[\left(\frac{Y(T)}{T S_c(T)}\right)^\alpha\right],
\end{equation}
where $\alpha$ is an exponent sometimes treated as a fitting parameter \cite{bir23}  (this is not universally accepted \cite{lub07} as $\alpha=1$ has been advocated for).

RFOT is appealing because it appears to resolve several issues simultaneously \cite{lub07,wol12,bir23}: (i) Eq. (\ref{eq:RFOT}) with $\alpha=1$ reproduces the classical Vogel--Fulcher form for the temperature dependence of viscosity, historically interpreted to diverge; (ii) RFOT predicts that dynamics correlate over the mosaic length $\xi$. $\xi$ can be measured using point-to-set correlations \cite{bou04,bir08} or estimated via non-linear susceptibilities \cite{Berthier2005} and indeed increases as temperature decreases; and (iii) RFOT yields a correlation between the jump in specific heat and the fragility. Also notable, but only tangentially related to this review, are recent successes in quantitatively predicting \cite{cha14,fra15} the marginal properties of sphere packing near jamming \cite{wya05,wya12,ler13}. 

\begin{figure}\begin{center}
      \includegraphics[width=8.5 cm]{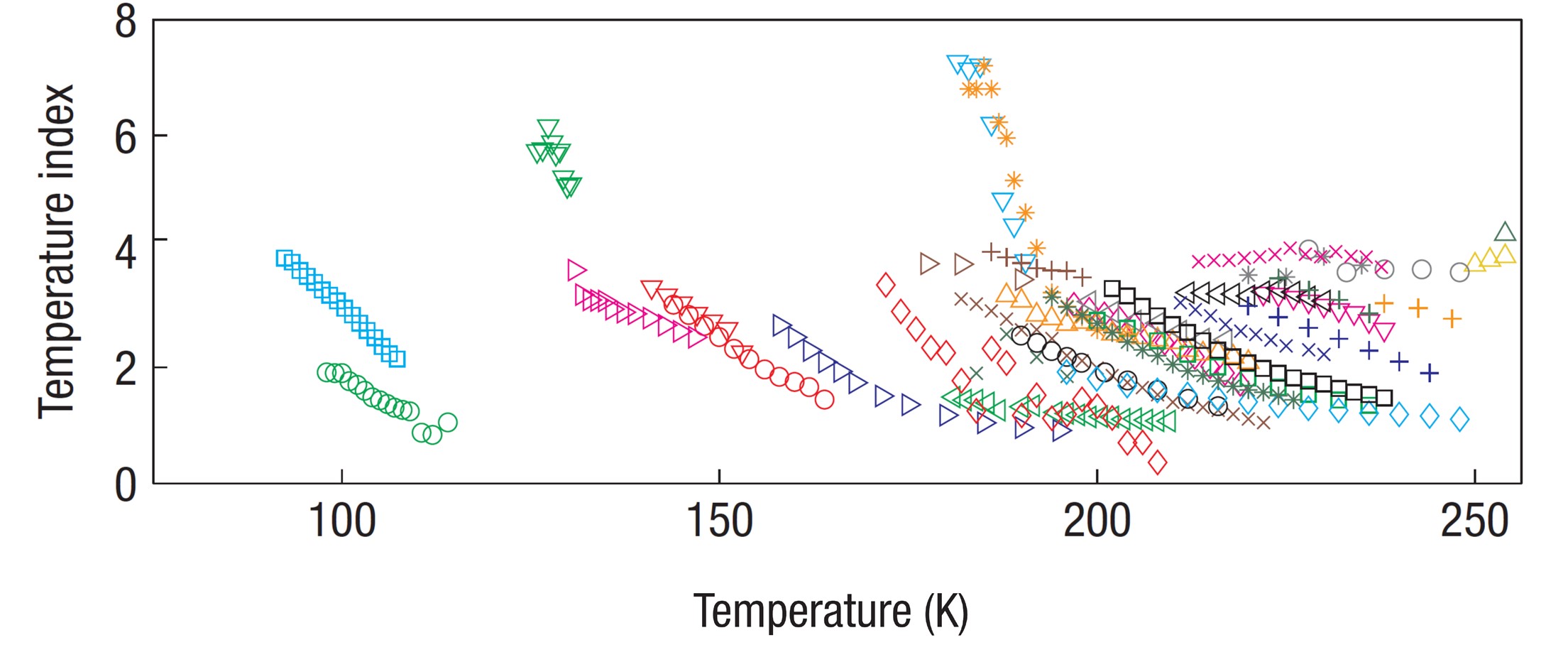}
\end{center}
        \caption{Temperature variation of the temperature index (\eq{eq:ti}) of the dielectric loss-peak frequency of 42 organic glass-forming liquids (reproduced from Ref. \onlinecite{hec08}, with permission). There is no indication that the index of any liquid diverges at a finite temperature. \label{fig:hec08} }
\end{figure}

In our view, however, the experimental and simulation support for these points remains limited. Regarding (i), no unambiguous experimental signature of a finite-temperature divergence of the relaxation time has been established. Dielectric data for many organic liquids have been analyzed in detail \cite{hec08}. The resulting temperature index $I$ (Eq. (\ref{eq:ti})) for 42 liquids is shown in Fig.~\ref{fig:hec08}. Although $I$ generally does increase upon cooling, which is not trivial, the data show no indication of a finite-temperature divergence of the relaxation time as would occur if $E_a \to \infty$. This conclusion was further reinforced by aging experiments on amber \cite{zha13}.

Regarding (ii), both earlier \cite{hed09} and more recent \cite{gui22} observations demonstrate that facilitation---where mobile regions induce mobility nearby -- plays a major role in dynamical heterogeneities. In response, some proponents of RFOT have incorporated facilitation into the framework, effectively generating correlations on a length scale larger than $\xi$ \cite{bir23,bou24} (see also \cite{Mai05,Mai08,bha08,char13}). This comes at a  cost, however, because by adding additional ingredients RFOT becomes less predictive.

In relation to (iii), although a correlation between dynamics and thermodynamics is empirically well established, such a correlation is generically expected. Indeed, if the jump in specific heat vanishes, the system continues to sample configurations of nearly identical inherent energy as temperature varies; it is thus unsurprising that activation barriers remain unchanged. For example in elastic models discussed below, where the high-frequency (plateau) shear modulus $G_\infty$ sets the activation energy, fragility scales with $\partial G_\infty/\partial T$, which in turn decomposes as
\[
\frac{\partial G_\infty}{\partial T}=\frac{\partial G_\infty}{\partial E}\,\frac{\partial E}{\partial T}
= \frac{\partial G_\infty}{\partial E} C_p,
\]
where $E$ is the inherent structure energy. This dependence on $C_p$ naturally leads to the observed correlations between thermodynamics and dynamics.

Taken together with numerical results in polydisperse systems---showing that kinetic rules strongly affect the dynamics but not the thermodynamics, and that the evolution of local barriers tracks that of the activation energy---these observations indicate that an alternative description is needed (at least for some materials).

\subsection{Locally favored structures}
Another influential theory of the glass transition, also assuming that thermodynamic properties of liquids control their dynamics, argues that the emergence of \emph{locally favoured structures} (LFS) is central.  LFS are short-range motifs that are energetically preferred in the liquid state but geometrically incompatible with global crystalline order~\cite{royall2015role,malins2013identification,coslovich2007characterization}. Being energetically stable, these structures are slow to relax and thought to underpin dynamic heterogeneity.

In some systems, such as metallic glasses or model binary mixtures, LFS correspond to well-defined local geometries—most notably icosahedral or fivefold-symmetric motifs. Bernal and later Nelson~\cite{nelson1983order} proposed that icosahedral order, while energetically favourable, is frustrated in Euclidean space, leading to the formation of finite domains that cannot grow indefinitely. This \emph{geometrical frustration} has been supported by experimental observations in supercooled metallic liquids~\cite{kelton2003icosahedral,shen2009icosahedral}, and by simulations showing enhanced populations of icosahedral motifs in slow regions of the liquid~\cite{coslovich2007characterization,malins2013identification}. These motifs often form extended networks or medium-range domains that appear to anchor slow dynamics~\cite{leocmach2012roles}.

However, the identification of LFS is not universal, and there may be many favored structures \cite{wei19}. Moreover, in some systems the structures that correlate with slow dynamics are not obviously energetically favored, nor do they correspond to a single geometric motif. This has led to the development of diverse structural metrics that aim to capture the relevant local environments. The multiplicity of approaches reflects the fact that the connection between structure and dynamics is system-dependent and often subtle.

Machine learning has recently emerged as a powerful tool to uncover hidden structural features that correlate with dynamics. Schoenholz et al.~\cite{schoenholz2016softness} introduced the concept of \emph{softness}, a learned structural quantity that predicts local rearrangements. Evidence supports that softness correlates with local activation energy \cite{rid24}. Boattini et al.~\cite{boattini2020,boattini2021averaging} extended this approach using unsupervised learning to identify structural heterogeneities from a single configuration. More recent studies~\cite{bapst2020gnn,alkemade2022mlcomparison,Jung2025} have benchmarked graph neural networks and simpler regression schemes, showing that physically informed descriptors often perform comparably to deep-learning models. Interesting generative models have also emerged \cite{mad26} to predict dynamical heterogeneities. These approaches support the relevance of LFS and reveal new structural fields that affect dynamics.

While correlations between local structure and dynamics are well established, the challenge is to quantitatively relate them to the slowdown of dynamics near the glass transition. The \emph{Frustration-Limited Domain} (FLD) theory~\cite{tarjus2005frustration} provides a conceptual framework that links geometrical frustration to dynamic scaling. It posits that LFS form domains whose growth is limited by frustration, leading to cooperative but spatially constrained dynamics. The theory predicts an avoided critical point at a temperature below which domain growth saturates~\cite{tarjus2000fld}. Similar ideas have been proposed by Tanaka and collaborators, who argue that dynamic heterogeneity is governed by a growing structural correlation length $\xi$ associated with orientational order~\cite{Tanaka2010b,rus18,tan24,ish25,tan12}, leading to activation barrier $E_a(T)$ scaling as $\xi^{d/2}$. Both approaches suggest that the spatial extent of structural correlations plays a key role in glassy dynamics.

In  some specifically chosen systems—such as two-dimensional liquids with hexatic order—large structural lengths have been observed that strongly correlate with dynamical heterogeneities ~\cite{Russo2015,tong2018hidden}. However, a persistent challenge for generic liquids is the apparent mismatch between structural and dynamic length scales. Simulations and experiments typically find that LFS domains span only $ 2{-}3\,\sigma$, with $\sigma$ the typical interparticle distance, while dynamic heterogeneities extend over $5{-}10\,\sigma $~\cite{leocmach2012roles,boattini2021averaging,lan25}. Moreover, as any thermodynamic-based theory, LFS does not explain why in some liquids the choice of kinetic rules drastically affects the length of dynamical correlation, while keeping any thermodynamic length unchanged \cite{gav24}. We return to the connection between LFS and local barriers in the discussion section. 

\subsection{Dynamical or mode-coupling transition}
We discuss here theoretical concepts that have been suggested to influence the dynamics of liquids at temperatures well above $T_{\rm g}$. Specifically, mean-field approaches predict a dynamical or mode-coupling transition. A recurring objection to such theories is that these  mean-field approximations do not survive in three-dimensional liquids. In particular, the absence of the true divergence of the relaxation time they predict, and the apparent crossover nature of the mode-coupling transition are frequently taken to imply that such concepts are irrelevant for the relaxation mechanisms operative near the glass transition. Although we agree that power-law fits of the relaxation time have little justification, we argue below that a dynamical transition affects local barriers, and thus remains highly relevant for understanding fragility in glass-forming liquids.\\
\\
We briefly recall the key features of landscape-based and mean-field descriptions of glassy dynamics. Goldstein~\cite{gol69} proposed the existence of an onset temperature $T_0$, such that for $T<T_0$ relaxation proceeds via activated barrier-crossing events, while for $T>T_0$ the system predominantly explores regions of the (free-) energy landscape near saddles. Closely related ideas emerged from mean-field, high-dimensional treatments of the glass transition, initially motivated by analogies with disordered spin models and later extended to realistic models of high-dimensional liquids~\cite{kir87,bou96,lub07,ber11,fra15_quasi,mai16a}. Within this framework, a dynamical transition occurs at a temperature $T_c$ with several distinctive features:

\begin{itemize}
 \item The relaxation time and viscosity diverge as a power law at $T_{\rm c}$, consistent with the predictions of mode-coupling theory (MCT)~\cite{got99}, which describes how the relaxation of density fluctuations becomes increasingly hindered upon cooling due to nonlinear feedback among different modes of motion.
\item The spectrum of the Hessian of the free-energy landscape is a semi circle that becomes stable below $T_{\rm c}$. The temperature $T_{\rm c}$ thus marks an elastic instability above which vibrational modes are unstable~\cite{bou96,fra15,mai16}.  
\item A finite plateau shear modulus appears for $T<T_{\rm c}$ and increases upon further cooling~\cite{lub07}. 
\item The dynamics is characterized by a length scale that diverges as $T$ approaches $T_{\rm c}$~\cite{fra00,bir06,franz2011field,don02} from both above and below.   
\end{itemize}

These predictions have limited success in describing the dynamics of three-dimensional liquids. Most importantly, the relaxation time does not diverge as a power-law. This fact is generally rationalized by noting that in three dimensions, hopping processes must exist that render the relaxation time finite: the dynamical transition is at best a crossover, and as a consequence $T_c$ is not very well-defined. In fact, fitting the relaxation time by a power-law divergence leads to an estimate of $T_c$ where nothing special occurs as far as activation is concerned, as illustrated in Fig. \ref{fig:Ea_measurement}. Also problematic, the nature of ``hopping processes'' remains unclear, except in fine-tuned models or very large spatial dimension \cite{cha14b}. Finally, describing the transition as a crossover does not explain why the observed dynamical length scale varies monotonically with temperature, rather than growing and then decreasing upon cooling, as predicted by MCT.

Despite these shortcomings, several observations support that (i) if an elastic instability does occur in amorphous materials, it is quite well captured by mean-field approaches describing a dynamical transition. (ii) The notion that an elastic instability does occur in some liquids at intermediate temperatures explains several important facts.  

Concerning (i), an elastic instability can be investigated in disordered three-dimensional spring networks under compression~\cite{ler14b}, in polydisperse systems using swap dynamics~\cite{ji20}, and theoretically within effective medium theory~\cite{deg14_pnas}. In these systems, the Hessian spectrum exhibits soft modes that stiffen upon entering the stable elastic phase, while the shear modulus increases and the length scale below which homogeneous elasticity breaks down decreases. These observations are very similar, sometimes quantitatively, to mean-field predictions. 

Concerning (ii), rapidly quenched soft or hard-sphere glasses often display  vibrational modes  that are marginally stable~\cite{bri06,bri09}. The fact that such systems lie close to an elastic instability suggests that the latter affects the dynamics. Another support for this notion comes from swap algorithms. Real-space~\cite{bri18}, mean-field~\cite{ike19}, and mode-coupling~\cite{sza19} analyses  predict that the dynamical transition shifts to lower temperatures when swap algorithms are used. Currently this is the only explanation for the remarkable efficiency of such algorithms.

We shall argue below that the apparent tension between mean-field predictions and three-dimensional liquid dynamics arises from a misidentification of the role of the dynamical transition. Rather than signaling a near-divergence of the relaxation time, the transition reflects the presence of an elastic instability that governs the structure and energy of the elementary excitations controlling activated dynamics. The growth of dynamical length scales then emerges from interactions among these excitations, whose sizes {\it shrink} upon cooling, in agreement with mean-field predictions.

\subsection{Kinetically constraint models}
Kinetically constrained models (KCMs) were introduced in 1984 by Fredrickson and Andersen \cite{fre84}. In this framework, a liquid contains mobile ``defects'' whose presence facilitates local relaxation events, thereby enabling surrounding defects  to move. Because these defects carry an energetic cost, their concentration decreases upon cooling and their average spacing $\xi$ increases, generating correlated dynamics on that growing length scale \cite{ton05}. Although KCMs generally display trivial thermodynamics, they exhibit highly non-trivial dynamics due to severe restrictions on the allowed transitions between configurations \cite{rit03}. Among the specific models designed to mimic realistic glass-formers, the East model is perhaps the most studied: its strongly anisotropic facilitation leads to super-Arrhenius relaxation \cite{gar02}, with an effective activation energy that grows logarithmically with $\xi$. In this picture, fragility and dynamical heterogeneities are thus tightly linked. Beyond minimal kinetically constrained models, a broader dynamic facilitation framework has been developed by Garrahan, Chandler and collaborators \cite{Gar03,cha10}. In this approach, the glass transition is viewed as a dynamical phenomenon controlled by mobility defects whose creation and annihilation obey local kinetic constraints, while thermodynamic quantities remain essentially featureless. A central idea is that space--time fluctuations, rather than static structure, encode the relevant critical behavior, and that dynamical heterogeneities arise from facilitation cascades in trajectory space. This perspective has led to a large body of analytical and numerical work, including space--time thermodynamic formalisms and large-deviation approaches to glassy dynamics. 

While dynamic facilitation successfully captures many qualitative features of heterogeneous relaxation and provides a powerful description of kinetic-rule dependence, it typically assumes that the local activation barriers themselves are either fixed or weakly temperature-dependent. In contrast, the measurements in poly-disperse numerical liquids reviewed in Sec.~III indicate that the dominant contribution to fragility arises from a systematic shift of the barrier distribution under cooling, suggesting that facilitation alone is not sufficient to account quantitatively for the temperature evolution of $E_a(T)$.

KCMs provide elegant minimal models that qualitatively reproduce facilitation and heterogeneous dynamics. However, they face several  limitations. (i) The physical nature of the putative ``defects'' in real liquids remains rather obscure (but see attempts in \cite{keys2011excitations,fra23}), and the kinetic rules invoked in models such as the East model, where the anisotropy of motion (toward ``East'') is key, are largely \textit{ad hoc}. (ii) These models predict only a weak connection between thermodynamics and dynamics \cite{bir05}. (iii) As argued below, direct measurements of local barriers indicate that the activation energy follows the magnitude of those, and is thus not controlled by a growing length scale---at least in some liquids---thereby challenging the core mechanism underlying KCM-based explanations of fragility.

\subsection{Summary}
In table \ref{tab:theory_comparison} we summarize these different frameworks, as well as those based on local barriers that are  discussed in details below.

\begin{table*}[t]
\centering
\caption{
Comparison between some theoretical frameworks for the glass transition, by order of appearance in the review. The columns summarize the proposed source of activation barriers, the role attributed to growing length scales, whether the framework makes quantitative predictions for the temperature evolution of the activation energy $E_a(T)$, and its compatibility with the strong kinetic-rule dependence observed in swap dynamics.
The last column lists representative references (reviews first when available).
}
\label{tab:theory_comparison}
\begin{tabular}{|p{2.5cm}|p{3.6cm}|p{3.4cm}|p{2.7cm}|p{2.7cm}|p{2.2cm}|}
\hline
\textbf{Theory} &
\textbf{Barrier source} &
\textbf{Role of length scale} &
\textbf{Predicts $E_a(T)$?} &
\textbf{Swap-compatible?} &
\textbf{Rep.~references} \\
\hline

RFOT &
Mosaic surface tension vs.\ configurational entropy &
Static mosaic length $\xi$ controls barrier growth &
$\left(\frac{Y(T)}{T S_c(T)}\right)^\alpha$ &
Challenged (kinetic rules strongly affect dynamics) &
\cite{lub07,wol12,bir23,kir89} \\
\hline

Locally Favored Structures &
Growth of specific structural motifs &
Structural correlation length controls slowdown and dynamical heterogeneities &
Framework specific. For orientational-order approaches $E_a\sim \xi^{d/2}$ &
Challenged (dynamical heterogeneities depend on kinetic rules, unlike structure) &
\cite{tan12,tan19,tarjus2005frustration,nelson1983order} \\
\hline

KCM &
Defect-facilitated dynamics (fixed local barriers) &
Growing defect spacing controls relaxation &
Yes (model-specific scaling laws) &
Yes (explicitly kinetic by construction) &
\cite{rit03,cha10,fre84,Gar03} \\
\hline

Shoving model &
Macroscopic shear modulus $G_\infty(T)$ &
None &
$E_a\sim G_\infty(T)$ &
Yes (elasticity depends on kinetic rules) &
\cite{dyr06,hec15a,dyr12} \\
\hline

Local elastic model &
Local linear elastic response $\kappa$ (local stiffness) &
None &
$E_a\sim \kappa$ &
Yes (elasticity depends on kinetic rules) &
\cite{dyr96,kap21,wag11} \\
\hline

Excitation-based theory &
Shift in spectrum of nonlinear local barriers &
Excitation length scale tied to dynamical transition, dynamical heterogeneities controlled by thermal avalanches &
$E_a(T)-E_a(T')=E_g(T)-E_g(T')$; absolute $E_a(T)$ empirically related to Debye--Waller factor &
Yes (excitations depend on kinetic rules) &
\cite{cia24,jic25,tah23,deg25,yan13} \\
\hline

\end{tabular}
\end{table*}

\begin{figure*}[ht!]
\centering
  \includegraphics[width = 1\textwidth]{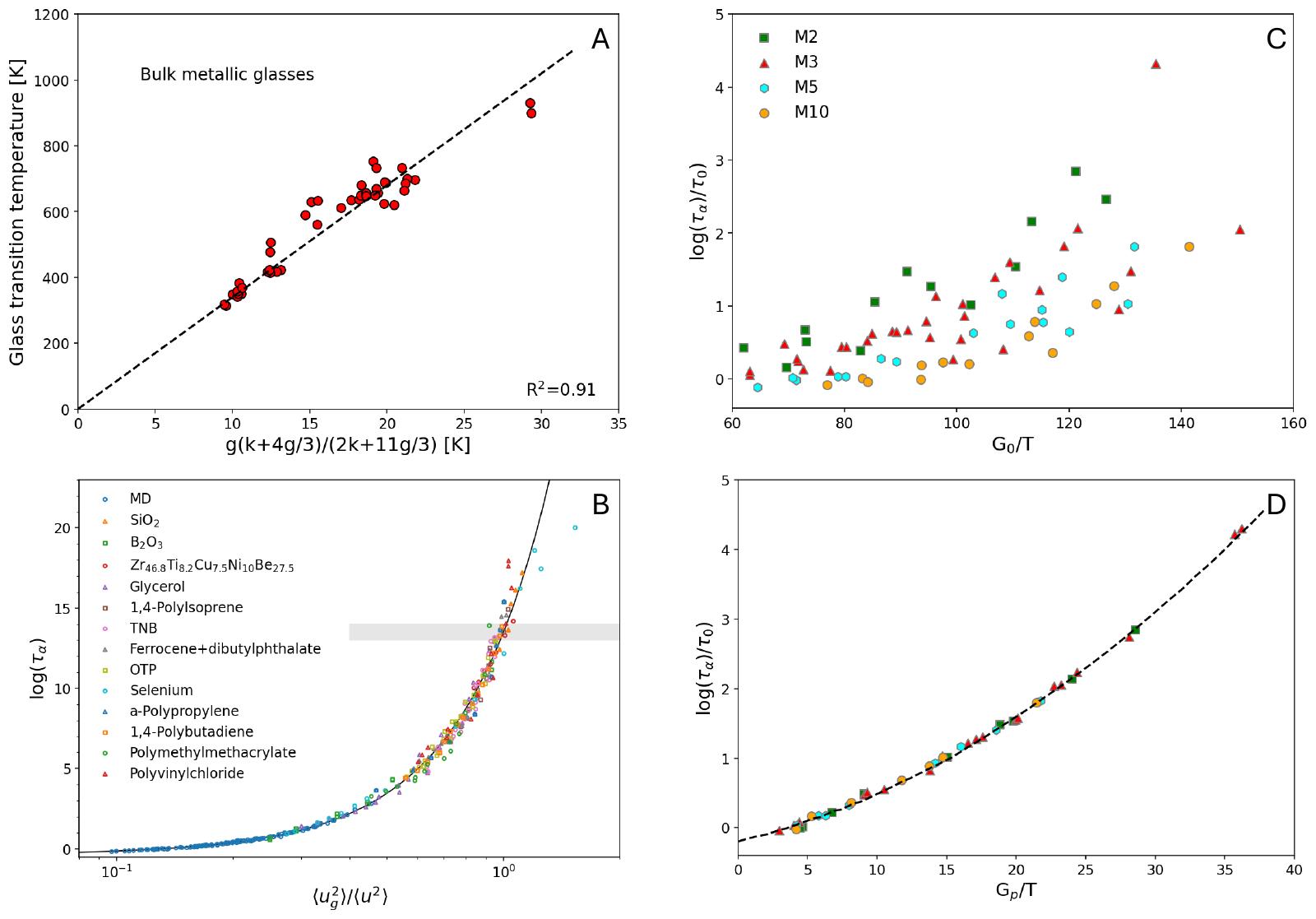}
  \vspace{-0.5cm}
  \caption{(A)  Linear correlation between the experimental glass transition temperature $T_{\rm g}$ and the $T_{\rm g}$ prediction involving the glass' bulk and shear moduli $G$ and $K$ of \eq{eq:Tg_pred} derived by assuming that the constant $C$ in \eq{eq:MSD} is a universal number times the average nearest-neighbor distance squared \cite{dyr04}. $g\propto G$ and $k\propto K$ are defined in the main text and the dashed line is the best-fit line through the origin, for 46 bulk metallic glasses. 
  (B) Experimental data for the average relaxation time relative to a typical microscopic time versus the logarithm of the vibrational mean-square displacement $\langle u^2\rangle$, the latter quantity in most cases deduced from the Debye-Waller factors of inelastic neutron scattering. The gray horizontal line marks the glass transition, the black curve represents a parabolic function of $\langle u_g^2\rangle/\langle u^2\rangle$ where subscript ``g'' denotes the value at the glass transition. Note that in this figure, time is rescaled to make curves collapse. (C,D) Simulation tests of the shoving model.  Plotting the primary ($\alpha$) relaxation time vs.~$G_0/T$ does not organize the data in any meaningful way (panel (C)), whereas plotting it vs.~$G_{\rm p}/T$ (panel (D)) leads to data collapse, see the text for further discussion.
  Panel (A) is adapted from Ref. \onlinecite{wan11a} (A), panel (B) from ~\onlinecite{lar08} (B), and panels (C) and (D) from \onlinecite{Puosi_jcp_2012}.  }
  \vspace{-0.3cm}\label{fig1e}
\end{figure*}

\section{First approach: fragility versus homogeneous  elasticity}
\label{sec:global}

\subsection{Historical perspective}
The proposition that energy barriers are controlled by the liquid's elastic properties is an old idea. For instance, Mooney in 1957 suggested that a glass-forming liquid ``not only could be but perhaps should be treated as an elastic continuum with a stress relaxation mechanism'' \cite{moo57}. Flow events take place on the nano/picosecond time scale, and elastic models are based on the conception that a glass-forming liquid's short-time properties determine its long-time relaxation. This crucially distinguishes elastic models from other models in the field, making viscous-liquid relaxation less exotic by proposing the physics is similar to that of defect diffusion in crystalline solids \cite{fly68}.  Elastic models exist in different versions but all have in common a focus on the individual flow events. This is in contrast to for instance RFOT that focuses on cooperativity. Elastic models thus ignore the question how flow events correlate, a question that, as shown below, must be addressed in any realistic model.

The two oldest and simplest elastic models are now summarized. A straightforward way of probing elastic properties is via the vibrational mean-square displacement, $\langle u^2\rangle$, which leads to the ansatz \cite{hal87,buc92,sta02,bor04,dyr04}

\be\label{eq:u2}
\tau=f(\langle u^2\rangle)\,.
\ee
Why should $\langle u^2\rangle$ have anything to do with the activation energy? If one imagines a jump between two minima, from simple Taylor expansions one expects the barrier to be higher, the higher is the curvature of the potential-energy function at the minima (for a fixed distance between the minima) \cite{hal87,dyr04}. This is an old idea that has been used, e.g., in theoretical electrochemistry. In Eq. (\ref{eq:u2}) super-Arrhenius behavior arises whenever $\langle u^2\rangle$ is not just proportional to temperature as in a harmonic solid, but decreases faster than this upon cooling, e.g., as an effect of contraction. 

\subsection{Shoving model \label{sec:showing}} A closely related elastic model is the shoving model in which the macroscopic high-frequency plateau shear modulus, $G_\infty=G_\infty(T)$, controls the relaxation time according to \cite{dyr96,dyr06,dyr06b,hec15a}

\be\label{eq:shoving}
\tau=t_0\,\exp\left({\frac{G_\infty(T) V_c}{k_BT}}\right)\,.
\ee
Here $V_c$ is a characteristic volume of order the molecular volume, a quantity that is usually taken to be temperature  independent.
$G_\infty$ increases upon cooling, giving rise to super-Arrhenius behavior. Considering the shear instead of the bulk modulus is justified by assuming that the barrier for a molecular rearrangement is lowered significantly if the density is decreased slightly by a thermal fluctuation; since the expansion of a sphere in an elastic solid results in a pure shear deformation in the surroundings ($\propto 1/r^2$ where $r$ is the distance to the flow event), the short-time (plateau) shear modulus $G_\infty$ is the relevant quantity controlling the activation energy \cite{dyr96,dyr98,dyr06,dyr06b}. Thus the main assumption of the shoving model is that the activation energy is mainly shear elastic energy located in the surroundings of the rearranging particles \cite{dyr06}. An overview of how the shoving model compares to a large amount of experimental data up to 2015 was given in Ref. \onlinecite{hec15a}. 

Although \eq{eq:u2} and \eq{eq:shoving} clearly differ, they are closely related in the simple approximation where all phonons are acoustic, with dispersion relation controlled by the macroscopic high-frequency (plateau) bulk and shear moduli. In this approximation, at least 92\% of the temperature index (\eq{eq:ti}) of $\langle u^2\rangle$ comes from that of $G_\infty(T)$ while at most 8\% derives from the short-time bulk modulus $K_\infty(T)$ \cite{dyr04}. This ``shear dominance'' reflects the facts that: i) two out of three phonons are transverse; ii) these are softer than the longitudinal phonon and thereby contribute more to $\langle u^2\rangle$; iii) the longitudinal phonon dispersion relation involves both the shear and bulk moduli. In regard to the temperature dependence, within the approximation of treating the material as a continuum \eq{eq:shoving} is thus virtually equivalent to writing \cite{dyr04}

\be\label{eq:MSD}
\tau=t_0 \exp(C/\langle u^2\rangle)\,,
\ee
which is a special case of \eq{eq:u2}. 

Suppose the constant $C$ in \eq{eq:MSD} is a universal fraction of the average nearest-neighbor distance squared. Then a prediction for $T_{\rm g}$ is arrived at if it is assumed that the vibrational MSD $\langle u^2\rangle$ can be calculated reliably from the above phonon argument involving two transverse and one longitudinal phonon for each wave vector, with dispersion relations determined by the macroscopic high-frequency bulk and shear moduli, $K_\infty$ and $G_\infty$ \cite{dyr04}. This leads to the following prediction in which the constant of proportionality is universal, $g\equiv G_\infty V_m/R$, and $k\equiv K_\infty V_m/R$ where $V_m$ the molar volume and $R$ is the gas constant) \cite{dyr04,wan11a,dyr12}
\be\label{eq:Tg_pred}
T_{\rm g}
\propto g\,\frac{k+4g/3}{2k+11g/3}\,.
\ee

\subsection{Empirical evidence}

\begin{figure}[ht!]
  \includegraphics[width = 0.3\textwidth]{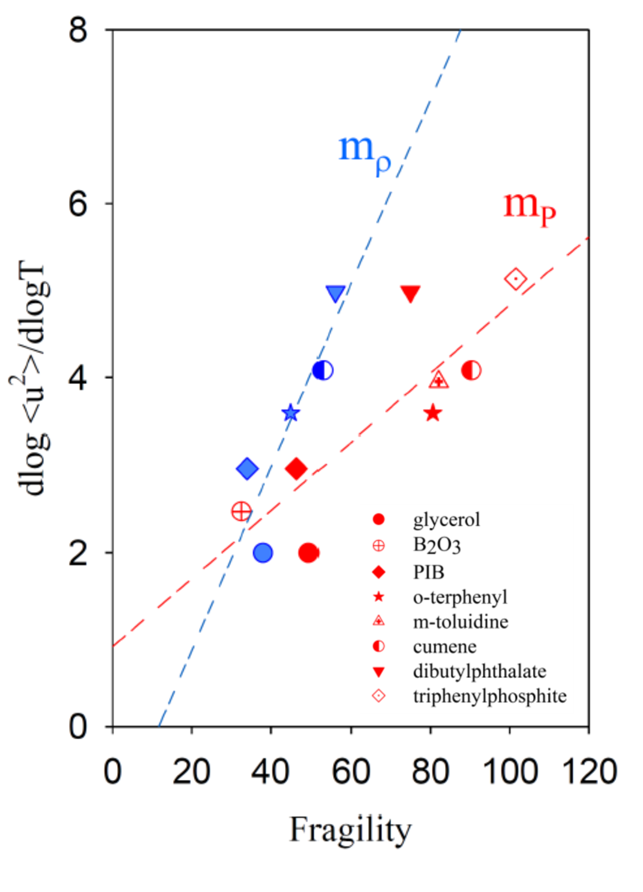}
  \vspace{-0.5cm}
  \caption{Correlation between the slope at $T_{\rm g}$ of the short-time vibrational MSD plateau measured by neutron scattering at the time 4 ns and the isobaric (red) and the isochoric (blue) fragility of different glass-forming liquids (reproduced from Ref. \onlinecite{alb23}, with permission).}
  \vspace{-0.3cm}\label{fig2e}
\end{figure}

\begin{figure*}[ht!]
\centering
  \includegraphics[width = 1\textwidth]{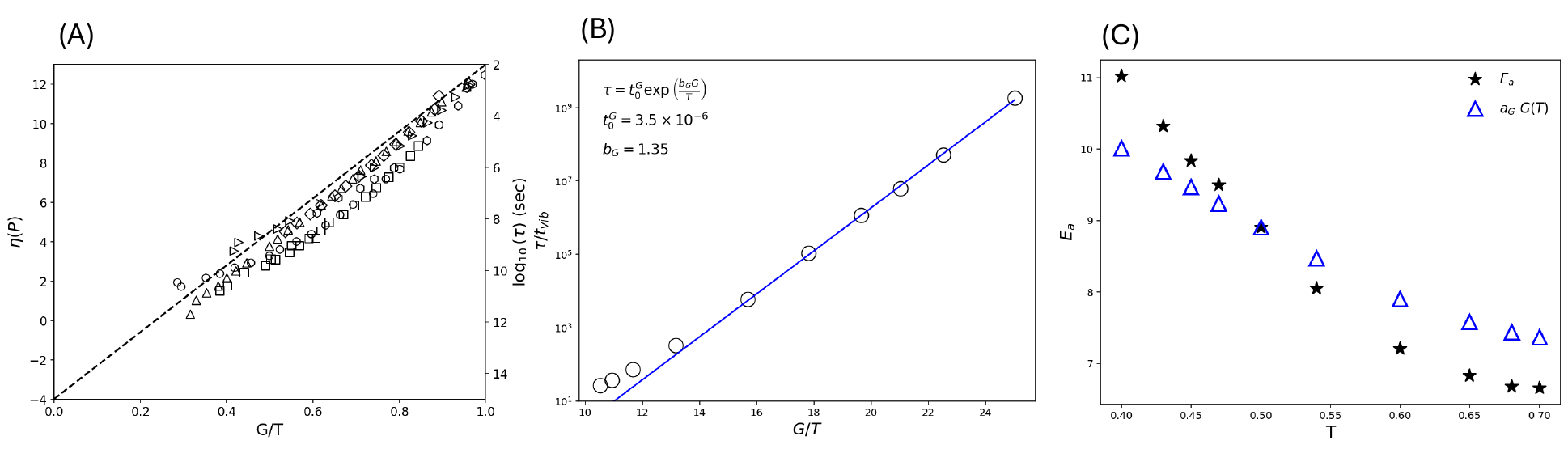}  
  \vspace{-0.5cm}
  \caption{(A) Logarithm of the viscosity (in Poise) as a function of $x\equiv G_\infty(T)/T$ normalized to unity at $T=T_{\rm g}$ for a silicone oil and five organic liquids (redrawn from Ref. \onlinecite{tor09}). The $x$ data extrapolate to a physically reasonable prefactor of same order of magnitude as the typical high-temperature viscosity of liquids at and above the melting temperature (reproduced from Ref. \onlinecite{dyr96}). (B) Numerical test of the shoving model for a polydisperse system of particles, for temperatures below the onset one. 
  The parameters $t_0^G$ and $b_G$ have been estimated via a numerical fit for temperatures below the critical one, corresponding to $G/T > 20$. $t_{\rm vib}$ correspond to the vibrational timescale. (C) The measured temperature dependence of the effective activation energy $E_a$ is compared with the predictions from the global elasticity, $a_G G(T)$. Here, $a_G\simeq 0.77$ is chosen so that the predictions are exact at the mode-coupling critical temperature $T_{\rm c} = 0.5$.   }
  \vspace{-0.3cm}\label{fig3e}
\end{figure*}

{\bf Glass transition temperature correlates with shear modulus:} \Eq{eq:Tg_pred} is tested in \fig{fig1e}(A) for 46 bulk metallic glasses, assuming that $G_\infty$ and $K_\infty$ freeze at the glass transition and therefore can be identified simply with the glass moduli, $G$ and $K$. Despite the many simplifying assumptions, there is a good overall agreement with \eq{eq:Tg_pred} with the same constant of proportionality for all glasses. Plotting instead $T_{\rm g} \propto G$ leads for the same data to the correlation coefficient $R^2=0.89$, which is only slightly lower than the $R^2=0.91$ of the figure; in contrast, testing $T_{\rm g} \propto K$ results in $R^2=0.41$ (data not shown).

{\bf Dynamics correlates to the Debye-waller factor:} Equation (\ref{eq:u2}) is tested in \fig{fig1e}(B). The dynamics of a wide variety of liquid collapses into a single curve; which is consistent with the notion that the short-time dynamics controls structural relaxation at much longer times.  \Fig{fig2e} provides further evidence for this connection. It shows that the fragility both at constant density and at constant pressure is linearly related to the relative change with temperature of $\langle u^2\rangle$, as predicted by \eq{eq:MSD}. Note that there is a significant difference between the constant-pressure fragility (blue data points) and the constant-volume fragility (red data points). This fact is generally understudied (but see Ref. \onlinecite{mei21}); within the shoving model this can be understood as an effect of increasing shear modulus upon decreasing volume.

{\bf The dynamics is tied to the  shear modulus- not to the bond energy scale:} In Refs.~\cite{Puosi_jcp_2012,Bernini_2017}, as described in Fig.~\ref{fig1e}(C,D) the authors employed a variety of glass-forming liquids and extracted from the stress-autocorrelation function: (i) the truly instantaneous shear modulus $G_0$, that would be obtained after an affine shear of the material (indicative of the bond strength) and (ii)  the actual (plateau) shear modulus obtained after vibrational modes could relax, here denoted by $G_{\rm p}$. The former would be measured on a femtosecond time scale upon a perfectly affine deformation, the latter would be measured on picosecond time scales or longer for which the system has time to relax to mechanical equilibrium. In Fig.~\ref{fig1e}(C) it is shown that plotting the $\alpha$ relaxation-time $\tau_\alpha$ against the ratio $G_0/T$ does not organize the data in any meaningful way. However, when plotted against $G_{\rm p}/T$ a striking data collapse emerges. These results robustly establish a relation between short-time elasticity and long-time structural relaxation, showing that their correlation does not simply stem from a common bond energy scale (which $G_0$ captures).

\subsection{Limitations of the shoving model}

Although the shoving model captures the tight connection, it does not predict quantitatively the activation energy $E_a$. This is apparent, for example, in Fig.\ref{fig1e}(D), where the plot shows a minor but clear curvature instead of the straight line predicted by Eq. (\ref{eq:shoving}). Note that this effect is also present in Fig.\ref{fig1e}(B) where departure from Eq.(\ref{eq:MSD}) is observed \cite{lar08} (although it is hard to visualize in this figure due to the logarithmic choice of the $x$-axis). Thus, the variation of $E_a$ with temperature is stronger than what the shoving model predicts. This effect can be quantified thanks to the measurement of activation energy provided in Sec. \ref{S2bis}. Fig. \ref{fig3e}(C) compares the prediction of the activation energy of the shoving model to observation. The shoving model has a fitting parameter that is fixed by imposing a correct prediction at some reference temperature.  The figure shows that the model captures only half of the variation in $E_a$ across the temperature range explored. We discuss the reason for this discrepancy below. 

Another point of caution concerns the manner in which the shoving model is often tested, by rescaling the x-axis of the Angell plot by $G$. \Fig{fig3e}(A) shows viscosity data of six molecular liquids plotted versus $x\propto G_\infty(T)/T$ (full symbols) normalized to unity at the glass transition temperature $T_{\rm g}$. The data follow roughly a line that, importantly, has a reasonable high-temperature limit (in the sense that the extrapolation of the relaxation time scale in that limit does not differ from the vibrational time scale by tens of orders of magnitude, which would be a red flag for a theory to apply). Even under these conditions, however, the linearity between $G$ and $E_a$ is not quantitatively guaranteed. This is illustrated in \Fig{fig3e}(B), which shows the same plot for the polydisperse numerical model used in panel (C). This panel once again indicates a qualitative success, as the curve is almost perfectly linear with a reasonable high-temperature limit. Yet we know that the prediction fails quantitatively,  as shown in Fig. \ref{fig3e}(C). The reason for this apparent paradox is simply that any affine dependence of $E_a$ with $T$ is actually not tested for by such a procedure. For example, if $E_a$ were purely affine, even without rescaling by $G$, the Angell plot would lead to a linear curve. A precise measurement of $t_0$ and $E_a$ is thus needed for a quantitative test.

\subsection{Elastically collective nonlinear Langevin equation theory}

In the elastically collective nonlinear Langevin equation (ECNLE) theory~\cite{mir14a,mir14b,mei20,zho20,mei21}, the energy scale governing structural relaxation is set by the free--energy cost associated with a particle escaping from its local cage and inducing an elastic distortion. This cost is evaluated for a system of hard spheres, with the equilibrium pair structure at packing fraction $\phi$ providing the primary input. The resulting activation barrier contains two coupled contributions: a local cage-escape term, $F_{\rm dyn}$, and an elastic shoving-model type contribution, $F_{\rm el}$, arising from the deformation of the surrounding 
particles.

The cage--escape process requires a jump of length $\Delta r$, which ECNLE predicts to increase with $\phi$. The elastic barrier is evaluated from the spatial integral of the energy density generated by the jump--induced displacement field in a surrounding Einstein model of a solid, yielding $F_{\rm el}(\phi)\simeq G(\phi)\,\Delta r(\phi)^{4}$.  The $\phi$--dependence of $\Delta r(\phi)$ causes the elastic contribution to grow  more rapidly upon cooling or densification than would be expected from elastic stiffening alone. While this is a key distinction between ECNLE and the shoving model, the growth of $\Delta r$ at high density appears weaker than that of $G$, suggesting an effective convergence between the two approaches.

ECNLE's predictions are connected to equilibrium structural properties through empirical correlations between $F(\phi)$ and the isothermal compressibility $S_0$ observed for hard spheres, and assumed to hold in molecular liquids as well. They take the form $F \propto S_0^{-q}$, with an exponent $q\simeq 1$ in the mildly supercooled regime, $q\simeq 3$ in the deeply supercooled regime, and $q\simeq 4.5$ at extremely deep supercooling beyond current experimental reach~\cite{mei25}.

\subsection{Summary}
To summarize this section, the experimental and numerical data do not uniquely identify an elastic model that universally explains the super-Arrhenius temperature dependence. Nevertheless, the data strongly suggest an important role of the mechanical moduli or, virtually equivalently, the vibrational MSD, at least in a zeroth-order approximation. Below, we take the ``local'' viewpoint of the elastic models as a starting point for a much more detailed approach to explaining physically the super-Arrhenius challenge.

\section{Second approach: fragility versus linear  local elasticity}
\label{sec:local}

\begin{figure}[ht!]
  \includegraphics[width = 0.5\textwidth]{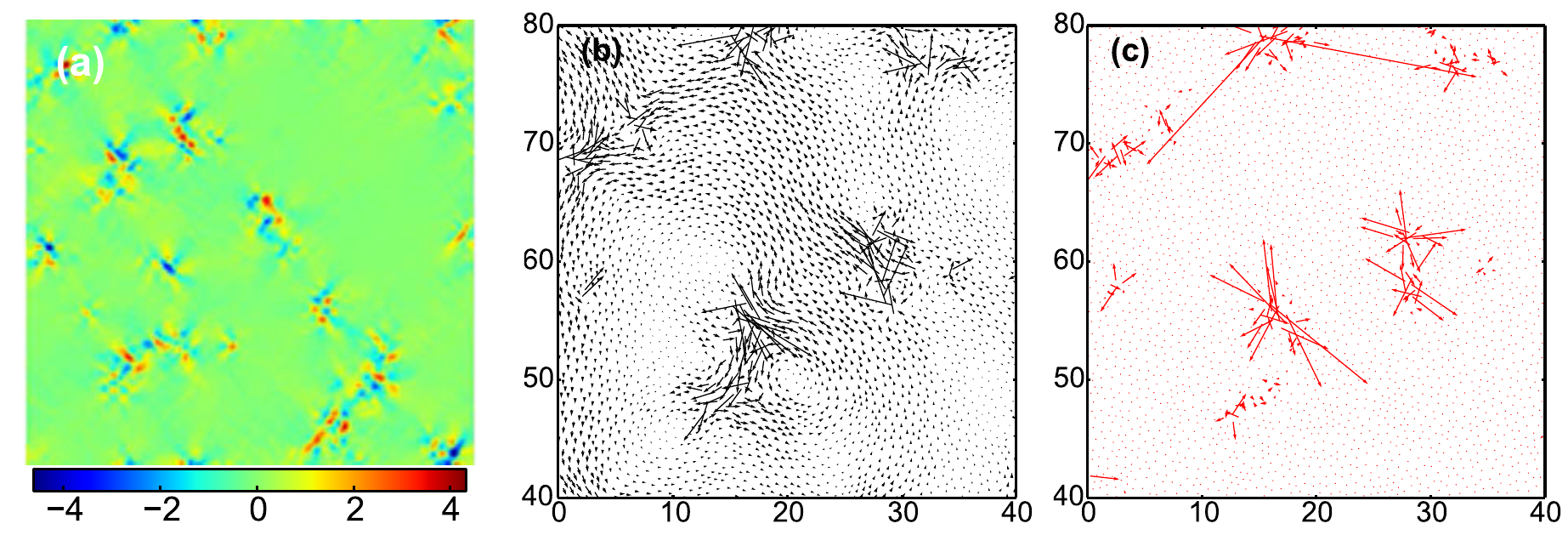}
  \vspace{-0.5cm}
  \caption{(a) Change in coarse-grained stress field over a time interval about 100 times shorter than the $\alpha$-relaxation time, extracted from a simulation of a 2D supercooled liquid, see~\cite{lem14} for details. (b) Displacement field accumulated over the same time interval. The localized cores resemble the quasilocalized excitation shown in Fig.~\ref{fig:kappa_fig}a below. (c) The linear force response to the displacement of panel (b) shows highly localized and nonaffine cores. Adapted with permissions from~\cite{lem14}.}
  \label{fig:lemaitre}
\end{figure}

While approaches relating relaxational dynamics to macroscopic elastic properties show promise as shown in the previoius section, several computational studies have established that structural relaxation events in supercooled liquids are largely localized in space and surrounded by highly non-affine motion of particles, as demonstrated in Fig.~\ref{fig:lemaitre} from~\cite{lem14}. This highly localized and non-affine character of relaxation events echoes the spatial structure of low-frequency soft vibrational modes seen in computer models of glassy solids. Indeed, following early works ~\cite{GurevichParshinSchober2003,ParshinSchoberGurevich2007,SchoberOligschleger1996,
MalandroLacks1999,TanguyEtAl2002} recent advances~\cite{ler21} have established that, in addition to phonons described by Debye's theory, glassy solids also host a population of soft, quasilocalized excitations. These excitations consist of a localized, highly nonaffine core of size between 5 and 10 particle diameters, decorated with affine algebraic decays away from the core, cf.~Fig.~\ref{fig:kappa_fig}(a) below. Their distribution over frequency $\omega$ has been shown to follow a universal quartic law $\sim\!\omega^4$ at low frequencies, independent of microscopic details, spatial dimension, or glass-formation history. At larger frequencies, even more soft modes can be present -- a cross-over that depends on the stability of the material \cite{ji20}. In Refs.~\onlinecite{wid08,bri09,Harrowell_jcp_2009} it was established that relaxation in supercooled liquids indeed correlates well with the loci and geometry of the soft quasilocalized modes. Similar correlations were reported in \cite{Tong_pre_2014}.

The observed similarity between relaxation events and soft localized excitations in glasses, and the observed correlations between the two suggest that relaxation in supercooled liquids is predominantly governed by heterogeneous elasticity -- rather than by homogeneous, continuum-like mechanics. It may explain why continuous elasticity predict too little variation of the activation energy with temperature. In turn, these observations led to efforts to treat heterogeneities explicitly, as discussed in this section and the next ones. This section focuses on linking local measures of linear elasticity to the relaxational dynamics of supercooled liquids.

\subsection{Definition of a local elastic modulus $\kappa$}

The highly localized and nonaffine relaxation events in supercooled liquids and the strikingly similar spatial structure of low-frequency nonphononic vibrational modes suggest that the two are intimately related. This might imply that, instead of macroscopic moduli controlling activation barriers in supercooled liquids, the characteristic stiffness of nonphononic quasilocalized modes does.

However, even in computer models, it is not straightforward to measure the characteristic stiffness of quasilocalized modes, see discussions in Refs.~\cite{new_variational_argument_epl_2016,lerner_jcp_2018,pinching_pnas}. These works suggest and verify numerically that the characteristic stiffness $\kappa$ associated with the glass' response to local force dipoles captures the characteristic stiffness of low-frequency, quasilocalized nonphononic vibrations, see panels (a) and (b) in Fig.~\ref{fig:kappa_fig} below. 

How is the stiffness $\kappa$ defined? Here, we denote by $\dv^{(ij)}$ a dipole vector applied to a pair of neighboring particles $i,j$ (the red arrows in Fig.~\ref{fig:kappa_fig}(b)), and the linear displacement response is given by
\begin{equation}
    \uv^{(ij)} = \calBold{H}^{-1}\cdot\dv^{(ij)}\,,
\end{equation}
where $\calBold{H}\!\equiv\!\frac{\partial^2U}{\partial\xv\partial\xv}$ is the Hessian matrix and $\cdot$ denotes a single contraction. 
A unit vector pointing in the same direction as $\uv^{(ij)}$ is obtained via
\begin{equation}
    \hat{\uv}^{(ij)} = \frac{\calBold{H}^{-1}\cdot\dv^{(ij)}}{\sqrt{\dv^{(ij)}\cdot\calBold{H}^{-2}\cdot\dv^{(ij)}}}\,.
\end{equation}
The stiffness $\kappa_{ij}$ associated with the direction $\hat{\uv}^{(ij)}$ is then
\begin{equation}
\label{kap}
    \kappa_{ij} = \hat{\uv}^{(ij)}\cdot\calBold{H}\cdot\hat{\uv}^{(ij)} = \frac{\dv^{(ij)}\cdot\calBold{H}^{-1}\cdot\dv^{(ij)}}{\dv^{(ij)}\cdot\calBold{H}^{-2}\cdot\dv^{(ij)}}\,.
\end{equation}
The characteristic stiffness $\kappa$ is obtained by averaging $\kappa_{ij}$ over many pairs $i,j$ of neighboring particles~\cite{lerner_jcp_2018,dipole_stiffness_jcp_2021}.

What is the relation between the mesoscopic stiffness $\kappa$ and the macroscopic stiffness $G$? In 
Refs.~\cite{lerner_jcp_2018,pinching_pnas,dipole_stiffness_jcp_2021} it was demonstrated that the dependence of the mesoscopic stiffness $\kappa(T)$ on the equilibrium parent temperature $T$ --- from which glasses were instantaneously quenched --- is stronger than the dependence of the macroscopic elastic stiffness, the shear modulus $G(T)$. Indeed, Ref.~\cite{dipole_stiffness_jcp_2021} reports a variability of 400\% of $\kappa(T)$ with supercooling, compared to a variability of just 55\% for the shear modulus.

Building on these aforementioned insights, it was proposed in Ref.~\cite{lerner_jcp_2018} that the average mesoscopic stiffness, $\kappa$, associated with the elastic response of inherent structures to local force dipoles might control activation barriers towards structural relaxation in the liquid states from which those inherent structures were quenched. Since the expected variability of $E_a(T)$ and of the mesoscopic stiffness $\kappa(T)$ with supercooling appear to match, Ref.~\onlinecite{kap21} suggested a simple relation of the form $E_a\!\propto\!\kappa$.

\subsection{Qualitative test}
\begin{figure}[ht!]
  \includegraphics[width = 0.5\textwidth]{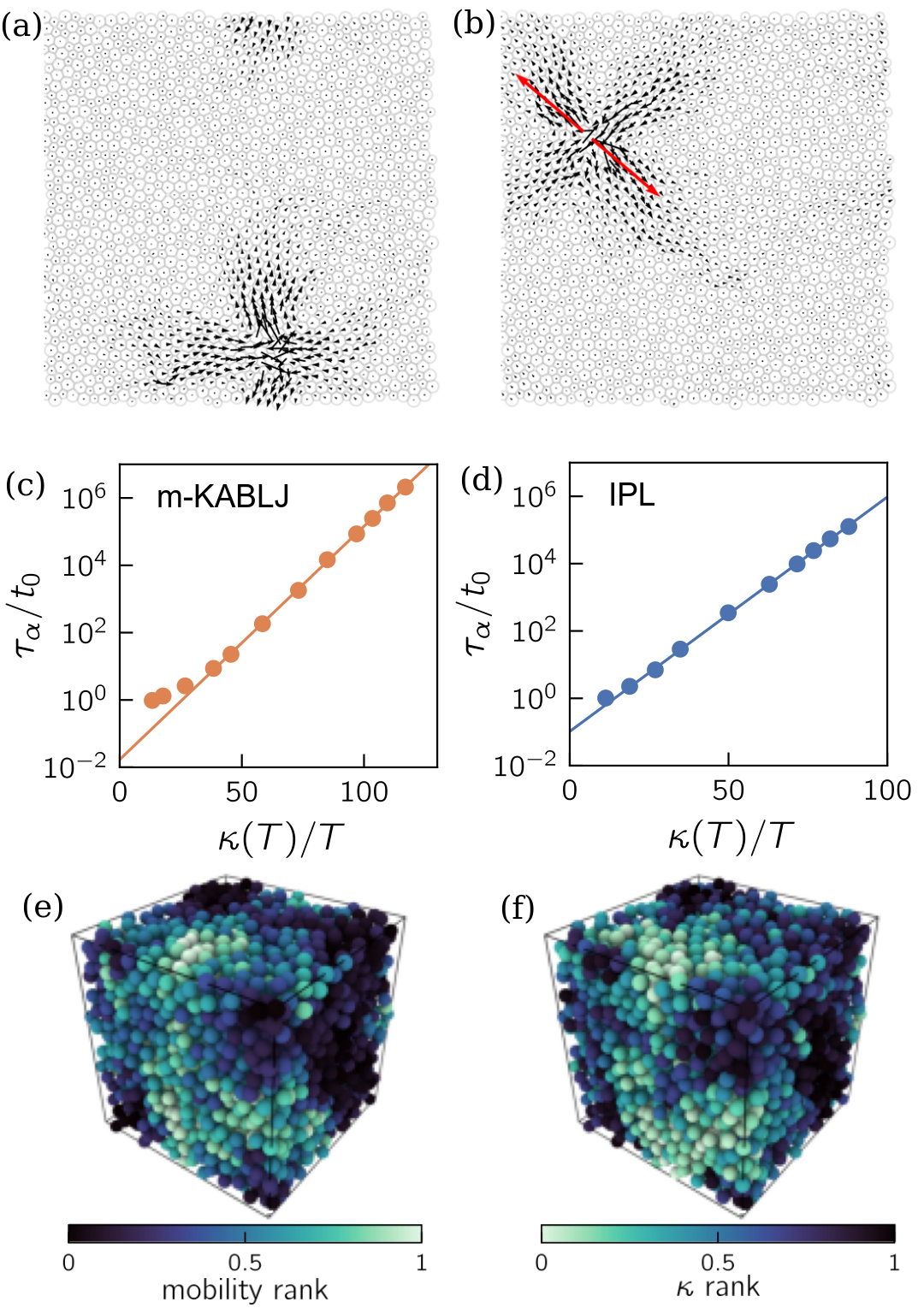}
  \vspace{-0.5cm}
  \caption{Relating structural relaxation of supercooled liquids to the mesoscopic stiffness $\kappa$. (a) A soft, quasilocalized vibrational mode measured in a 2D computer glass. (b) The response of the same glass to a local force dipole (in red) reveals a very similar spatial structure to that of soft, quasilocalized vibrations, suggesting that the characteristic stiffness of soft modes is captured by the stiffness $\kappa$ of responses to such force dipoles. Panels (c) and (d) plot the structural relaxation time $\tau_\alpha$ of the modified Kob-Andersen Binary Lennard-Jones model (panel (c)) and the IPL model (panel (d)) vs.~the ratio $\kappa/T$, see text for discussion. (e) Map of particle mobility of a supercooled liquid measured by the isoconfigurational ensemble~\cite{isoconfigurational_prl_2004}, see~\cite{kap21} for details. (f) The map of particle-wise mesoscopic stiffness $\kappa_i$ reveals strong correlation with mobility.
  Reproduced with permission from~\cite{kap21}.
  }
  \label{fig:kappa_fig}
\end{figure}

\textbf {Activation energy:} In Fig.~\ref{fig:kappa_fig}(c) the $\alpha$-relaxation time $\tau_\alpha$ of a slightly modified version of the Kob-Andersen binary Lennard Jones liquid~\cite{sch20}, is plotted on semilogarithmic scales against the ratio $\kappa/T$, where $\kappa$ was measured in inherent states quenched from equilibrium liquid configurations at temperature $T$. Panel (d) shows the same measurements for an inverse-power-law (IPL) model in which particles interact via a $1/r^{10}$ pairwise potential (see~\cite{kap21} for model details). The apparent alignment of the data with a straight line indicates that $E_a$ and $\kappa$ are correlated.  It was initially interpreted as supporting $E_a\!\propto\!\kappa$, yet as we will see below, systematic departures exist.

\begin{figure}[ht!]
  \includegraphics[width = 0.5\textwidth]{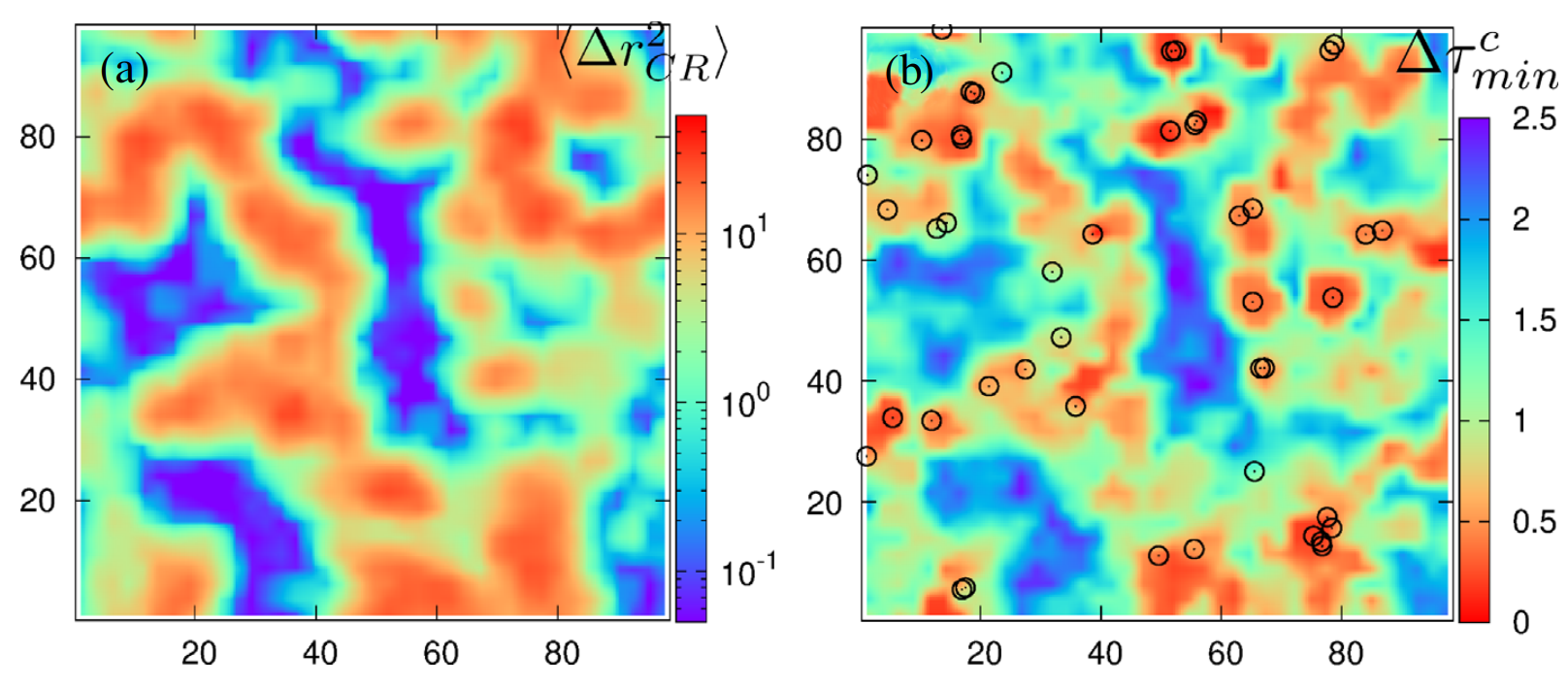}
  \vspace{-0.5cm}
  \caption{Relating structural relaxation of supercooled liquids to the local yield stress measured in underlying inherent structures. (a) Dynamical propensity map measured using the isoconfigurational ensemble~\cite{isoconfigurational_prl_2004} for a 2D Lennard-Jones glass former, see~\cite{ler22} for details. (b) The local yield stress $\Delta\tau^c_{\rm min}$ (reported is the minimum local yield stress over all possible deformation geometries) as obtained from the `frozen-matrix-method'~\cite{pat16}). The strong correlation between the two maps is obvious. The empty circle mark the loci of the first 50 thermally activated flow events. Reproduced with permission from Ref.~\cite{ler22}}
  \label{fig:sylvain_fig}
\end{figure}

\begin{figure}[ht!]
  \includegraphics[width = 0.45\textwidth]{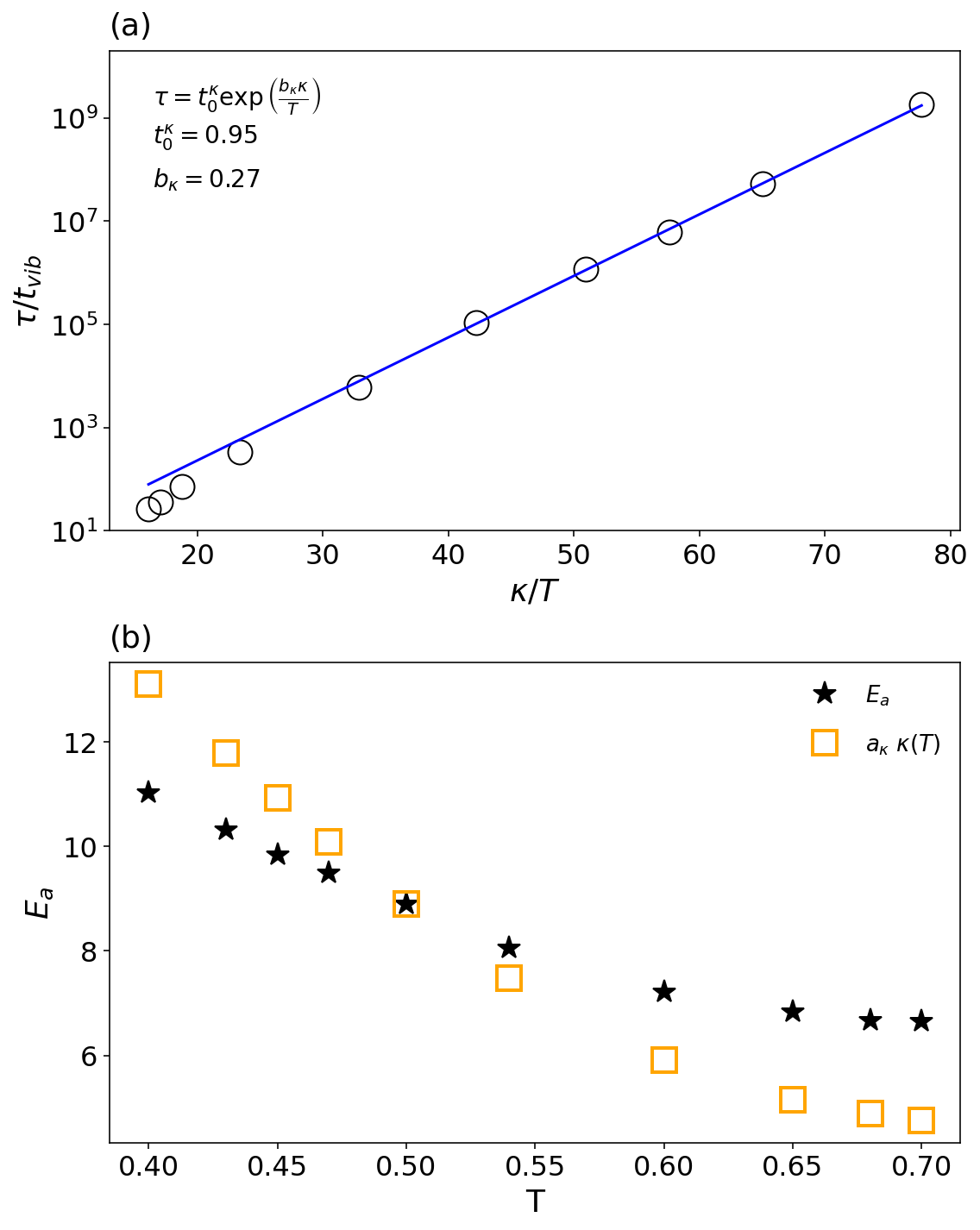}
  \caption{
  (a) Numerical test of the mesoscopic elasticity model for a polydisperse system of particles, for temperatures below the onset one. 
  The parameters $t_0^\kappa$ and $b_\kappa$ have been estimated via a numerical fit for temperatures below the critical one, corresponding to $\kappa/T > 20$.
  (b) Comparison of measured activation energy $E_a$ with prediction from local linear elasticity. Again the latter has one fitting parameter, fixed by matching the two quantities as some reference temperature. Predictions vary twice more than observations on the range of temperature explored.
  }
  \label{fig:K_polydisperse}
\end{figure}

\textbf{Propensity:}
Up to now we have discussed the connection between the \emph{average} mesoscopic stiffness $\kappa$ and the \emph{average} relaxation time $\tau_\alpha$. We next show that the correlation between these observables persists on the local/spatial level as well. In Fig.~\ref{fig:kappa_fig}(e) and (f) we show spatial maps of mobility (panel (e)) and of a particle-wise measure of the mesoscopic stiffness $\kappa_i\!\equiv\!\langle \kappa_{ij}\rangle_j$ (panel (f)), measured for a Lennard-Jones-like glass former. Not only is the average mesoscopic stiffness $\kappa$ correlates with the activation energy $E_a$, but also locally particles associated with smaller $\kappa_i$ values tend to be more mobile. 

Ref.~\cite{kap21} reviewed above examined the relation between the stiffness associated with the \emph{linear} elastic response of underlying inherent structures, and supercooled liquid dynamics. Moving beyond linear elasticity, Ref.~\cite{li22} investigated how local plastic properties correlate with slow, heterogeneous dynamics, highlighting the role of incipient yielding as a structural predictor of relaxation. In a similar spirit, Ref.~\cite{ler22} examined the correlation between local yield stress $\Delta\tau^c_{\rm min}$ (i.e.~the minimal yield stress over all possible deformation orientations) --- as extracted from loading the material locally using the so-called `frozen matrix method'~\cite{pat16} --- and relaxation patterns in the supercooled liquid. The main result of~\cite{ler22} is reproduced in Fig.~\ref{fig:sylvain_fig}. Panel (a) shows the dynamical propensity map as measured by the isoconfigurational ensemble~\cite{isoconfigurational_prl_2004} for a supercooled 2D Lennard-Jones glass former, see~\cite{ler22} for details about the model. Panel (b) displays the map of local yield stresses $\Delta\tau^c_{\rm min}$. 

\subsection{Quantitative test}
We can once again use the measurement of activation energy to compare the prediction $\kappa\sim E_a$ quantitatively by considering the continuous polydisperse system used in Sec. \ref{S2bis}. As shown in Fig. \ref{fig:K_polydisperse}(A), rescaling temperature by $\kappa$ indeed leads to a straight line in the Angell plot. However, as seen before, this does not imply success. As illustrated in Fig.\ref{fig:K_polydisperse}(B), local elasticity predicts a variation of $E_a$ which is twice too large in comparison with observations (and four times larger than predicted by continuous elasticity). This further underlines the danger of using the Angell plot with rescaled axes to quantitatively validate theories.

\subsection{Summary}
Considering local elasticity is an important step in the right direction, leading to precise predictions on propensity and qualitative predictions on activation energy. This approach predicts too much fragility, however, at least in systems where it can be quantitatively tested. We discuss why below.

\section{Third approach:   fragility versus spectrum of non-linear excitations}
\label{S4}

\subsection{Motivations}

The two previous sections sought to relate activation energies and local barriers to {\it averaged} and {\it linear} elastic properties. Here, instead, we adopt a qualitatively different viewpoint and describe recent approaches that seek to characterize local barriers directly, which are intrinsically non-linear objects. Rather than relying on averaged elastic quantities, these methods aim to characterize the full spectrum of local barriers, i.e., the spectrum of non-linear excitations of glasses. Measuring energy barriers in glassy landscapes has a long history~\cite{Denny2003,dol03a,Doliwa2003}, generally based on following the dynamics and performing frequent quenches to nearby local minima of the energy landscape, known as inherent structures. Among other results, these studies have shown, as we also document below, that elementary barriers are associated with the motion of progressively \textit{fewer} particles upon cooling. Here, however, our goal is different: we aim to measure a large number of barriers that allow escape from a given single metastable state and, denoting the barrier height by $E$, to characterize the corresponding density $N(E)$ of barriers at energy $E$.

\subsection{Algorithms}

The reference inherent structure, denoted IS$_{\rm r}$, is obtained by minimizing the energy of an equilibrated configuration at a parent temperature $T$. In both algorithms described below, once a catalog of unique inherent structures IS$_i$ near the reference one is constructed, the nudge-elastic-band method~\cite{neb1} is employed to reconstruct the energy profile of the IS$_{\rm r} \to$ IS$_i$ excitations and extract their activation energies $E^{i}$, defined as the difference between the energy at the top of the barrier and the energy of the reference inherent structure. An excitation is retained only if the energy path exhibits a single maximum, indicating that the two structures are adjacent in phase space.

{\bf SEER:} Ref.~\cite{cia24} introduced Systematic Excitation ExtRaction, an algorithm that reconstructs the catalog of inherent structures adjacent to IS$_{\rm r}$ through thermal cycling. In this approach, the reference configuration is evolved at a finite temperature, initially very low, until a transition to a novel inherent structure is detected by monitoring particle motion in the potential energy landscape. This procedure is commonly used to study plastic rearrangements or two level systems in glasses \cite{Doliwa2003,khomenko2021relationship,Ji2022}. Alone, it is unpractical to access more than a few barriers exiting a given meta-stable states. Indeed, since small barriers are much more likely to be triggered thermally, larger energy barriers are never triggered in isolation and thus cannot be studied. The key idea of SEER is that once excitations are discovered, the system is reinforced at that location such that this excitation cannot be triggered again.  In practice, the potential energy is  minimally perturbed to transform the discovered minimum into a saddle point- a perturbation that affects a single collective degree of freedom. A thermal cycle in this modified landscape uncovers a distinct inherent structure, which is added to the catalog only if it corresponds to a true minimum of the unperturbed potential energy. By iterating this procedure and gradually increasing the temperature of the thermal cycles, SEER builds a catalog of inherent structures close in phase space to the reference. For each reference structure, the procedure is repeated multiple times, and the resulting catalogs are merged to eliminate duplicates. Being based on thermal cycling, SEER is effective at reconstructing the low-energy portion of the spectrum, obtaining about $10^2$ excitations in systems of a few thousands particles.

{\bf ASEER:} Ref.~\cite{jic25} introduced Athermal SEER, an athermal algorithm designed to reconstruct more completely the spectrum of excitations. The underlying assumption is that relevant excitations induce structural relaxation by altering the local environment of particles through plastic processes. To trigger such events, the algorithm couples pairs of adjacent particles with a spring and quasistatically increases its rest length, effectively pushing the particles apart until a plastic event is detected. The spring is then removed, and the system is energy-minimized to reach a new inherent structure. By repeating this procedure for all pairs of nearby particles, this algorithm uncovers now thousands of excitations.

\subsection{ Main observations} 
\begin{figure}[t!]
\centering
\includegraphics[width=1\linewidth]{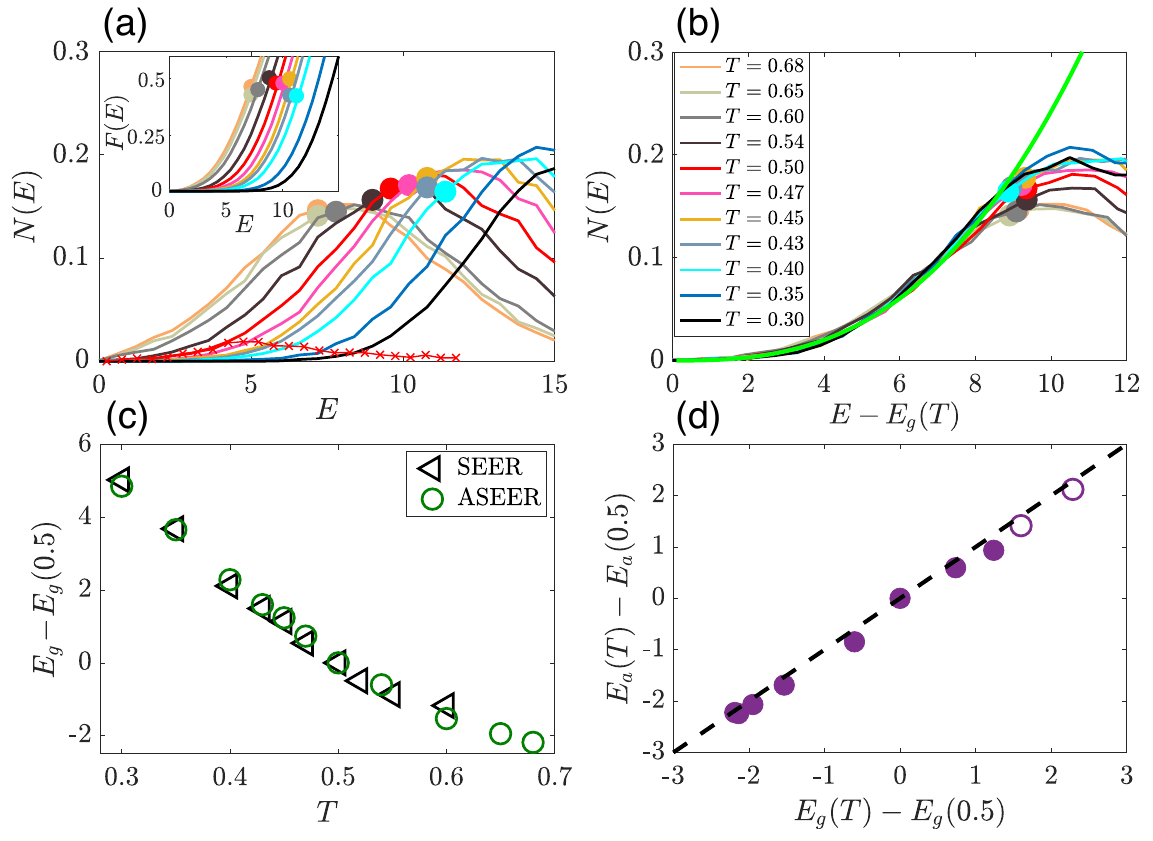}
\caption{
(a) Density of excitations $N(E)$,  normalized by the system size $\mathcal{N}$, and its cumulative distribution $F(E)$ (inset).
The solid dots mark the values of the activation energy~\cite{cia24}, $E_a(T)=T \log\left(\tau/t_0\right)$, and the cross-shaped line has been obtained with the SEER algorithm at $T=0.5$.
(b) The $N(E)$ curves collapse when the energy is shifted by $E_g(T)$. 
The thick green curves is $N(E)\!\approx g_1\!\times\!(E\!-\! E_g)^{2.7\pm0.1}$, where $g_1\!=\!\!(4.5\pm 0.5)\!\times\! 10^{-4}$.
(c) The shift of $E_g$ evaluated via SEER on cooling matches the shift in activation energy  \cite{cia24}. 
(d) The increase in $E_g$ matches the increase in activation energy measured from the relaxation dynamics, with $T=0.5$ an arbitrary reference temperature. 
Open circles correspond to the low-temperature values at which we use the time-temperature superposition to estimate the relaxation time and $E_a$.
All results refer to the poly-IPL10 model.
} 
\label{fig:N_E}
\end{figure}

The main observations obtained by these algorithms, as applied to the poly-IPL10 model, are shown in Fig.\ref{fig:N_E}:
\begin{itemize}
\item As exemplified in Fig.\ref{fig:N_E}.A; both methods give consistent density of excitations at low energy $N(E)$, supporting that detection is exhaustive in this energy range. However, ASEER leads to a tenfold improvement, and allows to obtain a smooth curve for $N(E)$ up to the activation energy $E_a$ (indicated with circles), suggesting that a large portion of excitations are captured in that range. This range is the relevant one, since energy higher than $E_a$ are unlikely to be thermally triggered while the liquid relaxes.
\item At lower temperatures, low barriers become less frequent as expected. But most remarkably, the effect of temperature is simple: it corresponds essentially to a shift of $N(E)$, as shown by the collapse of Fig.\ref{fig:N_E}.B. 
\item We find that:

\begin{equation}
\label{NE}
    N(E)\!\approx g_1\!\times\!(E\!-\! E_g)^{2.7\pm0.1}
    \end{equation}
    where $g_1$ is some constant. Below $E_g$, the density of excitation is not strictly zero, but very small. 
\end{itemize}

\subsection{Predicting the thermal evolution of activation energy}
$N(E)$ being a broad distribution, it is a priori difficult to extract the activation energy  from it. However, the remarkable fact that $N(E)$ simply shifts under cooling is a great simplification on this respect. It is statistically equivalent to the statement that all barriers increase by some given amount under cooling. If local barriers control the dynamics, then the activation energy must change according to this shift. In other words, it is possible to make the  prediction that for any pairs of temperature $T_1, T_2$ where this shift is observed that:
\begin{equation}
 E_a(T_1) - E_a(T_2) = E_g(T_1) - E_g(T_2)
\end{equation}
Note that this prediction of activation energy has no fitting parameter, which is possibly unique in this field. This prediction is tested in Fig.\ref{fig:N_E}.D, and is accurate: local barriers do control activation, at least in this liquid.

\subsection{How about the absolute value of $E_a$?}
It would seem reasonable to assume that the approximate gap $E_g$ that characterizes the distribution of excitations $N(E)$, as described by Eq.\ref{NE}, gives the activation energy directly. In fact, in elasto-plastic models of the glass transition reviewed in Sec.\ref{S8} below, this assumption is exact:  $N(E)$ develops a gap at low $T$ whose value is the activation energy.   

Why is it not true in actual liquids? One simplification made by elasto-plastic models is the assumption of irreversibility: once a region relaxes, it cannot return to its original state. By contrast, in super-cooled liquids the overall dynamics is reversible and the lowest-energy excitations are heavily biased: the difference between the energy of the two meta-stable states tend to be much larger than $k_BT$  \cite{cia24,jic25}, so that if activation takes place from the low-energy state, it most likely will snap back to the original configuration, which is useless for relaxation. As already intuited by Goldstein \cite{gol63} and later formalized by Heuer~\cite{Dol03}, the activation energy must relate to rearrangements that allow one to reach meta-stable states of similar energy as one started from. This will require having several interacting excitations to relax together. This view is equivalent to the droplet picture of spin glasses, where a number of interacting spins must flip together to eventually reach a state of energy similar to the starting one \cite{fis86}. For such collective events to take place, excitations cannot be spatially isolated, as interactions would then be negligible. Thus, a certain density $\rho^*$  of excitations is required.  In this view,  the activation energy implicitly follows:

\begin{equation}
\rho^*=\int_0^{E_a} N(E)dE.
\end{equation}
Our findings indicate that the thermal evolution of $\rho^*$ is negligible with respect to that of $N(E)$: the change of cooperativity the former corresponds to is negligible with respect to the change of local barriers. Yet predicting $\rho^*$ would be  necessary to obtain the absolute value of the activation energy. It may require to study the interaction between excitations. 

Such a study may also clarify what governs the value of the activation entropy $S_a$. For example if $n_0$ local barriers need to  activate to relax a given region toward a new state of similar energy, but the sequential order in which they do so has limited effects on the maximal energy barrier encountered, then $2^{n_0}$ paths for relaxation are available. That scenario corresponds to $S_a=n_0\ln(2)$.

\subsection{Excitation-based predictor for propensity}
\begin{figure}[t!]
\centering
\includegraphics[width=1\linewidth]{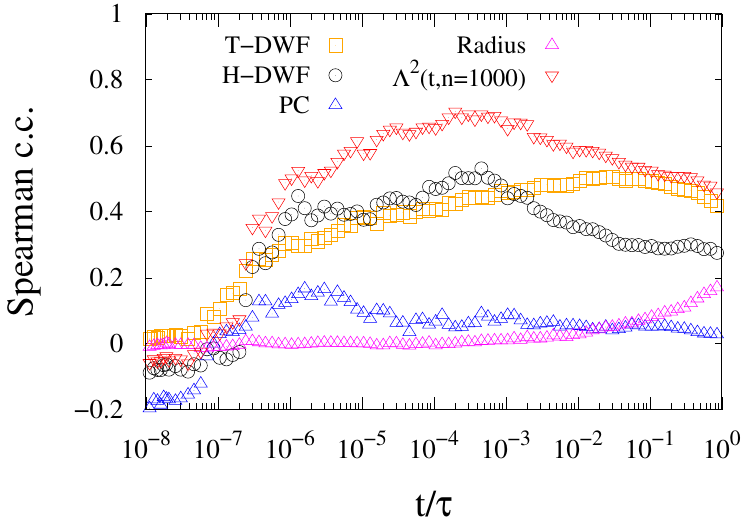}
\caption{
Time dependence of the Spearman correlation coefficient between the particle propensity and the Debye Waller factor (T-DWF), its Harmonic approximation (H-DWF), the packing capability (PC), and the excitation-based predictor. The radius of the particles is also indicated as a predictor.
All results refer to the poly-IPL10 model at $T=0.45$, below the estimated $T_c$. From Ref.~\cite{cia24}.
} 
\label{fig:propensity}
\end{figure}

Ref.~\cite{cia24} constructed a physically motivated, time-dependent predictor of structural relaxation from the excitations identified in the initial configuration. This predictor is simply based on the approximation that excitations are independent and relax according to their specific energy barrier. It weights particle displacements associated with these excitations by their time-dependent activation probabilities to yield a time-dependent predicted displacement field. As shown in Fig.~\ref{fig:propensity}, the excitation-based predictor $\Lambda^2$ for the square displacement of each particle exhibits the strongest correlation with the inherent-structure propensity over almost the entire time window, outperforming conventional structural indicators such as the finite-temperature Debye–Waller factor~\cite{wid08}, its harmonic approximation~\cite{Tong_pre_2014}, and local packing capability~\cite{Tong2019}. Except for the very short ballistic regime, no alternative physically motivated predictor achieves comparable accuracy. This further supports that excitations encode the key structural information governing relaxation dynamics.

\section{Comparing  theoretical descriptions of local barriers}
\label{comparison}

\begin{figure}[hbt!]
\centering
\includegraphics[width=1\linewidth]{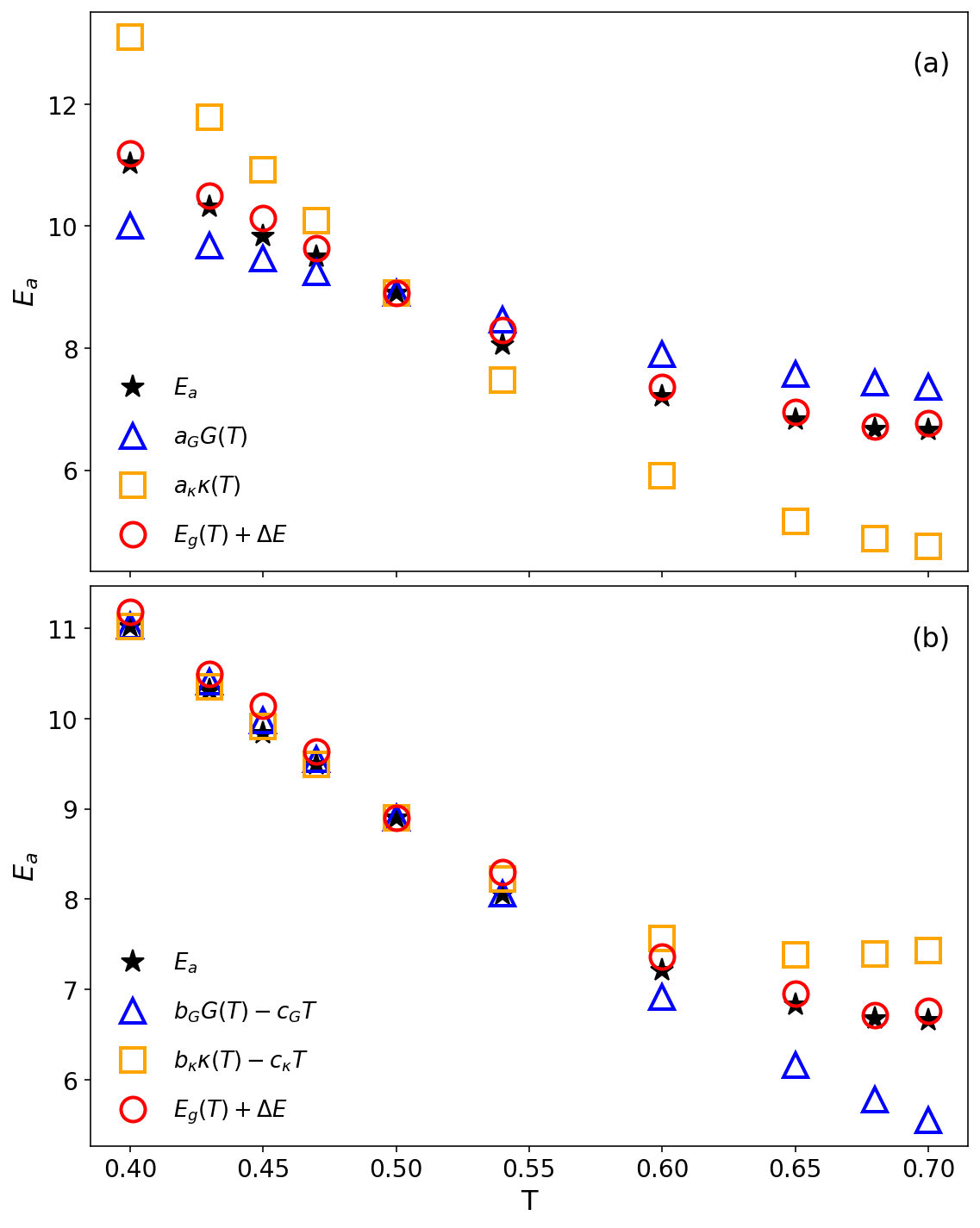}
\caption{
(a) The measured temperature dependence of the effective activation energy $E_a$ is compared with the predictions from the global elasticity, $a_G G(T)$, local elasticity, $a_\kappa \kappa(T)$, and the excitation description, $E_g(T)+\Delta E$. Here, $a_G\simeq 0.77$, $a_\kappa\simeq 0.42$, and $\Delta E\simeq 8.9$ are chosen so that the predictions are exact at the mode-coupling critical temperature $T_{\rm c} = 0.5$.
(b) The global and local elasticity predictions describe quantitatively the activation energy when corrected by a term scaling linearly with the temperature. Here, $b_\kappa = 0.275$, $c_\kappa = -6.18$, $b_G=1.05$, $c_G = 6.34$.\\
\label{fig:EaPredictions} 
}
\end{figure}

\subsection{Empirical comparison}  \Fig{fig:EaPredictions}.A compiles  data already shown in the last three sections for the the poly-disperse system of~\cite{cia24}. The predictions of elastic model $E_a(T) = a_G G(T)$; local elastic model  $E_a(T) = a_\kappa \kappa(T)$ and excitations-based theory $E_a(T) = E_g(T)+\Delta E$ are presented against observations. In terms of predicting the absolute value of $E_a$, each  approach has a single fitting parameter, chosen such that predictions match at some (arbitrary) reference temperature, here chosen to be the estimated mode-coupling temperature $T_{\rm c}\simeq 0.5$. The excitations-based theory is quite accurate, while elastic models predict too little fragility and local elastic models too much. 
 
The fact that the Angell plot appears linear both for elastic and local elastic models once temperature is rescaled by the respective modulus implies that these models predict well the activation energy {\it within a term linear in temperature}. This fact is confirmed in \Fig{fig:EaPredictions}.B; showing that with an added fitting parameter associated to such a linear term, these theories appear to match the data well (except at large temperatures where they still show deviations with observations).
  
\subsection{Proposed explanation for the difference between $G$ and $\kappa$}
The shift of the density of excitations $N(E)$ is reminiscent of the shift of the  spectrum of the Hessian $\calBold{H}$ predicted near the dynamical transition. We will argue in the next section that this correspondence is real, as it allows to predict in detail the architecture of excitations. Here we argue that this scenario explains why $G$ varies much less than $\kappa$ under cooling. Consider, as predicted by mean-field approaches, that the spectrum of the hessian is a semi-circle  whose lowest eigenvalue $\lambda^*$ crosses zero and becomes positive below some $T_c$, i.e. $\lambda^*\sim T_c-T$. It is well-known in these approaches \cite{wol12} that the shear modulus jumps to a finite value at $T_c$, and then grows with a square root singularity. Within RFOT, this effect is argued to be partially responsible for the material fragility, see e.g. \cite{rab13}. $G$ is indeed proportional to the inverse of the energy of the response to a force dipole, which reads schematically:

\begin{equation}
\label{gf}
    G\sim \frac{1}{\dv^{(ij)}\cdot\calBold{H}^{-1}\cdot\dv^{(ij)}}\sim \frac{1}{\int d\lambda  \frac{D(\lambda)}{\lambda}}\sim (\lambda^*)^0
    \end{equation}
where it is assumed that the dipole couples to all modes equally, and $D(\lambda)$ is the semi-circular spectrum. It can be expanded around $\lambda^*$ as  $D(\lambda)\sim \sqrt{\lambda-\lambda^*}$. Thus the integral in Eq.\ref{gf} is not singular even when $\lambda^*=0$: all modes contribute to $G$, which takes a finite value at $T_c$.

By contrast, from  the definition of $\kappa$ or Eq.\ref{kap} and the result just discussed, we get:
\begin{equation}
    \kappa \sim  \frac{\dv^{(ij)}\cdot\calBold{H}^{-1}\cdot\dv^{(ij)}}{\dv^{(ij)}\cdot\calBold{H}^{-2}\cdot\dv^{(ij)}}\sim  \frac{\int d\lambda  \frac{D(\lambda)}{\lambda}}{\int d\lambda D(\lambda) \frac{1}{\lambda^2}}\sim \sqrt{\lambda^*}
\end{equation}
where we used that the integral in the denominator diverges as $1/\sqrt{\lambda^*}$  at $T_c$. $\kappa$ is thus dominated by the lowest modes of the spectrum, and thus vanishes at $T_c$ as $\sqrt{\lambda^*}$. In mean-field, the relative variations of $\kappa$ are thus infinitely larger than those of $G$ near the transition.  In finite dimensions where singularities are avoided, we simply expect the variations of $\kappa$ to be much larger than those of $G$, as observed in our model liquid. 

\begin{figure}[ht!]
  \includegraphics[width = 0.5\textwidth]{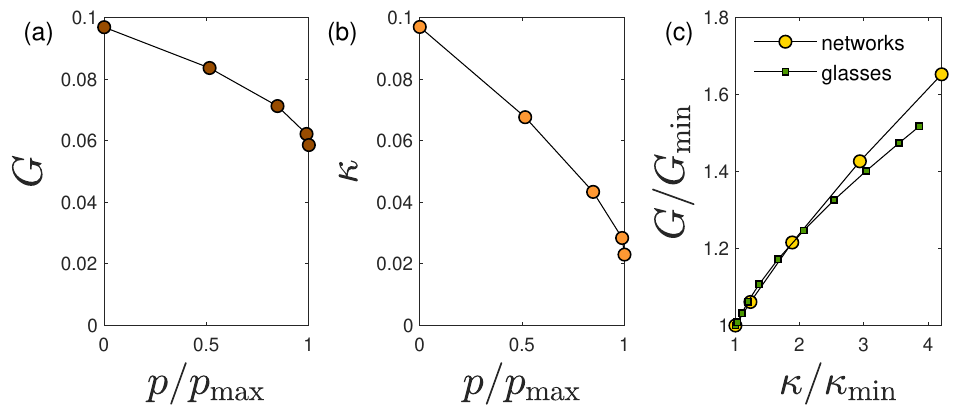}
  \caption{(a) Shear modulus $G$ and (b) local elastic modulus $\kappa$  v.s. $p/p_{max}$. (c) The relative variation $G/G_{min}$ v.s. $\kappa/\kappa_{min}$  confirms the prediction that $\kappa$ evolves much more than $G$ near an elastic instability. Remarkably, the same holds true quantitatively in the numerical liquids of ~\cite{dipole_stiffness_jcp_2021} as the temperature is varied, supporting that these liquids indeed undergo an elastic instability under heating. }
  \label{fig:frustrated_networks_fig}
\end{figure}

These views are tested in elastic networks approaching an elastic instability, which display various properties expected near a dynamical transition \cite{ler14b} as mentioned above. We employ networks of relaxed Hookean springs of unit stiffness, consisting of $N\!=\!16,000$ nodes derived from soft-sphere glasses as described in~\cite{anomalous_elasticity_soft_matter_2023}. The target coordination was set to $z\!=\!7.28$. The rest-lengths of the springs of the networks are incrementally and uniformly increased, each such variation is followed by a potential-energy minimization as described in~\cite{frustrated_networks_pre_2024}. Under this protocol, the pressure initially increases, until an elastic instability is encountered at a pressure denoted by $p_{\rm max}$. As shown in Fig.~\ref{fig:frustrated_networks_fig}, we confirm that in these networks $\kappa$ varies four fold more than $G$ approaching an elastic instability- numbers that even match the observations reported in~\cite{dipole_stiffness_jcp_2021} for liquids, supporting the proposed view.

\subsection{Proposed explanation for why $\kappa$ overestimates the variation $E_a$}
It seems at first  natural that the activation energy in liquids should relate to the characteristic lowest curvature of the landscape $\lambda^*$. In that light, the observable $\kappa$ is a very convenient tool to access this quantity. Yet as we have seen in the last section, the lowest energy scale of local barriers  $E_g$ does not equate the activation energy in liquids (although it does in elasto-plastic models). Instead,  $E_a=E_g + \Delta$ where $\Delta$ has negligible variation with temperature. Thus, in relative terms, $E_g$ varies much more than $E_a$ under cooling, which plausibly explains why it is also true for the lowest stiffness scale and $\kappa$. 

\subsection{Connecting the excitation theory to experiments via the Debye-waller factor}
\begin{figure}[ht!]
  \includegraphics[width = 0.48\textwidth]{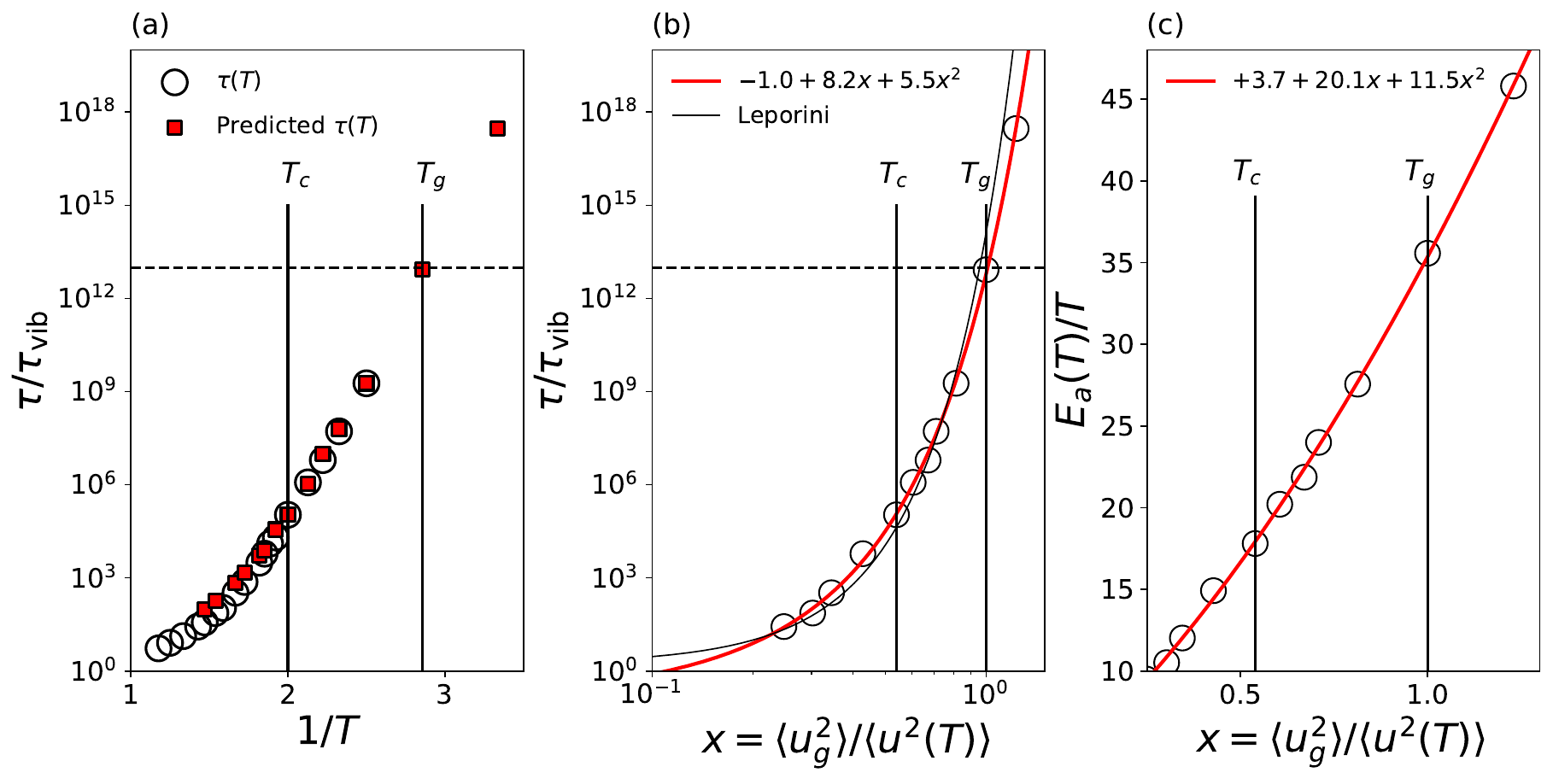}
  \caption{(a) Inverse–temperature dependence of the relaxation time and its theoretical prediction for the polydisperse model of Ref.~\cite{cia24}.
    (b,c) Dependence of the relaxation time and of the scaled activation energy on the normalized Debye--Waller factor, together with the corresponding parabolic fits.
    In panel (b), we also illustrate the prediction of Ref.~\cite{lar08}, which is vertically shifted following their protocol.
    In panels (b) and (c), the relaxation time for $T \le T_g$ is that estimated from panel (a).
  }
  \label{fig:Ea_DWF}
\end{figure}

The observation that the relaxation of liquids is tightly connected to the Debye-waller factor, the shear  and  local elastic moduli and the density of excitations is consistent with the view that these quantities are all controlled by a dynamical transition, as further supported by Fig.\ref{fig:frustrated_networks_fig} (see next section for our strongest observations supporting this view). Quantitatively however, the relaxation time is governed by the excitation spectrum as recalled in Fig.\ref{fig:Ea_DWF}.a, which cannot be accessed in experiments. 

To build an experimentally testable theory we thus relate the excitation spectrum to the Debye-waller factor in the controlled polydisperse model of Ref.~\cite{cia24}, hypothesizing a universal relation between these quantities. As a sanity check, we first confirm in Fig.\ref{fig:Ea_DWF}.b that this model indeed follows the universal behavior of liquids of  Fig.\ref{fig1e}.B. In this figure,  for $T\le T_c$ $\langle u^2\rangle$ is taken as the mean-square displacement in the plateau region, while for $T>T_c$ we follow Ref.~\cite{lar08} and evaluate $\langle u^2\rangle$ at the inflection point of the mean-square-displacement curve in a log--log representation.

Our  result central for experimental testing is then to plot the  activation energy rescaled by temperature $E_a/T$ as a function of $x=\langle u_g^2\rangle/\langle u^2\rangle$, as shown in Fig.\ref{fig:Ea_DWF}.c. We find:

\begin{equation}
\frac{E_a}{T}=3.7 + 20.1x + 11.5x^2
\label{eq:pred}
\end{equation}
A stringent test of the excitation-based theory is now possible, if the activation energy is  measured  following the protocol discussed in Sec.\ref{S2bis}.

\section{Architecture of excitations}
 \label{S7}

\subsection{Motivation}
Characterizing the properties of elementary excitations  is a fundamental question in condensed-matter. The algorithms described in Sec.\ref{S4} give a new handle to study this question in glasses. This point is important for the glass transition: it was argued in Sec.\ref{S4} that a shift in the density of excitations governs the fragility of liquids, yet no mechanism causing this shift was discussed so far. This section provides evidence that a dynamical transition causes this shift, as this hypothesis naturally explains how the architecture of excitations evolves under cooling. This provides an explanation for the known observation that elementary barriers involve less particles under cooling \cite{dol03a}, and also gives a simple rule of thumb for where excitations can become string-like, a phenomenon that has received significant attention \cite{don98,zha11,bet18,Oligschleger1999Collective,Yu2017JGstrings,ste06}.
The treatment below follows \cite{Ji2022,jic25}.

\subsection{Observations}
An excitation corresponds to a displacement field ${\bf r}$ between two meta-stable states, as well as an energy profile connecting these two minima. From the latter, one defines an energy $E$ from the bottom well to the barrier top. From the former, one can extract: (i) the norm $|d{\bf r}|$ of the displacement field, (ii) the characteristic displacement $\delta$ of the particles (here chosen to be that of the particle moving the most), (iii) the number $V$ of particles involved (technically, it is extracted from the participation ratio as usually done) and (iv) the characteristic correlation length $\ell$ of the displacement. Note that we expect trivially (as observed) that:
\begin{equation}
\label{rr}
|d r|^2\sim V \delta^2.
\end{equation}

\textbf{Minimal energy excitation:}
In \cite{Ji2022}, the lowest energy excitation in systems of a few thousand particles is studied, as the parent temperature $T$ is varied. A remarkable finding is that architectural properties of minimal energy excitations scale with their energy, as already apparent in the green stars of Fig.~\ref{fig:core}(a--c). One finds more generally:
\begin{equation}
\label{del}
\begin{aligned}
V_{\min}(T) &\sim E_{\min}(T)^{-1/3},\\
\delta_{\min}(T) &\sim E_{\min}(T)^{1/3},\\
|dr_{\min}(T)|^{2} &\sim E_{\min}(T)^{1/3},\\
\ell_{\min}(T) &\sim E_{\min}(T)^{-1/6}.
\end{aligned}
\end{equation}
In words, under cooling, excitations become less extended,  involve fewer particles, but move each of them more, and cost more energy.

\textbf{Higher-energy excitation:} The algorithms of \cite{cia24,jic25} allow one to study how the architecture of excitations depends both on the parent temperature $T$ and on the excitation energy $E$. As shown in Fig.~\ref{fig:core}(a--c), one still finds a simple relation for the maximal displacement, which only depends on energy: 
\begin{equation}
\label{core}
\delta(T,E)\sim E^{1/3}.
\end{equation}
However, other quantities, such as the excitation norm, depend on both $T$ and $E$.
At fixed energy, the norm is larger the less stable the system is (i.e., larger $T$).

Finally, Ref.\cite{jic25} observes that the probability for excitations to be string-like is governed by the particle scale displacement $\delta$: when it becomes of order of the inter-particle distance, that probability becomes significant, as particles can start exchanging position.  Since activation energy grows under cooling,  strings become more predominant at low temperatures.

\subsection{Scaling theory and dynamical transition}

\textbf{Minimal energy excitation:}
We follow the treatment of ~\cite{Ji2022}. We model the lowest-energy excitations as a symmetric double well (relaxing the symmetry does not change the scaling results), whose curvature $\omega_{min}$ characterizes the distance to a dynamical transition:
\begin{equation}
\label{eqsm}
    E(X)=-\omega^2_{min} X^2+\alpha X^4
\end{equation}
where $\alpha$ characterizes the quartic non-linearity along a soft mode. It is straightforward to check by considering the minima of $E(X)$ that: (i) the curvature in each of the two minima is of order $\omega^2_{min}$ independently of $\alpha$, (ii) the norm square of the excitation (corresponding to $X_c^2$ where $X_c$ minimizes Eq.\ref{eqsm} ) follows $|dr|^{2}\sim \omega^2_{min}/\alpha$, (iii) the barrier follows $E\sim \omega^4_{min}/\alpha$. 

To obtain the scaling forms Eqs. \ref{del} we must now compute how the quartic term $\alpha$ depends on the proximity of the dynamical transition  and $\omega_{min}$. Intuitively, the argument below quantifies the idea that if a mode is extended, non-linear terms are weaker, as locally particles move less at constant excitation's norm.  Specifically, $\alpha$ can  be estimated building on the mean-field framework~\cite{franz2000non,bir06,franz2011field}, re-interpreting the scaling results for frequency scale, length scale, and correlation function near a dynamical transition as characterizing excitations near that transition. These works predicts a diverging length scale near a dynamical transition going as $\ell_{min}\sim 1/\sqrt{\omega_{min}}$,  associated to a correlation volume $V_{\min}\sim 1/\omega_{min}$ (see \cite{Ji2022} for a detailed discussion). 
 
Ultimately, the quartic term comes from the quartic non-linearity  $\tilde \alpha$ of the particle interaction potential- a constant that does not depend on the proximity of the dynamical transition embodied by $\omega_{min}$. Writing that the quartic energy is the sum of $V_{min}$ terms, each quartic in the local displacement, implies that  $\alpha |dr|^{4}\sim V_{min} \tilde \alpha \delta^4 $.  Injecting Eq.\ref{rr} in this expression, one gets $\alpha\sim 1/V_{min}$. Using (ii) and (iii), one readily obtains the scaling observations Eqs. \ref{del}.

\begin{figure}[t!]
\centering
\includegraphics[width=0.9\linewidth]{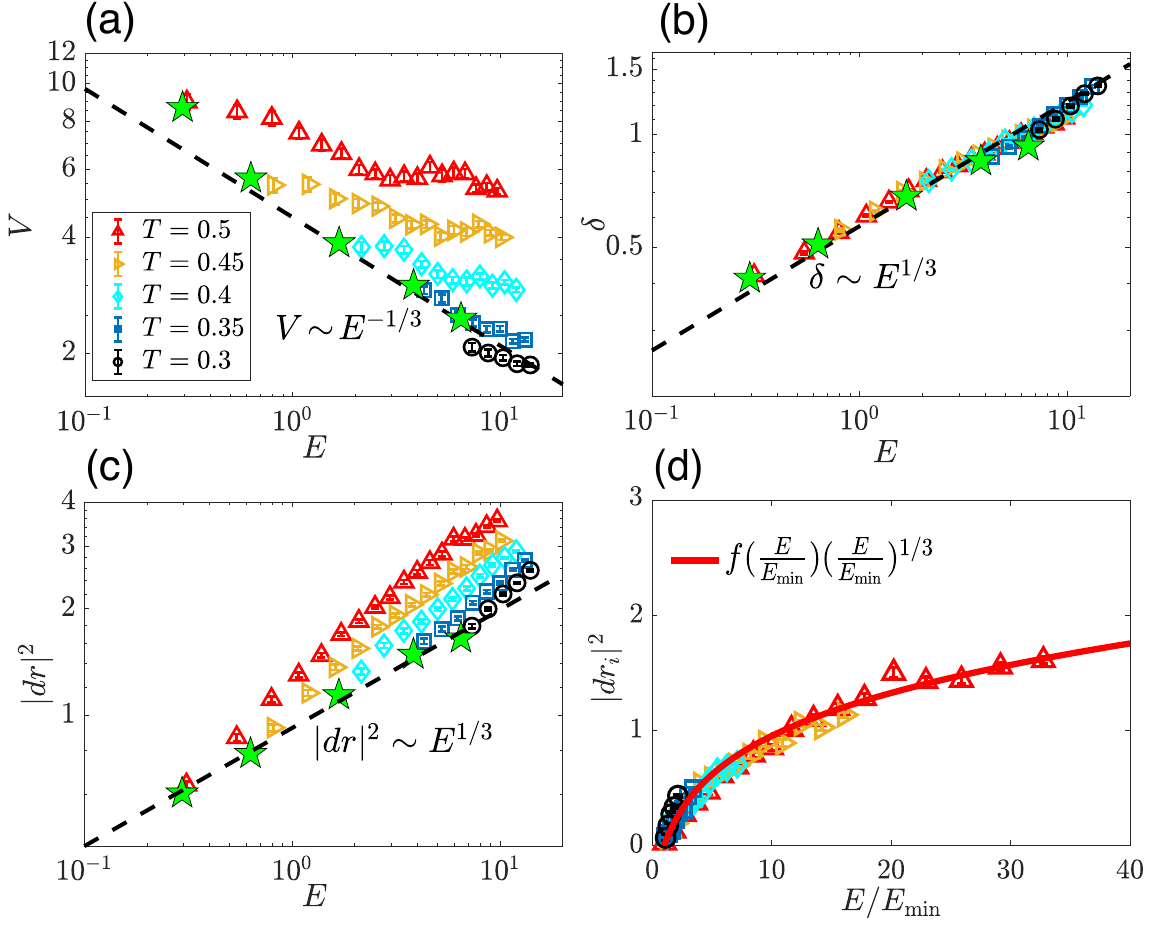}
\caption{
Dependence of (a) Number of particles involved $V$, (b) largest particle displacement $\delta$, and (c) squared norm $|dr|^2$ of the displacement field on the excitation's activation energy $E$, for various temperatures.
The black line $c_0E^{1/3}$ corresponds to the mean field prediction for the excitations with the smallest energy, illustrated in green.
In panel (d), the square norm of the induced displacement field, $|dr_{\rm i}|^2 = |dr|^2-|dr_p|^2$ are collapsed by a model postulating that the excitations have a primary field inducing a $T$-dependent far-field displacement. 
Here, $|dr_p|^2= c_0 E^{1/3}$ where $c_0$ is estimated as $|dr|^2(E=E_{\min})/E_{\rm min}^{1/3}$.
In all panels,  median values are plotted. From \cite{jic25}.
\label{fig:core}
}
\end{figure}

\textbf{Higher energy excitation:}
Ref.\cite{jic25} argues that for $E>E_{min}$, excitations are composed of a ``core'' and a ``corona''.  The core only depends on $E$, and fixes the maximal displacement $\delta$- which thus does not depend on $T$, as observed in Eq.\ref{core}. This core acts as a dipole, triggering a response of extension $\ell_{min}(T)$ that only depends on the stability of the material and thus on $T$, as observed in \cite{jic25}. Scaling for the amplitude of the corona can also be obtained in this picture \cite{jic25}, leading to predictions tested in Fig.~\ref{fig:core}(d).

\textbf{Summary:} A detailed description of excitations in glasses is now available. Under cooling, they involve less particles but move them more, overall costing more energy. Remarkably, the geometry of excitations scale with their energy. These observations can be explained assuming the presence of an underlying dynamical transition. 

Let us recall that the role of a dynamical transition is also consistent with observations of other sections, including the observed shift in the density of excitations $N(E)$ and the decoupling between local and global elasticity.

\section{Excitation-based  theory of Dynamical heterogeneities  }
\label{S8}

\subsection{Equilibrated supercooled liquids}

We now review recent results showing that there is no contradiction between the fact that local barriers control the dynamics and the fact that the latter is heterogeneous. The viewpoint below allows one to accurately predict the coarsening length $\ell_{\rm c}(T,t)$.

\begin{figure}
\centering
\includegraphics[width=\linewidth]{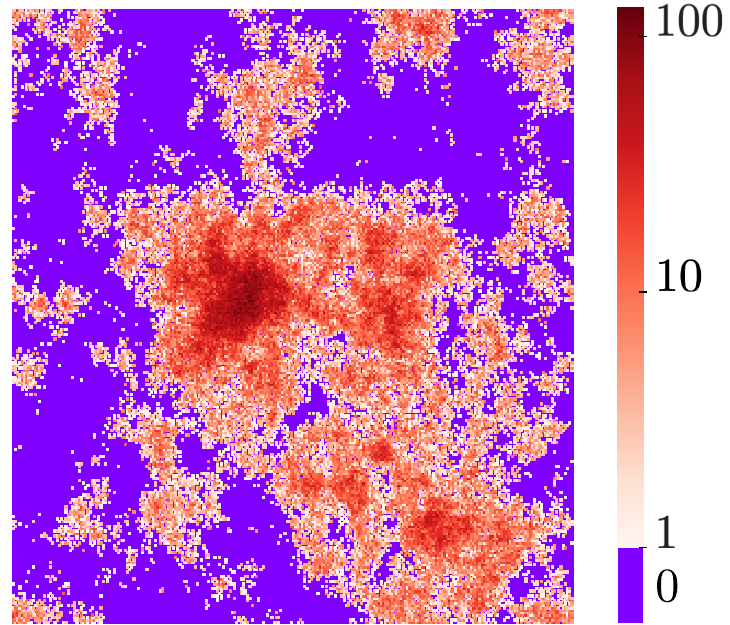}
\caption{
Snapshots of an avalanche formation for the extremal dynamics of the tensorial model with $L=256$.
Event-based avalanche size is $S \simeq 2.7 \times 10^5$.
Purple shows immobile sites (zero event), and the colorbar shows the number of relaxation events in mobile sites. From \cite{tah23}.
}
\label{fig:event_and_site_based_avalanches}
\end{figure}

These ideas relate to the notion of \emph{facilitated dynamics}—that is, the fact that structural relaxation in one region of a supercooled liquid can facilitate relaxation elsewhere. This intuition motivated kinetically constrained models \cite{rit03,fre84}, in which dynamics is facilitated by the presence of nearby defects. Here we consider what we view as a more realistic mesoscopic viewpoint, in which the system is divided into blocks. Each block is characterized by the activation energy $E_i$ of its weakest local barrier, as well as a local stress $\sigma_i$. When a local barrier $i$ rearranges—on the time scale $\exp(E_i/k_BT)$—it transmits a stress kick $G_{ij}$ to each block $j$. This affects the local barrier energy $E_j$, which depends on the stress at that block. Concerning the block that yielded, a new variable $E_i$ is randomly sampled from some distribution.  To our knowledge, the first model of the glass transition formulated in these terms is the elasto-plastic model of Bulatov and Argon \cite{Bul94}. Observations of elastic interaction in liquids abound \cite{lem14,cha21,klo22,wu15}. More recent and systematic studies consider a scalar \cite{oza23} or tensorial stress description \cite{tah23}. All these models exhibit dynamical heterogeneities, as illustrated in Fig.~\ref{fig:event_and_site_based_avalanches}. Simplified models \cite{reh10b}, in which stress conservation is neglected, also display dynamical heterogeneities. We expect all these models to lead to the same scaling picture \footnote{When stress is not conserved, the exponents are expected to correspond to those of directed percolation.} described below, although the precise numerical values of the exponents will differ. 

\begin{figure*}[t!]
\centering
\includegraphics[width=1\linewidth]{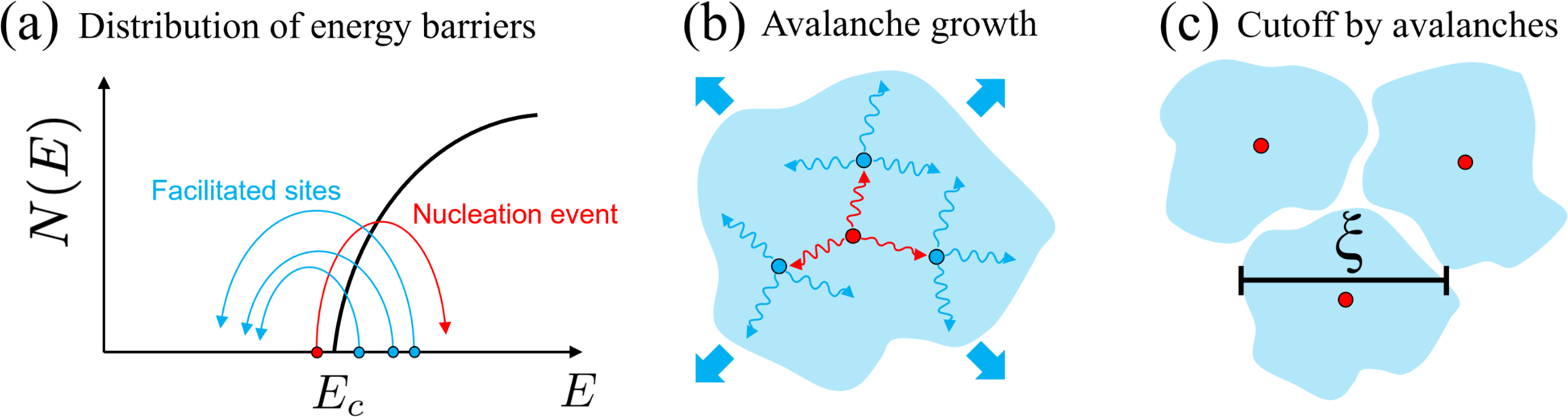}
\caption{Description of relaxation in liquids based on thermal avalanches. 
(a) At low temperatures, the distribution of activation energy barriers $N(E)$ exhibits a gap below a characteristic energy $E_c$. On a time scale $\tau \sim \exp(E_c/k_BT)$, a site with a barrier near $E_c$ relaxes (red arrow), which we refer to as a “nucleation event.” Due to elastic interactions, other sites may then experience lower barriers $E < E_c$ (blue arrows), which we call “facilitated sites.” These sites can require activation and take a significant time to relax, but this time is  much shorter than $\tau$, producing a  cascade of activated events—a thermal avalanche. 
(b) In real space, a nucleation event (red circle) triggers facilitated sites (blue circles) via elastic interactions (wavy arrows), which in turn activate additional sites, leading to avalanche growth.
(c) Different regions of the liquid experience thermal avalanches at different rates, giving rise to a dynamic correlation length $\xi$ that can be estimated from a finite-size scaling argument. From \cite{tah23}.}
\label{thermalavalanche}
\end{figure*}

\textit{Activation energy $E_a$ and the gap in the excitation density $N(E)$:}  
At low temperatures, the density of local barriers $N(E)$ in elasto-plastic models develops an approximate gap of magnitude $E_c$, which governs the relaxation time, implying $E_a = E_c$. Thus, in these models—as in the simple liquids discussed in Sec.~\ref{S4}—local barriers set the time scale of structural relaxation. This result can be obtained by mean-field (infinite-dimensional) analysis \cite{pop20,oza23}. The  physical origin of the gap however is clear: at sufficiently low temperatures, the dynamics of thermal elasto-plastic models approach an extremal dynamics where the weakest site always relaxes first. Such dynamics have been extensively studied in the context of self-organized criticality (SOC) \cite{pac96} and indeed the depletion of sites at small activation energy leads to a gap in finite dimensions.

\textit{Regular avalanches:}  
Elasto-plastic models are known to exhibit avalanches at zero temperature \cite{nic18}, corresponding to cascades of instabilities where certain sites are always unstable ($E < 0$). A common feature observed in disordered systems is their scale-free intermittent response to continuous driving forces. In such cases the dynamics exhibit a `bursty' nature, corresponding to internal rearrangements involving numerous  excitations. The size of these avalanches usually follows a power-law distribution, giving rise to  crackling noise \cite{set01, mul14,bud17,ros22}. Noteworthy examples encompass the strain-stress response of disordered materials \cite{sal14,bar13}, crack propagation \cite{bar18}, the response of magnets to slowly varying magnetic fields (Barkhausen noise) \cite{kim03, per95}, dynamics of vortex matter in type-II superconductors \cite{alt04}, spin glass dynamics \cite{paz99}, bursts  of neuronal activity \cite{beg03,fri12, mor11}, frictional interfaces \cite{fry22}, and earthquakes \cite{gut54,wes94, cor04}.

\textit{Dynamical heterogeneities as thermal avalanches:}  At finite temperature, the notion of avalanche is murkier- see the next section for some review of existing literature. 
In \cite{tah23}, the concept of thermal avalanches was developed, by considering sequences of events involving sites lying within the gap of $N(E)$, instead of sequence of unstable sites with $E<0$. This process is illustrated schematically in Fig.~\ref{thermalavalanche}.

\textit{Length scale $\xi$:}  
Avalanches—thermal or otherwise—are finite-size phenomena. At finite $T$ and in the thermodynamic limit, these cannot be sharply defined as many avalanches occur simultaneously. Reference \cite{deg25} identifies the dynamical correlation length $\xi$ or characteristic size of avalanches using a finite-size scaling argument. For a finite system of size $L$, $E_c$ cannot be uniquely defined but exhibits sample-to-sample fluctuations. Averaging over configurations, the distribution $P_L(E)$ of $E_c$ has a standard deviation $\Delta E_c$ that decays algebraically with $L$ as

\begin{equation}
    \Delta E_c \sim L^{-1 / \nu}.
    \label{eq:nu}
\end{equation}
Consider several finite systems at temperature $T$. If $\Delta E_c \ll T$, fluctuations of the gap are negligible and all systems relax at the same rate. In contrast, if $\Delta E_c \gg T$, systems of different sizes relax at significantly different rates. This defines a temperature-dependent length scale:

\begin{equation}
\label{xi}
    \xi \sim T^{-\nu},
\end{equation}
below which fluctuations in the dynamics are significant. In an infinite system, regions up to size $\xi$ relax at different rates, making $\xi$ the dynamic correlation length. This scaling behavior has been confirmed in both short-range \cite{deg25} and long-range propagator models \cite{tah23}, although the corresponding exponent $\nu$ differs (in the short-range case it is simply the well-known depinning exponent).

\begin{figure}[t!]
\centering
\includegraphics[width=1\linewidth]{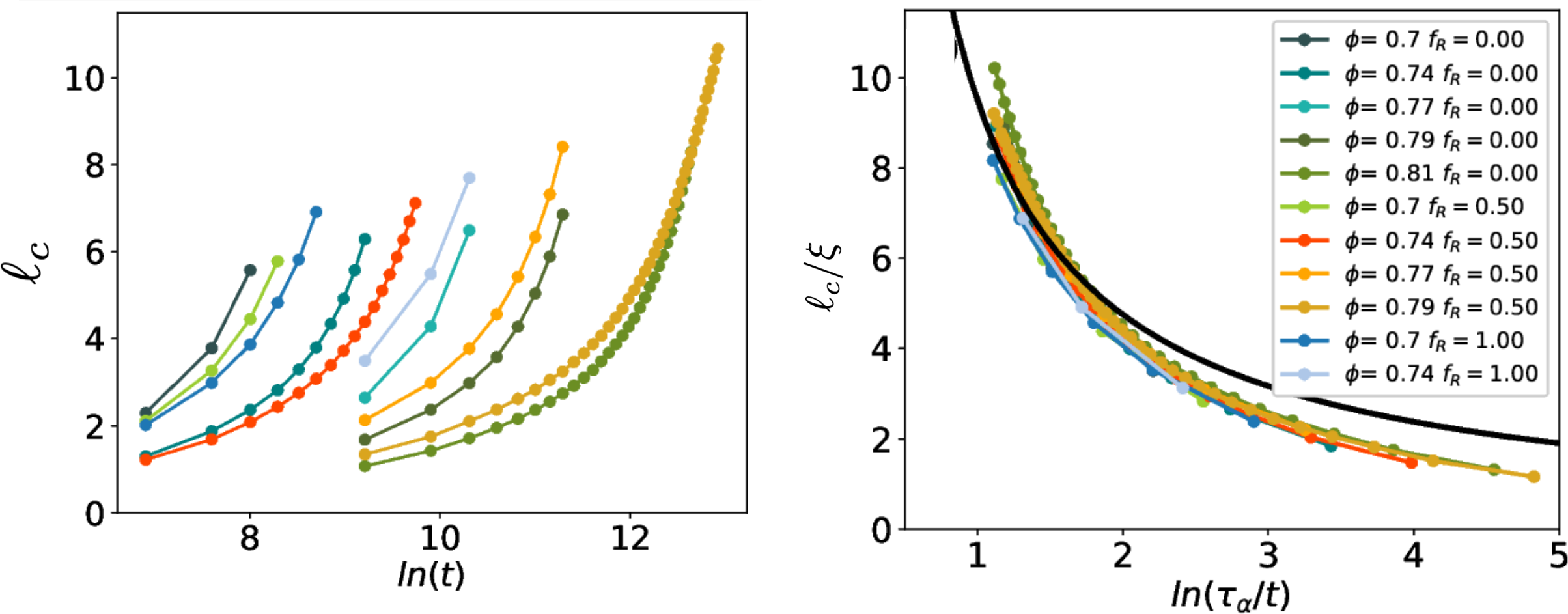}
\caption{Left: Coarsening length $\ell_{\rm c}$ for $d=2$ as a function of $\ln(t)$ for various packing fractions $\phi$ and swapping fractions $f_R$ (see legend in panel {\bf (d)}), with no particles constrained to planes ($f_P = 0$) and $t \le \tau/3$. 
Right: Collapse of $\ell_{\rm c}$ upon axis rescaling, as predicted by Eq.~\ref{eq:ellc}. The solid black line shows the theoretical prediction $l_c = a \, \xi [\ln(\tau_{\alpha}/t)]^{-1}$ with $a = 9.5$ and $\nu \approx 1$ extracted from elasto-plastic models \cite{tah23}. Reproduced with permission from \cite{gav24}.}
\label{fig:collapsecoarseninig}
\end{figure}

\textit{Coarsening length $\ell_{\rm c}(t,T)$:}  
At vanishing temperatures, the dynamics are extremal: the weakest site always relaxes first. Avalanches can then be defined as sequences of $S$  fast events where the most unstable site $E_{\min}$ always satisfies $E_{\min} < E_0$, with $E_0$ a chosen threshold \cite{pac96,pur17}. In the thermodynamic limit, the sequence never stops if $E_0 > E_c$ and is critical at $E_0 = E_c$. For $E_0 < E_c$, avalanches terminate when they reach a spatial extent $\ell_{\rm c}$. This length $\ell_{\rm c}$ corresponds to the scale at which local fluctuations of the gap, $\Delta E_c(\ell_{\rm c})$, become comparable to the difference $E_c - E_0$; on smaller scales, this difference is irrelevant. Using \eq{eq:nu}, one obtains:

\begin{equation}
    \ell_{\rm c} \sim (E_c - E_0)^{-\nu}.
    \label{eq:result3}
\end{equation}
At finite temperatures, the same picture applies, but dynamics now occur on finite time scales. At time $t$, an avalanche stops when the corresponding threshold $E_0$ satisfies $t \sim \exp(E_0/T)$, leading to an avalanche extent given by \eq{eq:result3}. Note that $E_0$ is smaller than the critical value $E_c$, which is reached only at much longer times $\tau_\alpha \sim \exp(E_c/T)$, when relaxation percolates throughout the system. Inserting these relations into \eq{eq:result3} yields the growth of avalanche extent for $t \leq \tau_\alpha$:

\begin{equation}
    \label{eq:ellc}
    \ell_{\rm c} (t, T) \sim T^{-\nu} \left[ \ln (\tau_\alpha / t) \right]^{-\nu} \sim \xi(T) \left[ \ln (\tau_\alpha / t) \right]^{-\nu}.
\end{equation}

\textit{Comparison with molecular dynamics:}  
\Eq{eq:ellc} for the spatio-temporal nature of thermal avalanches predicts that dynamical heterogeneities evolve via a logarithmically slow coarsening process. This prediction was tested in Ref.~\cite{tah23} using existing molecular dynamics (MD) data, with $\nu$ values extracted from elasto-plastic models. The results are shown in Fig.~\ref{figglass1}. The agreement is good, yet a temperature-dependent prefactor was introduced as a fitting parameter in \eq{eq:ellc}. A more stringent test was performed in \cite{gav24}, where hard spheres were simulated at various packing fractions and with different kinetic rules (as illustrated in Fig.~\ref{swap}). The resulting $\ell_{\rm c}(t,\phi)$ curves are shown in Fig.~\ref{fig:collapsecoarseninig} (left). \Eq{eq:ellc} predicts that these curves should collapse upon proper axis rescaling, which is indeed observed in Fig.~\ref{fig:collapsecoarseninig} (right), along with close agreement with the theoretical prediction using $\nu$ values from elasto-plastic models. This constitutes arguably the most quantitative description of dynamical heterogeneities in MD simulations to date.

Note that Eq.\ref{eq:ellc} and Fig.\ref{fig:collapsecoarseninig} are a quantitative incarnation of the observation \cite{ghi24} that both kinetic rules and temperature crucially affect the pace of the dynamics while not affecting the relationship between certain observables. It is explained here  for $\ell_c/\xi$ v.s. $\tau_\alpha/t$.

\subsection{Creep flow and  aging phenomena}

{\it Creep phenomena:}
A variety of disordered systems display analogous behaviors when subjected to comparable driving conditions. Following a perturbation, many of these systems exhibit creep—a slow, logarithmic relaxation extending over a vast range of time scales, from seconds to days or even years. Logarithmic relaxations have been reported in the conductivity of disordered electronic systems and devices \cite{ham93,vak00,gre07,ami11}, in the magnitude of “static” friction \cite{ben10, rub06}, in the shear flow of ice \cite{duv78}, in flux dynamics of certain superconductors \cite{and62}, and in granular media \cite{bou97,kni95,ben96,nic00}.
The same phenomenon also occurs in the volume or stiffness of crumpled sheets \cite{mat02,lah17}, and in the motion of pinned elastic interfaces \cite{bus07,pur17,kol05} below the threshold force required for spontaneous flow. In yield-stress materials, creep can precede sudden fluidization and failure \cite{gio16,bau06,div11,sie12}. Remarkably, pronounced dynamical heterogeneities during creep act as precursors of failure \cite{cab19,liu21,vas22}, highlighting the importance of understanding the mechanisms governing them. In this section, we review evidence that these heterogeneities share the same physical origin as those observed in supercooled liquids.

\begin{figure*}[hbt!]
\centering
\includegraphics[width=1\linewidth]{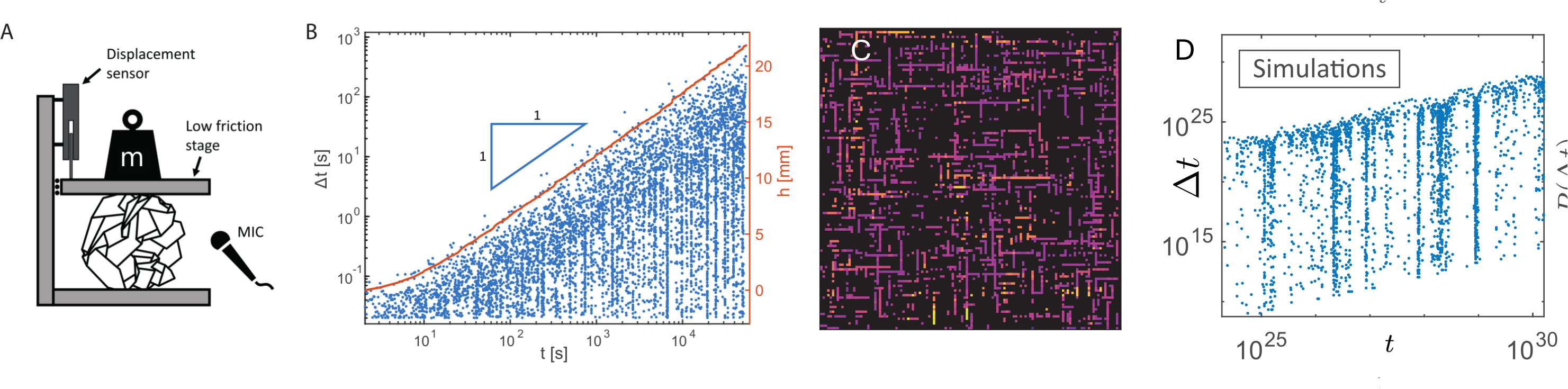}
\caption{
\textbf{a.} Schematic illustration of the experimental setup in \cite{sho23}. A crumpled sheet is compressed by a load $m$. A sensor records the displacement $h(t)$, while a microphone detects acoustic emissions from the sheet.
\textbf{b.} The displacement $h(t)$ (red curve; linear scale, right axis) is plotted alongside the acoustic activity (blue circles; logarithmic scale, left axis), showing a logarithmic creep response. For each acoustic pulse, the waiting time since the previous pulse $\Delta t$ is plotted as a function of $t$. The cutoff of $\Delta t$ increases linearly with $t$, indicating substantial slowing down during creep. Yet, activity occurs in bursts—clusters of simultaneous acoustic events represented by vertical groupings of blue points. (c) visualization of the elasto-plastic model activity over a strain interval of magnitude $0.4$. Instability
locations are colored by the log of their occurrence time, revealing many line-like avalanches. (d) Same quantity as (b) for this elasto-plastic model, where $\Delta t$ is the time interval between the rearrangement of two blocks. Reproduced from https://arxiv.org/pdf/2306.00567 and \cite{kor25}.
}
\label{paper}
\end{figure*}

{\it Experiments on thermal avalanches during creep:}
The intermittent dynamics characteristic of creep can be directly observed in table-top experiments on crumpled sheets, as illustrated in Fig.~\ref{paper} A,B. These studies reveal that, during creep, the system evolves through discrete events whose characteristic time scales become increasingly long as the system ages. This is evidenced by measurements of the time intervals between acoustic emissions, which signal local rearrangements. At late times, the dynamics consists of bursts of very slow events—behavior fully consistent with the picture of thermal avalanches introduced in Sec.~\ref{S8}. A recent study on ultrathin magnetic films reports a similar scenario \cite{dur23}.

{\it Theory of intermittency in creep flow:}
From a numerical standpoint, avalanches occurring at low applied forces and temperatures have been extensively studied in pinned elastic interfaces \cite{cha00,bus07,pur17,kol05}. Algorithms have been developed to identify the irreversible rearrangements that constitute these avalanches \cite{kol06,kol09,fer21}. Renormalization Group (RG) arguments \cite{cha00} predict the emergence of avalanches. However, as emphasized in \cite{cha00}, determining how the correlation length depends simultaneously on temperature and on avalanche duration remains a challenge within the RG framework.

\begin{figure}[htp]
    \centering
    \includegraphics[width=\linewidth]{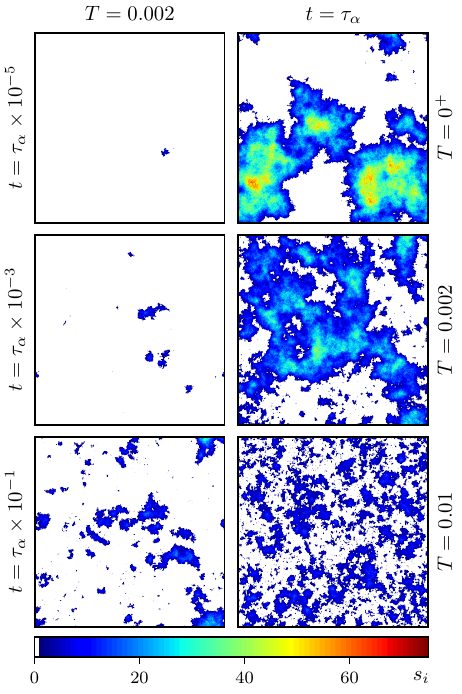}
    \caption{
        Left column: number of times $s_i$ that a block failed during a time interval $t$ (increasing from top to bottom) for our lowest temperature $T = 0.002$ and our biggest system $L = 2^9 = 512$.
        Right column: same quantity for different temperatures (increasing from top to bottom) at $t = \tau_\alpha$, defined as the time for which half of the sites failed at least once. Reproduced with permission from \cite{deg25}.
    }
    \label{fig:snapshots}
\end{figure}

The concept of thermal avalanches, where mechanical and thermal noise compete, is expected to apply broadly to materials that exhibit creep. Beyond liquids, the theoretical framework presented in the previous section has been successfully tested in several systems:

(i) In a mesoscopic model of a pinned elastic interface with short-range interactions, a force $f < f_c$ was applied at finite $T$ until a stationary state was reached. The interface roughness was found to be cut off beyond a characteristic length $\xi \sim T^{-\nu}$  \cite{deg25}. This length scale is also visible by considering the motion of the interface after half of the sites have relaxed, as shown in Fig.\ref{fig:snapshots}, Right. Moreover, a coarsening mechanism very similar to that of liquids appear, see Fig.\ref{fig:snapshots}, Left. Its length $\ell(t,T)$ follows Eq.~\ref{eq:ellc} \cite{deg25}.

(ii) In an elasto-plastic model driven at a stress $\Sigma$ below the yield stress, in the aging regime where strain increases logarithmically with time \cite{kor25}. Time intervals between plastic events become extremely large, but as for the crumpled paper, they still cluster into bursts as shown in Fig.~\ref{paper} C,D.   Both in this model and in the crumpled sheet, dynamical heterogeneities appear to be governed by thermal avalanches, whose size follows a relation of the form of Eq.~\ref{eq:ellc}.

In these set-ups, logarithmic aging was shown to be controlled by the slow growth of a gap in the excitation density $N(E)$. This scenario is reminiscent of the “exhaustion” theory proposed for metals half a century ago \cite{cot52}. Note that the gap remains effectively constant over the time scale of a single avalanche, so their statistics during aging or in a stationary state are very similar.

(iii) In glasses at low temperatures, the statistics of rearrangements follows that of thermal avalanches, with exponents relatively close to those found in elasto-plastic models \cite{takaha2025avalanche}.

\section{Excitation-based theory for the fragility of different materials}
\label{S9}

Many competing theories of the glass transition have been proposed. By contrast, explicit explanations of which structural features of a glass control the liquid’s fragility remain comparatively scarce. Here we review the work of Ref.~\cite{yan13}, which considers that relaxation in liquids is governed by local excitations whose interactions are set by the material’s rigidity—rather than by the cruder approximation adopted in the previous section, where the excitation–excitation kernel $G_{ij}$ was taken from continuum elasticity.  Such a view leads to an explanation for an intriguing empirical observation of super-cooled network liquids: glasses near a rigidity threshold, known to host an abundance of soft modes~\cite{phi85,deg14_pnas}, are strong and display a small specific-heat jump at the glass transition. In contrast, both under-constrained and over-constrained networks are more fragile and exhibit a larger specific-heat jump. 

\begin{figure*}[t!]
\centering
\includegraphics[width=1\linewidth]{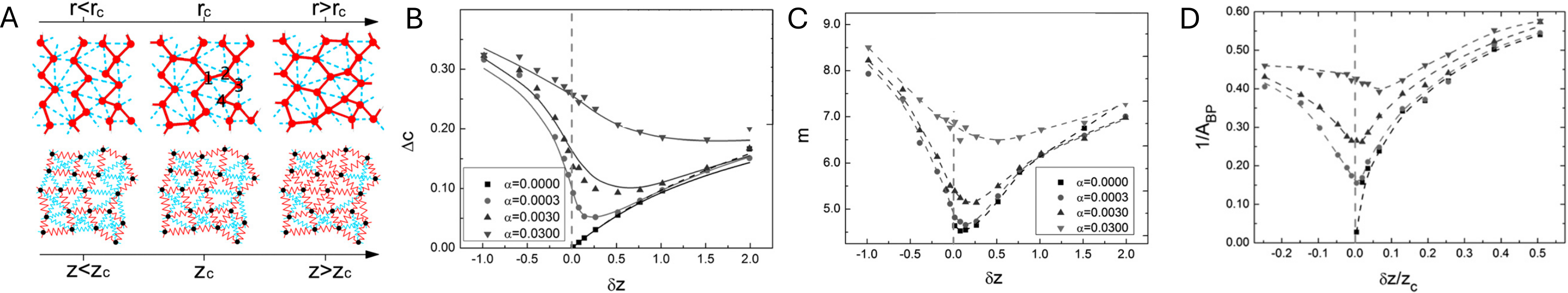}
\caption{
A/Sketches of covalent networks with different mean valence $r$ around the valence $r_c$ . Red solid lines represent covalent bonds; cyan dashed lines represent van der Waals interactions. (Lower) Sketch of our elastic network model with varying coordination number $z$ (defined as the average number of strong springs in red) around Maxwell threshold $z_c$; cyan springs have a much weaker stiffness and model weak interactions. 
B/ Jump of specific heat $\Delta c$ versus excess coordination $\delta z$ in $d = 2$ for different strength of weak springs $\alpha$, as indicated in the legend. Solid lines are mean-field predictions. In both cases the specific heat is computed at the numerically obtained temperature $T_{\rm g}$ . C/ Fragility $m$ vs. excess coordination $\delta z$ for different strength of weak interactions $\alpha$ as indicated in the legend, in $d = 2$. Dashed lines are guide to the eyes and reveal the non-monotonic behavior of $m$ near the rigidity transition.
D/ Inverse boson peak amplitude $1/A_{BP}$ vs. excess coordination $\delta z$ in a $d=3$ elastic network model, for different weak interaction strengths as indicated in the legend. Dashed lines are drawn to guide one’s eyes. From \cite{yan13}.
\label{fig:yan}
} 
\end{figure*}

In Ref.~\cite{yan13} a disordered elastic-network model is introduced to represent covalent networks, with three essential ingredients. (i) \emph{Strong springs} model covalent bonds; they have stiffness $k$, mean coordination $z$, and an associated Maxwell rigidity threshold at $z_c=2d$. We denote the excess coordination by $\delta z \equiv z - z_c$. (ii) \emph{Weak interactions} (van der Waals–like) are represented by springs of stiffness $k_w$ and large coordination $z_w$. The dimensionless parameter $\alpha \equiv k_w/k$ controls their relative strength. (iii) \emph{Nonlinear excitations} are modeled by allowing strong springs to switch between two rest lengths. The resulting \emph{linear} elastic response—whose properties depend sensitively on $\delta z$ and $\alpha$~\cite{deg14_pnas,ler14b}—couples different excitations. A schematic of the model is shown in Fig.~\ref{fig:yan}A.

As illustrated in Figs.~\ref{fig:yan}B–D, this model reproduces central empirical trends for liquids: both the fragility and the specific-heat jump vary non-monotonically with coordination and attain a minimum at the rigidity transition, leading to an overall anti-correlation between the abundance of soft modes and liquid fragility.

Theoretically, the model maps to a spin-glass–like description, since each strong spring can occupy two states that may be encoded by an Ising variable. In mean-field and annealed treatments (solid lines in Fig.~\ref{fig:yan}B), one arrives at a simple physical picture: near the rigidity threshold, soft modes are plentiful and any configuration of spring states can elastically \emph{adapt} to lower its energy. Consequently, the effective number of degrees of freedom that \emph{cost} energy—and hence contribute to the specific heat—is small. This elastic adaptability near isostaticity underlies both the reduced specific-heat jump and the strong (nearly Arrhenius) behavior observed in that regime.

\section{Summary}\label{S12}
We have presented a theoretical framework for the glass transition in which \emph{local barriers} (or \emph{excitations}) govern the relaxation of supercooled liquids. The two most central ingredients are (a) a dynamical or Goldstein crossover at high temperature. It was argued previously \cite{wol12} to increase the shear modulus under cooling; here we argue it leads to an even more pronounced stiffening of a ``local modulus''. Importantly, it also corresponds to an upward shift on the excitation spectrum and affects the architecture of these excitations. (b) Interactions among these excitations generate dynamical heterogeneities, most naturally viewed as \emph{thermal avalanches}. (c) The material’s rigidity or connectivity controls excitation–excitation interactions, thereby influencing the specific heat and, in turn,  fragility.

This perspective accounts for a broad set of observations:
\begin{itemize}
\item The empirical correlation between the thermal evolution of the high-frequency shear modulus and the local modulus with liquid fragility;
\item Numerical evidence that the shift in the density of excitations quantitatively match changes in the activation energy;
\item The observation that excitations \emph{shrink} upon cooling yet displace particles more strongly;
\item The quantitative evolution of the coarsening length $\ell(T,t)$ characterizing dynamical heterogeneities;
\item The close analogy between dynamical heterogeneities in liquids and in creep flow;
\item The impact of swap algorithms—and, more generally, of altered kinetic rules—which modify the dynamical transition and the excitation density;
\item Correlations between fragility and the presence of excess soft elastic modes.

\end{itemize}

\section{Outlook}\label{S11}

\subsection{Universality and experimental tests}

A limitation of the results presented here is their focus on numerical “simple” liquids, in particular polydisperse systems for which swap moves are highly effective. In our view, these results largely settle the question of the dominant relaxation mechanism in such systems. The observed thermal dependence of the high-frequency shear modulus suggests that this framework could also apply, at least qualitatively, to a variety molecular liquids. It is primordial to know if it is the case.  As we emphasize in this review, to make progress in this field and distinguish theories  more empirical quantitative analysis must be performed. In particular, it is   crucial to measure the activation energy in experiments via reheating measurements, since that measurement removes a fitting parameter for all theories. It allowed us here to distinguish between different models of local barriers; more generally we expect this measure to discard a variety of approaches. The excitation-based framework developed here could be tested by comparing the measured activation energy to that obtained from the Debye-waller factor Eq.(\ref{eq:pred}).

Direct comparisons between  the coarsening length that characterizes dynamical heterogeneities in liquids or during creep and the theoretical predictions reviewed here would also be extremely informative.

\subsection{Universality and numerical tests}

It is clear that some systems will display specific features that should affect their glassy dynamics. Mildly polydisperse two-dimensional systems, for example,  develop long-range hexatic order~\cite{kaw07}, which contributes to dynamical heterogeneities and modifies local barriers. Similar caveats arise near liquid–liquid phase transitions or in systems with pronounced locally favored structures. As long as these static correlations remain shorter than the dynamical heterogeneities predicted here, however, the latter will not be affected \cite{bou24}. Rapidly evolving locally favored structures will likely alter local barriers \cite{lan25} and contribute to fragility.  Such effects may complicate the temperature dependence of the excitation density, which may not be simply described by a shift—even if that density continues to control the activation energy.

The question of universality is thus central \cite{nis18}. We hope this review will stimulate efforts to classify liquid behaviors systematically. On the numerical side, several tools discussed here are well suited for quantitative studies, including precise measurements of both activation energies and excitation densities, as well as the controlled manipulation of kinetic rules to rule out scenarios purely based on thermodynamic activation barriers. 

 \subsection{Energy vs.\ Free-Energy Landscapes}
Most of the frameworks reviewed above ultimately aim to characterize local free-energy barriers, because free energy---not the bare potential energy---governs activated dynamics at finite temperature. For gently varying interaction potentials this distinction is often of minor importance, but when the potential is strongly non-linear the difference between energy and free-energy barriers can be substantial. Important examples include hard spheres, soft spheres in the vicinity of jamming, and interaction potentials with narrow attractive wells. In such cases, barrier estimates based solely on the $T=0$ energy landscape can be quantitatively misleading.

This mismatch has practical consequences for testing theories. Approaches that relate relaxation to global elastic response, e.g., those using the high-frequency shear modulus, are less affected, since the relevant moduli are defined and measured directly at finite $T$. By contrast, theories built on local elasticity or on the statistics of local excitations require finite-$T$ constructs to be predictive across a broad class of interactions. Much of the present focus has therefore been on ``gentle'' potentials where athermal proxies work reasonably well.

For highly non-linear interactions, observables should be defined at finite temperature. In practice, one can construct effective vibrational modes and associated local stiffness fields directly from thermal fluctuations, both for hard spheres~\cite{bri07,deg14_pnas} and for soft spheres near jamming~\cite{deg15b}. The same protocols can be used to infer local elastic moduli. Estimating the density of excitations in these settings is more involved and calls for finite-$T$ rare-event tools (e.g., finite-temperature path sampling or string-method variants) to determine local barrier heights in a thermodynamically consistent way. Together, these steps provide a route to testing excitation-based and local-elasticity theories beyond the realm of gentle potentials.

\subsection{Stretching exponents and Stokes--Einstein breakdown}

Two additional hallmarks of liquid dynamics are (i) stretched-exponential relaxation~\cite{ang00} and (ii) the decoupling between particle diffusion and structural relaxation time, commonly termed the Stokes--Einstein (SE) violation~\cite{tar95,edi00,swa09,swa11,sen13,cha14b,kaw17}. Physically, SE breakdown arises from the spatially heterogeneous nature of relaxation: highly mobile regions host an intense accumulation of rearrangements that enhances the diffusion constant $D$ relative to the structural relaxation time because fast-moving particles contribute very much to $D$ but little to the overall structural relaxation.

In elasto-plastic models of the glass transition, SE violation is present and can be traced to the roughness of thermal avalanches: within a large avalanche, individual sites typically relax multiple times~\cite{tah23}, as apparent in Fig.\ref{fig:event_and_site_based_avalanches}. For the parameter choices explored here, however, this decoupling is modest and the relaxation remains close to single-exponential---features characteristic of strong liquids. An important open question is whether more fragile-like behavior can be obtained by tuning parameters within elasto-plastic models.

A positive indication comes from facilitated trap models of glassy dynamics, which though cruder than elasto-plastic descriptions (site interactions are not mediated by stress) capture detailed aspects of experimental relaxation spectra~\cite{sca21}. In these models, a broad distribution of trap energies---interpretable as a broad distribution of local barriers across blocks---naturally generates stretched relaxation and SE decoupling: some blocks relax much more slowly than others. Such heterogeneity may effectively model the influence of locally favored structures that are slow to relax, providing a route to more fragile-like phenomenology.


\bibliography{merged_unique}

@article{henot2024crossing,
  title={Crossing the frontier of validity of the material time approach in the aging of a molecular glass},
  author={H{\'e}not, Marceau and Nguyen, Xuan An and Ladieu, Fran{\c{c}}ois},
  journal={The Journal of Physical Chemistry Letters},
  volume={15},
  number={11},
  pages={3170--3177},
  year={2024},
  publisher={ACS Publications}
}

@article{Berthier2005,
  author  = {Berthier, Ludovic and Biroli, Giulio and Bouchaud, Jean-Philippe and Cipelletti, Luca and El Masri, Dalila and L'H\^ote, Damien and Ladieu, Fran\c{c}ois and Pierno, Matteo},
  title   = {Direct Experimental Evidence of a Growing Length Scale Accompanying the Glass Transition},
  journal = {Science},
  volume  = {310},
  number  = {5755},
  pages   = {1797--1800},
  year    = {2005},
  doi     = {10.1126/science.1117086}
}

@article{takaha2025avalanche,
  title={Avalanche criticality emerges by thermal fluctuation in a quiescent glass},
  author={Takaha, Yuki and Mizuno, Hideyuki and Ikeda, Atsushi},
  journal={Physical Review E},
  volume={112},
  number={4},
  pages={L043401},
  year={2025},
  publisher={APS}
}

@article{mad26,
  title={MAPping dynamic heterogeneity in supercooled glass-formers},
  author={Madanchi, Ata and Simine, Lena},
  journal={The Journal of Chemical Physics},
  volume={164},
  number={7},
  year={2026},
  publisher={AIP Publishing}
}

@article{bha08,
  title={Facilitation, complexity growth, mode coupling, and activated dynamics in supercooled liquids},
  author={Bhattacharyya, Sarika Maitra and Bagchi, Biman and Wolynes, Peter G},
  journal={Proceedings of the National Academy of Sciences},
  volume={105},
  number={42},
  pages={16077--16082},
  year={2008},
  publisher={National Academy of Sciences}
}

@article{ada65,
  author = {G. Adam and J. H. Gibbs},
  doi = {10.1063/1.1696442},
  journal = {J. Chem. Phys.},
  pages = {139--146},
  title = {On Temperature Dependence of Cooperative Relaxation Properties in Glass-Forming Liquids},
  volume = {43},
  year = {1965}
}

@article{ang00,
  author = {C. A. Angell and K. L. Ngai and G. B. McKenna and P. F. McMillan and S. W. Martin},
  doi = {10.1063/1.1286035},
  journal = {J. Appl. Phys.},
  pages = {3113-3157},
  title = {Relaxation in glassforming liquids and amorphous solids},
  volume = {88},
  year = {2000}
}

@article{ber11,
  author = {L. Berthier and G. Biroli},
  doi = {10.1103/RevModPhys.83.587},
  journal = {Rev. Mod. Phys.},
  pages = {587--645},
  title = {Theoretical Perspective on the Glass Transition and Amorphous Materials},
  volume = {83},
  year = {2011}
}

@article{boh14a,
  author = {R. B{\"o}hmer and C. Gainaru and R. Richert},
  doi = {http://dx.doi.org/10.1016/j.physrep.2014.07.005},
  journal = {Phys. Rep.},
  pages = {125 - 195},
  title = {Structure and dynamics of monohydroxy alcohols -- {Milestones} towards their microscopic understanding{, 100}\, years after {Debye}},
  volume = {545},
  year = {2014}
}

@book{bra85,
  author = {S. Brawer},
  publisher = {American Ceramic Society, Columbus, OH},
  title = {Relaxation in Viscous Liquids and Glasses},
  year = {1985}
}

@article{bri06,
  author = {C. Brito and M. Wyart},
  journal = {EPL (Europhysics Letters)},
  pages = {149},
  title = {On the rigidity of a hard-sphere glass near random close packing},
  volume = {76},
  year = {2006}
}

@article{bri07,
  author = {C. Brito and M. Wyart},
  journal = {J. Stat. Mech.},
  number = {08},
  pages = {L08003},
  title = {Heterogeneous dynamics, marginal stability and soft modes in hard sphere glasses},
  volume = {2007},
  year = {2007}
}

@article{bri09,
  author = {C. Brito and M. Wyart},
  journal = {J. Chem. Phys.},
  pages = {024504-1 - 024504-14},
  title = {Geometric interpretation of previtrification in hard sphere liquids},
  volume = {131},
  year = {2009}
}

@article{buc92,
  author = {U. Buchenau and R. Zorn},
  doi = {10.1209/0295-5075/18/6/009},
  journal = {Europhys. Lett.},
  pages = {523--528},
  title = {A relation between fast and slow motions in glassy and liquid selenium},
  volume = {18},
  year = {1992}
}

@article{cha14,
  article-number = {{3725}},
  author = {P. Charbonneau and J. Kurchan and G. Parisi and P. Urbani and F. Zamponi},
  doi = {{10.1038/ncomms4725}},
  journal = {Nat. Commun.},
  pages = {3725},
  title = {{Fractal free energy landscapes in structural glasses}},
  volume = {{5}},
  year = {{2014}}
}

@book{deb96,
  author = {P. G. Debenedetti},
  publisher = {Princeton University Press},
  title = {Metastable Liquids: Concepts and Principles},
  year = {1996}
}

@article{rid24,
  title={The dynamics of machine-learned" softness" in supercooled liquids describe dynamical heterogeneity},
  author={Ridout, Sean A and Liu, Andrea J},
  journal={arXiv preprint arXiv:2406.05868},
  year={2024}
}

@article{deg14_pnas,
  author = {DeGiuli, Eric and Lerner, Edan and Brito, Carolina and Wyart, Matthieu},
  doi = {10.1073/pnas.1415298111},
  journal = {Proc. Natl. Acad. Sci. (USA)},
  pages = {17054--17059},
  title = {Force distribution affects vibrational properties in hard-sphere glasses},
  volume = {111},
  year = {2014}
}

@article{dol03,
  author = {B. Doliwa and A. Heuer},
  journal = {J. Phys.: Condens. Matter},
  pages = {S849-S858},
  title = {Finite-size effects in a supercooled liquid},
  volume = {15},
  year = {2003}
}

@article{doo51,
  author = {A. K. Doolittle},
  doi = {10.1063/1.1699894},
  journal = {J. Appl. Phys.},
  pages = {1471--1475},
  title = {Studies in {Newtonian} Flow. II. {The} Dependence of the Viscosity of Liquids on Free‐Space},
  volume = {22},
  year = {1951}
}

@article{dyr04,
  author = {Dyre, Jeppe C. and Olsen, Niels Boye},
  doi = {10.1103/PhysRevE.69.042501},
  issue = {4},
  journal = {Phys. Rev. E},
  pages = {042501},
  title = {Landscape equivalent of the shoving model},
  volume = {69},
  year = {2004}
}

@article{dyr06,
  author = {J. C. Dyre},
  doi = {10.1103/RevModPhys.78.953},
  journal = {Rev. Mod. Phys.},
  pages = {953-972},
  title = {The Glass Transition and Elastic Models of Glass-Forming Liquids},
  volume = {78},
  year = {2006}
}

@article{dyr09,
  author = {J. C. Dyre and T. Hecksher and K. Niss},
  doi = {10.1016/j.jnoncrysol.2009.01.039},
  journal = {J. Non-Cryst. Solids},
  pages = {624--627},
  title = {A brief critique of the {Adam-Gibbs} entropy model},
  volume = {355},
  year = {2009}
}

@article{dyr12,
  author = {J. C. Dyre and W. H. Wang},
  doi = {10.1063/1.4724102},
  journal = {J. Chem. Phys.},
  pages = {224108},
  title = {The Instantaneous Shear Modulus in the Shoving Model},
  volume = {136},
  year = {2012}
}

@article{dyr95a,
  author = {Dyre, J. C.},
  doi = {10.1103/PhysRevB.51.12276},
  journal = {Phys. Rev. B},
  pages = {12276--12294},
  title = {Energy master equation: A low-temperature approximation to {B\"a}ssler's random-walk model},
  volume = {51},
  year = {1995}
}

@article{dyr96,
  author = {J. C. Dyre and N. B. Olsen and T. Christensen},
  doi = {10.1103/PhysRevB.53.2171},
  journal = {Phys. Rev. B},
  pages = {2171--2174},
  title = {Local elastic expansion model for viscous-flow activation energies of glass-forming molecular liquids},
  volume = {53},
  year = {1996}
}

@article{edi00,
  author = {M. D. Ediger},
  doi = {10.1146/annurev.physchem.51.1.99},
  journal = {Annu. Rev. Phys. Chem.},
  pages = {99--128},
  title = {Spatially heterogeneous dynamics in supercooled liquids},
  volume = {51},
  year = {2000}
}

@article{edi96,
  author = {M. D. Ediger and C. A. Angell and S. R. Nagel},
  doi = {10.1021/jp953538d},
  journal = {J. Phys. Chem},
  pages = {13200--13212},
  title = {Supercooled liquids and glasses},
  volume = {100},
  year = {1996}
}

@article{fly68,
  author = {C. P. Flynn},
  doi = {10.1103/PhysRev.171.682},
  journal = {Phys. Rev.},
  pages = {682--698},
  title = {Atomic migration in monatomic crystals},
  volume = {171},
  year = {1968}
}

@article{fra00,
  author = {S. Franz and M. A. Virasoro},
  journal = {J. Phys. A: Math. Gen.},
  number = {5},
  pages = {891-905},
  title = {Quasi-equilibrium interpretation of ageing dynamics},
  volume = {33},
  year = {2000}
}

@article{Micoulaut2022,
  author  = {Micoulaut, M.},
  title   = {Topology and rigidity of silicate melts and glasses},
  journal = {Reviews in Mineralogy and Geochemistry},
  year    = {2022},
  note    = {Discusses constraint counting and isostatic conditions in silica glass networks},
}

@article{swa11,
  title={Self-diffusion of the amorphous pharmaceutical indomethacin near T g},
  author={Swallen, Stephen F and Ediger, MD},
  journal={Soft Matter},
  volume={7},
  number={21},
  pages={10339--10344},
  year={2011},
  publisher={Royal Society of Chemistry}
}

@article{swa09,
  title={Self-diffusion of supercooled tris-naphthylbenzene},
  author={Swallen, Stephen F and Traynor, Katherine and McMahon, Robert J and Ediger, MD and Mates, Thomas E},
  journal={The Journal of Physical Chemistry B},
  volume={113},
  number={14},
  pages={4600--4608},
  year={2009},
  publisher={ACS Publications}
}

@article{fra15,
  title={Universal spectrum of normal modes in low-temperature glasses},
  author={Franz, Silvio and Parisi, Giorgio and Urbani, Pierfrancesco and Zamponi, Francesco},
  journal={Proceedings of the National Academy of Sciences},
  volume={112},
  number={47},
  pages={14539--14544},
  year={2015},
  publisher={National Academy of Sciences}
}

@article{fra15_quasi,
  author = {S. Franz and G. Parisi and P. Urbani},
  journal = {J. Phys. A: Math. Theor.},
  number = {19},
  pages = {19FT01},
  title = {Quasi-equilibrium in glassy dynamics: a liquid theory approach},
  volume = {48},
  year = {2015}
}

@article{gol63,
  author = {M. Goldstein},
  doi = {10.1063/1.1734202},
  journal = {J. Chem. Phys.},
  pages = {3369--3374},
  title = {Some Thermodynamic Aspects of the Glass Transition: Free Volume, Entropy, and Enthalpy Theories},
  volume = {39},
  year = {1963}
}

@article{gol69,
  author = {M. Goldstein},
  doi = {10.1063/1.1672587},
  journal = {J. Chem. Phys.},
  pages = {3728-3739},
  title = {Viscous liquids and the glass transition: A potential energy barrier picture},
  volume = {51},
  year = {1969}
}

@article{hal87,
  author = {R. W. Hall and P. G. Wolynes},
  doi = {10.1063/1.452045},
  journal = {J. Chem. Phys.},
  pages = {2943--2948},
  title = {The aperiodic crystal picture and free energy barriers in glasses},
  volume = {86},
  year = {1987}
}

@book{har76,
  author = {G. Harrison},
  publisher = {Academic, New York},
  title = {The Dynamic Properties of Supercooled Liquids},
  year = {1976}
}

@article{hec08,
  author = {T. Hecksher and A. I. Nielsen and N. B Olsen and J. C. Dyre},
  doi = {10.1038/nphys1033},
  journal = {Nat. Phys.},
  pages = {737--741},
  title = {Little evidence for dynamic divergences in ultraviscous molecular liquids},
  volume = {4},
  year = {2008}
}

@article{hec15a,
  author = {T. Hecksher and J. C. Dyre},
  doi = {10.1016/j.jnoncrysol.2014.08.056},
  journal = {J. Non-Cryst. Solids},
  pages = {14-22},
  title = {A review of experiments testing the shoving model},
  volume = {407},
  year = {2015}
}

@article{nis18,
  author = {K. Niss and T. Hecksher},
  journal = {J. Chem. Phys.},
  pages = {230901},
  title = {Perspective: Searching for simplicity rather than universality in glass-forming liquids},
  volume = {149},
  year = {2018}
}

@article{hed09,
  author = {Hedges, Lester O. and Jack, Robert L. and Garrahan, Juan P. and Chandler, David},
  doi = {10.1126/science.1166665},
  journal = {Science},
  pages = {1309-1313},
  title = {Dynamic Order-Disorder in Atomistic Models of Structural Glass Formers},
  volume = {323},
  year = {2009}
}

@article{I,
  author = {N. P. Bailey and U. R. Pedersen and N. Gnan and T. B. Schr{\o}der and J. C. Dyre},
  journal = {J. Chem. Phys.},
  pages = {184507},
  title = {Pressure-energy correlations in liquids. {I. Results} from computer simulations},
  volume = {129},
  year = {2008}
}

@article{kar14,
  author = {S. Karmakar and C. Dasgupta and S. Sastry},
  doi = {10.1146/annurev-conmatphys-031113-133848},
  journal = {Annu. Rev. Cond. Mat. Phys.},
  pages = {255--284},
  title = {Growing Length Scales and Their Relation to Timescales in Glass-Forming Liquids},
  volume = {5},
  year = {2014}
}

@article{kaw07,
  author = {Kawasaki, Takeshi and Araki, Takeaki and Tanaka, Hajime},
  doi = {10.1103/PhysRevLett.99.215701},
  journal = {Phys. Rev. Lett.},
  pages = {215701},
  title = {Correlation between Dynamic Heterogeneity and Medium-Range Order in Two-Dimensional Glass-Forming Liquids},
  volume = {99},
  year = {2007}
}

@article{kob97,
  author = {W. Kob and C. Donati and S. J. Plimpton and P. H. Poole and S. C. Glotzer},
  issue = {15},
  journal = {Phys. Rev. Lett.},
  pages = {2827-2830},
  title = {Dynamical Heterogeneities in a Supercooled Lennard-Jones Liquid},
  volume = {79},
  year = {1997}
}

@article{lar08,
  author = {L. Larini and A. Ottochian and C. De Michele and D. Leporini},
  doi = {10.1038/nphys788},
  journal = {Nat. Phys.},
  pages = {42-45},
  title = {Universal scaling between structural relaxation and vibrational dynamics in glass-forming liquids and polymers},
  volume = {4},
  year = {2008}
}

@article{lem14,
  author = {A. Lemaitre},
  doi = {10.1103/PhysRevLett.113.245702},
  journal = {Phys. Rev. Lett.},
  pages = {245702},
  title = {Structural Relaxation is a Scale-Free Process},
  volume = {113},
  year = {2014}
}

@article{lub07,
  author = {V. Lubchenko and P. G. Wolynes},
  doi = {10.1146/annurev.physchem.58.032806.104653},
  journal = {Annu. Rev. Phys. Chem.},
  pages = {235--266},
  title = {{Theory of Structural Glasses and Supercooled Liquids}},
  volume = {{58}},
  year = {{2007}}
}

@article{mai16,
  author = {T. Maimbourg and J. Kurchan},
  doi = {10.1209/0295-5075/114/60002},
  journal = {{EPL}},
  pages = {60002},
  title = {Approximate scale invariance in particle systems: {A} large-dimensional justification},
  volume = {114},
  year = {2016}
}

@article{mai16a,
  author = {T. Maimbourg and J. Kurchan and F. Zamponi},
  doi = {10.1103/PhysRevLett.116.015902},
  issue = {1},
  journal = {Phys. Rev. Lett.},
  numpages = {6},
  pages = {015902},
  title = {Solution of the Dynamics of Liquids in the Large-Dimensional Limit},
  volume = {116},
  year = {2016}
}

@article{mar01,
  author = {L.-M. Martinez and C. A. Angell},
  journal = {Nature},
  pages = {663--667},
  title = {A thermodynamic connection to the fragility of glass-forming liquids},
  volume = {410},
  year = {2001}
}

@article{mir14a,
  author = {Mirigian, Stephen and Schweizer, Kenneth S.},
  doi = {10.1063/1.4874842},
  eid = {194506},
  journal = {J. Chem. Phys.},
  number = {19},
  pages = {194506},
  title = {Elastically cooperative activated barrier hopping theory of relaxation in viscous fluids. {I. General} formulation and application to hard sphere fluids},
  volume = {140},
  year = {2014}
}

@article{mir14b,
  author = {Mirigian, Stephen and Schweizer, Kenneth S.},
  doi = {10.1063/1.4874843},
  eid = {194507},
  journal = {J. Chem. Phys.},
  number = {19},
  pages = {194507},
  title = {Elastically cooperative activated barrier hopping theory of relaxation in viscous fluids. {II. Thermal} liquids.},
  volume = {140},
  year = {2014}
}

@article{nin17,
  author = {A. Ninarello and L. Berthier and D. Coslovich},
  doi = {10.1103/PhysRevX.7.021039},
  journal = {Phys. Rev. X},
  pages = {021039},
  title = {Models and algorithms for the next generation of glass transition studies},
  volume = {7},
  year = {2017}
}

@article{sen13,
  author = {S. Sengupta and T. B. Schr{\o}der and S. Sastry},
  journal = {Eur. Phys. J. E},
  pages = {113},
  title = {{Density-Temperature Scaling of the Fragility in a Model Glass-Former}},
  volume = {36},
  year = {2013}
}

@article{sta02,
  author = {F. W. Starr and S. Sastry and J. F. Douglas and S. C. Glotzer},
  doi = {10.1103/PhysRevLett.89.125501},
  journal = {Phys. Rev. Lett.},
  pages = {125501},
  title = {What do we learn from the local geometry of glass-forming liquids?},
  volume = {89},
  year = {2002}
}

@article{tan12,
  author = {H. Tanaka},
  journal = {Eur. Phys. J. E},
  pages = {113},
  title = {Bond orientational order in liquids: Towards a unified description of water-like anomalies, liquid-liquid transition, glass transition, and crystallization},
  volume = {35},
  year = {2012}
}

@article{tar95,
  author = {G. Tarjus and D. Kivelson},
  doi = {10.1063/1.470495},
  journal = {J. Chem. Phys.},
  pages = {3071-3073},
  title = {Breakdown of the {Stokes–Einstein} relation in supercooled liquids},
  volume = {103},
  year = {1995}
}

@article{wag11,
  author = {H. Wagner and D. Bedorf and S. K{\"u}chemann and M. Schwabe and B. Zhang and W. Arnold and K. Samwer},
  doi = {10.1038/NMAT3024},
  journal = {Nat. Mater.},
  pages = {439-442},
  title = {Local elastic properties of a metallic glass},
  volume = {10},
  year = {2011}
}

@article{wid06,
  author = {Widmer-Cooper, Asaph and Harrowell, Peter},
  doi = {10.1103/PhysRevLett.96.185701},
  journal = {Phys. Rev. Lett.},
  pages = {185701},
  title = {Predicting the Long-Time Dynamic Heterogeneity in a Supercooled Liquid on the Basis of Short-Time Heterogeneities},
  volume = {96},
  year = {2006}
}

@book{wol12,
  author = {P. G. Wolynes and V. Lubchenko},
  publisher = {Wiley, New York},
  title = {Structural Glasses and Supercooled Liquids: Theory, Experiment, and Applications},
  year = {2012}
}

@article{wya12,
  author = {M. Wyart},
  doi = {10.1103/PhysRevLett.109.125502},
  issue = {12},
  journal = {Phys. Rev. Lett.},
  numpages = {5},
  pages = {125502},
  title = {Marginal Stability Constrains Force and Pair Distributions at Random Close Packing},
  volume = {109},
  year = {2012}
}

@article{yan13,
  author = {Yan, Le and D{\"u}ring, Gustavo and Wyart, Matthieu},
  doi = {10.1073/pnas.1300534110},
  journal = {PNAS},
  number = {16},
  pages = {6307--6312},
  title = {Why glass elasticity affects the thermodynamics and fragility of supercooled liquids},
  volume = {110},
  year = {2013}
}

@article{zha11,
  author = {H. Zhang and P. Kalvapalle and J. F. Douglas},
  journal = {J. Phys. Chem. B},
  pages = {14068-14076},
  title = {String-Like Collective Atomic Motion in the Melting and Freezing of Nanoparticles},
  volume = {115},
  year = {2011}
}

@article{zha13,
  author = {J. Zhao and S. L. Simon and G. B. McKenna},
  doi = {10.1038/ncomms2809},
  journal = {Nature Commun.},
  pages = {1783},
  title = {Using 20-million-year-old amber to test the super-{Arrhenius} behaviour of glass-forming systems},
  volume = {4},
  year = {2013}
}

@article{mck17,
  author = {G. B. McKenna and S. L. Simon},
  doi = {10.1021/acs.macromol.7b01014},
  journal = {Macromolecules},
  pages = {6333-6361},
  title = {50th Anniversary Perspective: {Challenges} in the Dynamics and Kinetics of Glass-Forming Polymers},
  volume = {50},
  year = {2017}
}

@article{hec19,
  author = {T. Hecksher and N. B. Olsen and J. C. Dyre},
  doi = {10.1073/pnas.1904809116},
  journal = {Proc. Natl. Acad. Sci. (USA)},
  pages = {16736--16741},
  title = {Fast contribution to the activation energy of a glass-forming liquid},
  volume = {116},
  year = {2019}
}

@article{ber05,
  author = {M.N. Berberan-Santos and E.N. Bodunov and B. Valeur},
  doi = {https://doi.org/10.1016/j.chemphys.2005.04.006},
  journal = {Chem. Phys.},
  pages = {171 -- 182},
  title = {Mathematical functions for the analysis of luminescence decays with underlying distributions 1. {Kohlrausch} decay function (stretched exponential)},
  volume = {315},
  year = {2005}
}

@article{reh10,
  author = {C. Rehwald and N. Gnan and A. Heuer and T. Schr\o{}der and J. C. Dyre and G. Diezemann},
  doi = {10.1103/PhysRevE.82.021503},
  journal = {Phys. Rev. E},
  pages = {021503},
  title = {Aging effects manifested in the potential-energy landscape of a model glass former},
  volume = {82},
  year = {2010}
}

@article{wei19,
  author = {D. Wei and J. Yang and M.-Q. Jiang and L.-H. Dai and Y.-J. Wang and J. C. Dyre and I. Douglass and P. Harrowell},
  doi = {10.1063/1.5064531},
  journal = {J. Chem. Phys.},
  pages = {114502},
  title = {Assessing the utility of structure in amorphous materials},
  volume = {150},
  year = {2019}
}

@article{sch20,
  author = {T. B. Schr{\o}der and J. C. Dyre},
  doi = {10.1063/5.0004093},
  journal = {J. Chem. Phys.},
  pages = {141101},
  title = {Solid-like mean-square displacement in glass-forming liquids},
  volume = {152},
  year = {2020}
}

@article{cha14b,
  author = {P. Charbonneau and Y. Jin and G. Parisi and F. Zamponi},
  doi = {10.1073/pnas.1417182111},
  journal = {Proc. Natl. Acad. Sci. {(USA)}},
  pages = {15025--15030},
  title = {Hopping and the {Stokes-Einstein} relation breakdown in simple glass formers},
  volume = {111},
  year = {2014}
}

@article{phi85,
  author = {J. C. Phillips and M. F. Thorpe},
  doi = {https://doi.org/10.1016/0038-1098(85)90381-3},
  journal = {Solid State Commun.},
  pages = {699--702},
  title = {Constraint theory, vector percolation and glass formation},
  volume = {53},
  year = {1985}
}

@article{don98,
  author = {Donati, C. and Douglas, J. F. and Kob, W. and Plimpton, S. J. and Poole, P. H. and Glotzer, S. C.},
  doi = {10.1103/PhysRevLett.80.2338},
  journal = {Phys. Rev. Lett.},
  pages = {2338--2341},
  title = {Stringlike Cooperative Motion in a Supercooled Liquid},
  volume = {80},
  year = {1998}
}

@article{don02,
  author  = {Donati, Claudio and Franz, Silvio and Glotzer, Sharon C. and Parisi, Giorgio},
  title   = {Theory of non-linear susceptibility and correlation length in glasses and liquids},
  journal = {Journal of Non-Crystalline Solids},
  volume  = {307--310},
  pages   = {215--224},
  year    = {2002},
  doi     = {10.1016/S0022-3093(02)01461-8}
}

@article{cic96,
  author = {Cicerone, M. T. and Ediger, M. D.},
  doi = {10.1063/1.471433},
  journal = {J. Chem. Phys.},
  pages = {7210--7218},
  title = {Enhanced translation of probe molecules in supercooled o-terphenyl: {Signature} of spatially heterogeneous dynamics?},
  volume = {104},
  year = {1996}
}

@article{yam98,
  author = {Yamamoto, R. and Onuki, A.},
  doi = {10.1103/PhysRevLett.81.4915},
  journal = {Phys. Rev. Lett.},
  pages = {4915--4918},
  title = {Heterogeneous Diffusion in Highly Supercooled Liquids},
  volume = {81},
  year = {1998}
}

@article{kap21,
  author = {Kapteijns, G. and Richard, D. and Bouchbinder, E. and Schr{\o}der, T. B. and Dyre, J. C. and Lerner, E.},
  doi = {10.1063/5.0051193},
  journal = {J. Chem. Phys.},
  pages = {074502},
  title = {Does mesoscopic elasticity control viscous slowing down in glassforming liquids?},
  volume = {155},
  year = {2021}
}

@article{pat16,
  author = {S. Patinet and D. Vandembroucq and M. L. Falk},
  doi = {10.1103/PhysRevLett.117.045501},
  journal = {Phys. Rev. Lett.},
  pages = {045501},
  title = {Connecting Local Yield Stresses with Plastic Activity in Amorphous Solids},
  volume = {117},
  year = {2016}
}

@article{ler22,
  author = {M. Lerbinger and A. Barbot and D. Vandembroucq and S. Patinet},
  doi = {10.1103/PhysRevLett.129.195501},
  journal = {Phys. Rev. Lett.},
  pages = {195501},
  title = {Relevance of Shear Transformations in the Relaxation of Supercooled Liquids},
  volume = {129},
  year = {2022}
}

@article{bar18,
  author = {A. Barbot and M. Lerbinger and A. Hernandez-Garcia and R. Garcia-Garcia and M. L. Falk and D. Vandembroucq and S. Patinet},
  doi = {10.1103/PhysRevE.97.033001},
  journal = {Phys. Rev. E},
  pages = {033001},
  title = {Local yield stress statistics in model amorphous solids},
  volume = {97},
  year = {2018}
}

@article{oza23,
  author = {M. Ozawa and G. Biroli},
  doi = {10.1103/PhysRevLett.130.138201},
  journal = {Phys. Rev. Lett.},
  pages = {138201},
  title = {Elasticity, Facilitation, and Dynamic Heterogeneity in Glass-Forming Liquids},
  volume = {130},
  year = {2023}
}

@article{cha21,
  author = {R. N. Chacko and F. P. Landes and G. Biroli and O. Dauchot and A. J. Liu and D. R. Reichman},
  doi = {10.1103/PhysRevLett.127.048002},
  journal = {Phys. Rev. Lett.},
  pages = {048002},
  title = {Elastoplasticity Mediates Dynamical Heterogeneity Below the Mode Coupling Temperature},
  volume = {127},
  year = {2021}
}

@article{moo57,
  author = {M. Mooney},
  doi = {10.1122/1.548809},
  journal = {Trans. Soc. Rheol.},
  pages = {63--94},
  title = {A theory of the viscosity of a {Maxwellian} elastic liquid},
  volume = {1},
  year = {1957}
}

@article{bir23,
  author = {G. Biroli and J. P. Bouchaud},
  doi = {10.5802/crphys.136},
  journal = {C. R. Phys.},
  pages = {10.5802/crphys.136},
  title = {The {RFOT} Theory of Glasses: {Recent} Progress and Open Issues},
  volume = {24},
  year = {2023}
}

@article{tor09,
  author = {D. H. Torchinsky and J. A. Johnson and K. A. Nelson},
  doi = {10.1063/1.3072476},
  journal = {J. Chem. Phys.},
  pages = {064502},
  title = {A direct test of the correlation between elastic parameters and fragility of ten glass formers and their relationship to elastic models of the glass transition},
  volume = {130},
  year = {2009}
}

@article{bet18,
  author = {B. A. P. Betancourt and F. W. Starr and J. F. Douglas},
  doi = {10.1063/1.5009442},
  journal = {J. Chem. Phys.},
  pages = {104508},
  title = {String-like collective motion in the $\alpha$- and $\beta$-relaxation of a coarse-grained polymer melt},
  volume = {148},
  year = {2018}
}

@article{cia24,
  author = {M. P. Ciamarra and W. Ji and M. Wyart},
  doi = {10.1073/pnas.2400611121},
  journal = {PNAS},
  pages = {e2400611121},
  title = {Local vs. cooperative: {Unraveling glass transition mechanisms with {SEER}}},
  volume = {121},
  year = {2024}
}

@article{tah23,
  author = {A. Tahaei and G. Biroli and M. Ozawa and M. Popovic and M. Wyart},
  doi = {10.1103/PhysRevX.13.031034},
  journal = {Phys. Rev. X},
  pages = {031034},
  title = {Scaling Description of Dynamical Heterogeneity and Avalanches of Relaxation in Glass-Forming Liquids},
  volume = {13},
  year = {2023}
}

@article{dyr98,
  author = {J. C. Dyre},
  doi = {10.1016/S0022-3093(98)00502-X},
  journal = {J. Non-Cryst. Solids},
  pages = {142--149},
  title = {Source of {non-Arrhenius} average relaxation time in glass-forming liquids},
  volume = {235},
  year = {1998}
}

@article{sca22,
  author = {C. Scalliet and B. Guiselin and L. Berthier},
  doi = {10.1103/PhysRevX.12.041028},
  journal = {Phys. Rev. X},
  pages = {041028},
  title = {Thirty Milliseconds in the Life of a Supercooled Liquid},
  volume = {12},
  year = {2022}
}

@article{nic18,
  author = {A. Nicolas and E. E. Ferrero and K. Martens and J.-L. Barrat},
  doi = {10.1103/RevModPhys.90.045006},
  journal = {Rev. Mod. Phys.},
  pages = {045006},
  title = {Deformation and flow of amorphous solids: {Insights} from elastoplastic models},
  volume = {90},
  year = {2018}
}

@article{dol03a,
  author = {B. Doliwa and A. Heuer},
  doi = {10.1103/PhysRevLett.91.235501},
  journal = {Phys. Rev. Lett.},
  pages = {235501},
  title = {What Does the Potential Energy Landscape Tell Us about the Dynamics of Supercooled Liquids and Glasses?},
  volume = {91},
  year = {2003}
}

@article{ber19,
  author = {L. Berthier and P. Charbonneau and A. Ninarello and M. Ozawa and S. Yaida},
  doi = {10.1038/s41467-019-09512-3},
  journal = {Nat. Commun.},
  pages = {1508},
  title = {Zero-temperature glass transition in two dimensions},
  volume = {10},
  year = {2019}
}

@article{sca21,
  author = {C. Scalliet and B. Guiselin and L. Berthier},
  doi = {10.1063/5.0060408},
  journal = {J. Chem. Phys.},
  pages = {064505},
  title = {{Excess wings and asymmetric relaxation spectra in a facilitated trap model}},
  volume = {155},
  year = {2021}
}

@article{li22,
  author = {Y.-W. Li and Y. Yao and M. P. Ciamarra},
  doi = {10.1103/PhysRevLett.128.258001},
  journal = {Phys. Rev. Lett.},
  pages = {258001},
  title = {Local Plastic Response and Slow Heterogeneous Dynamics of Supercooled Liquids},
  volume = {128},
  year = {2022}
}

@article{hur95,
  author = {M. M Hurley and P. Harrowell},
  doi = {10.1103/PhysRevE.52.1694},
  journal = {Phys. Rev. E},
  pages = {1694--1698},
  title = {Kinetic structure of a two-dimensional liquid},
  volume = {52},
  year = {1995}
}

@book{parisibog,
  author = {G. Parisi and P. Urbani and F. Zamponi},
  doi = {10.1017/9781108120494},
  publisher = {Cambridge University Press},
  title = {Theory of Simple Glasses},
  year = {2020}
}

@article{klo22,
  author = {L. Klochko and J. Baschnagel and J. P. Wittmer and H. Meyer and O. Benzerara and A. N. Semenov},
  doi = {10.1063/5.0085800},
  journal = {J. Chem. Phys.},
  pages = {164505},
  title = {Theory of length-scale dependent relaxation moduli and stress fluctuations in glass-forming and viscoelastic liquids},
  volume = {156},
  year = {2022}
}

@article{ler21,
  author = {E. Lerner and E. Bouchbinder},
  doi = {10.1063/5.0069477},
  journal = {J. Chem. Phys.},
  pages = {200901},
  title = {Low-energy quasilocalized excitations in structural glasses},
  volume = {155},
  year = {2021}
}

@article{tan19,
  author = {H. Tanaka and H. Tong and R. Shi and J. Russo},
  doi = {10.1038/s42254-019-0053-3},
  journal = {Nat. Rev. Phys.},
  pages = {333--348},
  title = {Revealing key structural features hidden in liquids and glasses},
  volume = {1},
  year = {2019}
}

@article{wid08,
  author = {A. Widmer-Cooper and H. Perry and P. Harrowell and D. R. Reichman},
  doi = {10.1038/nphys1025},
  journal = {Nat. Phys.},
  pages = {711--715},
  title = {Irreversible reorganization in a supercooled liquid originates from localized soft modes},
  volume = {4},
  year = {2008}
}

@article{alb22,
  author = {C. Alba-Simionesco and G. Tarjus},
  doi = {10.1016/j.nocx.2022.100100},
  journal = {J. Non-Cryst. Solids X},
  pages = {100100},
  title = {A perspective on the fragility of glass-forming liquids},
  volume = {14},
  year = {2022}
}

@article{alb23,
  author = {C. Alba-Simionesco},
  doi = {10.5802/crphys.148},
  journal = {C. R. Phys.},
  pages = {10.5802/crphys.148},
  title = {Organic Glass-Forming Liquids and the Concept of Fragility},
  volume = {24},
  year = {2023}
}

@article{gui22,
  author = {B. Guiselin and C. Scalliet and L. Berthier},
  doi = {10.1038/s41567-022-01508-z},
  journal = {Nat. Phys.},
  pages = {468--472},
  title = {Microscopic origin of excess wings in relaxation spectra of supercooled liquids},
  volume = {18},
  year = {2022}
}

@inbook{ang85,
  author = {C. A. Angell},
  editor = {K. L. Ngai and G. B. Wright},
  pages = {3--11},
  publisher = {U.S. GPO},
  title = {Strong and fragile liquids},
  year = {1985}
}

@article{sch91a,
  author = {K. Schmidt-Rohr and H. W. Spiess},
  doi = {10.1103/PhysRevLett.66.3020},
  journal = {Phys. Rev. Lett.},
  pages = {3020--3023},
  title = {Nature of nonexponential loss of correlation above the glass transition investigated by multidimensional {NMR}},
  volume = {66},
  year = {1991}
}

@article{wen00,
  author = {H. Wendt and R. Richert},
  doi = {10.1103/PhysRevE.61.1722},
  journal = {Phys. Rev. E},
  pages = {1722--1728},
  title = {Heterogeneous relaxation patterns in supercooled liquids studied by solvation dynamics},
  volume = {61},
  year = {2000}
}

@article{ric02,
  author = {R. Richert},
  doi = {10.1088/0953-8984/14/23/201},
  journal = {J. Phys. Cond. Mat.},
  pages = {R703--R738},
  title = {Heterogeneous dynamics in liquids: fluctuations in space and time},
  volume = {14},
  year = {2002}
}

@article{rit03,
  author = {F. Ritort and P. Sollich},
  doi = {10.1080/0001873031000093582},
  journal = {Adv. Phys.},
  pages = {219--342},
  title = {Glassy dynamics of kinetically constrained models},
  volume = {52},
  year = {2003}
}

@article{coh59,
  author = {M. H. Cohen and D. Turnbull},
  doi = {10.1063/1.1730566},
  journal = {J. Chem. Phys.},
  pages = {1164--1169},
  title = {Molecular Transport in Liquids and Glasses},
  volume = {31},
  year = {1959}
}

@inbook{gre81,
  author = {G. S. Grest and M. H. Cohen},
  chapter = {Adv. Chem. Phys., Vol. 48},
  doi = {10.1002/9780470142684},
  editor = {I. Prigogine and S. A. Rice},
  pages = {455--525},
  publisher = {Wiley},
  title = {Liquids, glasses, and the glass transition: {A} free-volume approach},
  year = {1981}
}

@article{fre84,
  author = {G. H. Fredrickson and H. C. Andersen},
  doi = {10.1103/PhysRevLett.53.1244},
  journal = {Phys. Rev. Lett.},
  pages = {1244--1247},
  title = {Kinetic {Ising} Model of the Glass Transition},
  volume = {53},
  year = {1984}
}

@article{gar02,
  author = {J. P. Garrahan and D. Chandler},
  doi = {10.1103/PhysRevLett.89.035704},
  journal = {Phys. Rev. Lett.},
  pages = {035704},
  title = {Geometrical Explanation and Scaling of Dynamical Heterogeneities in Glass Forming Systems},
  volume = {89},
  year = {2002}
}

@article{bor04,
  author = {P. Bordat and F. Affouard and M. Descamps and K. L. Ngai},
  doi = {10.1103/PhysRevLett.93.105502},
  journal = {Phys. Rev. Lett.},
  pages = {105502},
  title = {Does the Interaction Potential Determine Both the Fragility of a Liquid and the Vibrational Properties of Its Glassy State?},
  volume = {93},
  year = {2004}
}

@article{sil99,
  author = {H. Sillescu},
  doi = {https://doi.org/10.1016/S0022-3093(98)00831-X},
  journal = {J. Non-Cryst. Solids},
  pages = {81--108},
  title = {Heterogeneity at the glass transition: a review},
  volume = {243},
  year = {1999}
}

@article{mei21,
  author = {B. Mei and Y. Zhou and K. S. Schweizer},
  doi = {10.1073/pnas.2025341118},
  journal = {PNAS},
  pages = {e2025341118},
  title = {Experimental test of a predicted dynamics–structure–thermodynamics connection in molecularly complex glass-forming liquids},
  volume = {118},
  year = {2021}
}

@article{zho20,
  author = {Y. Zhou and B. Mei and K. S. Schweizer},
  doi = {10.1103/PhysRevE.101.042121},
  journal = {Phys. Rev. E},
  pages = {042121},
  title = {Integral equation theory of thermodynamics, pair structure, and growing static length scale in metastable hard sphere and {Weeks-Chandler-Andersen} fluids},
  volume = {101},
  year = {2020}
}

@article{mei20,
  author = {B. Mei and Y. Zhou and K. S. Schweizer},
  doi = {10.1021/acs.jpcb.0c03613},
  journal = {J. Phys. Chem. B},
  pages = {6121--6131},
  title = {Thermodynamics-Structure-Dynamics Correlations and Nonuniversal Effects in the Elastically Collective Activated Hopping Theory of Glass-Forming Liquids},
  volume = {124},
  year = {2020}
}

@article{wan11a,
  author = {J. Q. Wang and W. H. Wang and Y. H. Liu and H. Y. Bai},
  doi = {10.1103/PhysRevB.83.012201},
  journal = {Phys. Rev. B},
  pages = {012201},
  title = {Characterization of activation energy for flow in metallic glasses},
  volume = {83},
  year = {2011}
}

@article{gio16,
  author = {V. M. Giordano and B. Ruta},
  doi = {10.1038/ncomms10344},
  journal = {Nat. Commun.},
  pages = {10344},
  title = {Unveiling the structural arrangements responsible for the atomic dynamics in metallic glasses during physical aging},
  volume = {7},
  year = {2016}
}

@article{ghi24,
  author = {F. Ghimenti and L. Berthier and J. Kurchan and F. {van Wijland}},
  journal = {arXiv:2409.17121},
  title = {What do clever algorithms for glasses do? {Time} reparametrization at work},
  year = {2024}
}

@article{ber23,
  author = {L. Berthier and D. R. Reichman},
  doi = {10.1038/s42254-022-00548-x},
  journal = {Nat. Rev. Phys.},
  pages = {102--116},
  title = {Modern computational studies of the glass transition},
  volume = {5},
  year = {2023}
}

@article{mei24,
  author = {B. Mei and K. S. Schweizer},
  doi = {10.1021/acs.jpcb.4c05488},
  journal = {J. Phys. Chem. B},
  pages = {11293--11312},
  title = {Medium-Range Structural Order as the Driver of Activated Dynamics and Complexity Reduction in Glass-Forming Liquids},
  volume = {128},
  year = {2024}
}

@article{ish25,
  author = {S. Ishino and Y.-C. Hu and H. Tanaka},
  doi = {10.1038/s41563-024-02068-8},
  journal = {Nat. Mater.},
  pages = {268--277},
  title = {Microscopic structural origin of slow dynamics in glass-forming liquids},
  volume = {24},
  year = {2025}
}

@article{bir08,
  title={Thermodynamic signature of growing amorphous order in glass-forming liquids},
  author={Biroli, GBJP and Bouchaud, J-P and Cavagna, Andrea and Grigera, Tom{\'a}s S and Verrocchio, Paolo},
  journal={Nature Physics},
  volume={4},
  number={10},
  pages={771--775},
  year={2008},
  publisher={Nature Publishing Group UK London}
}

@article{bou04,
  title={On the Adam-Gibbs-Kirkpatrick-Thirumalai-Wolynes scenario for the viscosity increase in glasses},
  author={Bouchaud, Jean-Philippe and Biroli, Giulio},
  journal={The Journal of chemical physics},
  volume={121},
  number={15},
  pages={7347--7354},
  year={2004},
  publisher={American Institute of Physics}
}

@article{char13,
  title={Decorrelation of the static and dynamic length scales in hard-sphere glass formers},
  author={Charbonneau, Patrick and Tarjus, Gilles},
  journal={Physical Review E—Statistical, Nonlinear, and Soft Matter Physics},
  volume={87},
  number={4},
  pages={042305},
  year={2013},
  publisher={APS}
}

@article{Mai05,
  author  = {Sarika Maitra Bhattacharyya and Biman Bagchi and Peter G. Wolynes},
  title   = {Bridging the Gap Between the Mode Coupling and the Random First Order Transition Theories of Structural Relaxation in Liquids},
  journal = {The Journal of Chemical Physics},
  year    = {2005},
  note    = {arXiv:cond-mat/0505030},
  doi     = {10.1063/1.1914642}
}

@article{Mai08,
  author  = {Sarika Maitra Bhattacharyya and Biman Bagchi and Peter G. Wolynes},
  title   = {Hopping induced continuous diffusive dynamics below the non-ergodic transition},
  journal = {The Journal of Chemical Physics},
  year    = {2008},
  note    = {arXiv:0807.0998},
  doi     = {10.1063/1.2967614}
}

@article{tan24,
  author  = {Hajime Tanaka},
  title   = {Structural Origin of Dynamic Heterogeneity in Supercooled Liquids},
  journal = {Journal of Physical Chemistry B},
  year    = {2024},
  doi     = {10.1021/acs.jpcb.4c06392}
}

@article{rus18,
  author  = {John Russo and Flavio Romano and Hajime Tanaka},
  title   = {Glass Forming Ability in Systems with Competing Orderings},
  journal = {Physical Review X},
  volume  = {8},
  pages   = {021040},
  year    = {2018},
  doi     = {10.1103/PhysRevX.8.021040}
}

@article{nov22,
  author = {V. N. Novikov and A. P. Sokolov},
  doi = {10.3390/e24081101},
  journal = {Entropy},
  pages = {1101},
  title = {Temperature Dependence of Structural Relaxation in Glass-Forming Liquids and Polymers},
  volume = {24},
  year = {2022}
}

@article{kir87,
  title={Stable and metastable states in mean-field Potts and structural glasses},
  author={Kirkpatrick, TR and Wolynes, PG},
  journal={Physical Review B},
  volume={36},
  number={16},
  pages={8552},
  year={1987},
  publisher={APS}
}

@article{rab13,
  title={Microscopically based calculations of the free energy barrier and dynamic length scale in supercooled liquids: The comparative role of configurational entropy and elasticity},
  author={Rabochiy, Pyotr and Wolynes, Peter G and Lubchenko, Vassiliy},
  journal={The Journal of Physical Chemistry B},
  volume={117},
  number={48},
  pages={15204--15219},
  year={2013},
  publisher={ACS Publications}
}

@article{ste06,
  title={The shapes of cooperatively rearranging regions in glass-forming liquids},
  author={Stevenson, Jacob D and Schmalian, J{\"o}rg and Wolynes, Peter G},
  journal={Nature Physics},
  volume={2},
  number={4},
  pages={268--274},
  year={2006},
  publisher={Nature Publishing Group UK London}
}

@article{mei25,
  author = {B. Mei and K. S. Schweizer},
  doi = {10.1103/66zz-y23m},
  journal = {Phys. Rev. Lett.},
  pages = {256101},
  title = {Medium-Range Order, Density Fluctuations, and Activated Relaxation in the Equilibrated Deep Glass Regime},
  volume = {134},
  year = {2025}
}

@article{dal07,
  author = {C. Dalle-Ferrier and C. Thibierge and C. Alba-Simionesco and L. Berthier and G. Biroli and J.-P. Bouchaud and F. Ladieu and D. L'H{\^o}te and G. Tarjus},
  doi = {10.1103/PhysRevE.76.041510},
  journal = {Phys. Rev. E},
  pages = {041510},
  title = {Spatial correlations in the dynamics of glassforming liquids: Experimental determination of their temperature dependence},
  volume = {76},
  year = {2007}
}

@article{ber19b,
  author = {L. Berthier and G. Biroli and J.-P. Bouchaud and G. Tarjus},
  doi = {10.1063/1.5086509},
  journal = {J. Chem. Phys.},
  pages = {094501},
  title = {Can the glass transition be explained without a growing static length scale?},
  volume = {150},
  year = {2019}
}

@article{dyr06b,
  author = {J. C. Dyre and T. Christensen and N. B. Olsen},
  doi = {10.1016/j.jnoncrysol.2006.02.173},
  journal = {J. Non-Cryst. Solids},
  pages = {4635--4642},
  title = {Elastic models for the non-{Arrhenius} viscosity of glass-forming liquids},
  volume = {352},
  year = {2006}
}

@article{wan06,
  author = {L.-M. Wang and C. A. Angell and R. Richert},
  doi = {10.1063/1.2244551},
  journal = {J. Chem. Phys.},
  pages = {074505},
  title = {Fragility and thermodynamics in nonpolymeric glass-forming liquids},
  volume = {125},
  year = {2006}
}

@article{boh98a,
  author = {R. B{\"o}hmer},
  doi = {10.1080/01411599808209288},
  journal = {Phase Transitions},
  pages = {211--231},
  title = {Non-exponential relaxation in disordered materials: {Phenomenological} correlations and spectrally selective experiments},
  volume = {65},
  year = {1998}
}

@article{nis06,
  author = {K. Niss and C. Alba-Simionesco},
  doi = {10.1103/PhysRevB.74.024205},
  journal = {Phys. Rev. B},
  pages = {024205},
  title = {Effects of density and temperature on correlations between fragility and glassy properties},
  volume = {74},
  year = {2006}
}

@article{loi25,
  author = {A. Loidl and P. Lunkenheimer and K. Samwer},
  doi = {10.1103/PhysRevE.111.035407},
  journal = {Phys. Rev. E},
  pages = {035407},
  title = {{Prigogine-Defay} ratio of glassy freezing scales with liquid fragility},
  volume = {111},
  year = {2025}
}

@article{anomalous_elasticity_soft_matter_2023,
  author = {Lerner, Edan and Bouchbinder, Eran},
  doi = {10.1039/D2SM01253G},
  issue = {6},
  journal = {Soft Matter},
  pages = {1076-1080},
  publisher = {The Royal Society of Chemistry},
  title = {Anomalous linear elasticity of disordered networks},
  url = {http://dx.doi.org/10.1039/D2SM01253G},
  volume = {19},
  year = {2023}
}

@article{frustrated_networks_pre_2024,
  author = {Pettinari, Tommaso and During, Gustavo and Lerner, Edan},
  doi = {10.1103/PhysRevE.109.054906},
  issue = {5},
  journal = {Phys. Rev. E},
  month = {May},
  numpages = {9},
  pages = {054906},
  publisher = {American Physical Society},
  title = {Elasticity of self-organized frustrated disordered spring networks},
  url = {https://link.aps.org/doi/10.1103/PhysRevE.109.054906},
  volume = {109},
  year = {2024}
}

@article{royall2015role,
  author = {Royall, C. Patrick and Williams, Stephen R.},
  doi = {10.1016/j.physrep.2014.11.004},
  journal = {Physics Reports},
  pages = {1--75},
  title = {The role of local structure in dynamical arrest},
  volume = {560},
  year = {2015}
}

@article{malins2013identification,
  author = {Malins, Alex and Williams, Stephen R. and Eggers, Jens and Royall, C. Patrick},
  doi = {10.1063/1.4832897},
  issn = {0021-9606},
  journal = {The Journal of Chemical Physics},
  month = {12},
  number = {23},
  pages = {234506},
  title = {Identification of structure in condensed matter with the topological cluster classification},
  url = {https://doi.org/10.1063/1.4832897},
  volume = {139},
  year = {2013}
}

@article{coslovich2007characterization,
  author = {Coslovich, D. and Pastore, G.},
  doi = {10.1063/1.2773716},
  issn = {0021-9606},
  journal = {The Journal of Chemical Physics},
  month = {09},
  number = {12},
  pages = {124504},
  title = {Understanding fragility in supercooled Lennard-Jones mixtures. I. Locally preferred structures},
  url = {https://doi.org/10.1063/1.2773716},
  volume = {127},
  year = {2007}
}

@article{leocmach2012roles,
  author = {Leocmach, Mathieu and Tanaka, Hajime},
  doi = {10.1038/ncomms1974},
  journal = {Nature Communications},
  pages = {974},
  title = {Roles of icosahedral and crystal-like order in the hard spheres glass transition},
  volume = {3},
  year = {2012}
}

@article{keys2011excitations,
  author = {Aaron S. Keys and Lester O. Hedges and Juan P. Garrahan and Sharon C. Glotzer and David Chandler},
  doi = {10.1103/PhysRevX.1.021013},
  issue = {2},
  journal = {Physical Review X},
  month = {11},
  pages = {021013},
  title = {Excitations Are Localized and Relaxation Is Hierarchical in Glass-Forming Liquids},
  volume = {1},
  year = {2011}
}

@article{schoenholz2016softness,
  author = {Schoenholz, Samuel S. and Cubuk, Ekin D. and Sussman, Daniel M. and Kaxiras, Efthimios and Liu, Andrea J.},
  doi = {10.1038/nphys3644},
  journal = {Nature Physics},
  pages = {469--471},
  title = {A structural approach to relaxation in glassy liquids},
  volume = {12},
  year = {2016}
}

@article{bapst2020gnn,
  author = {Bapst, Victor and Keck, Thomas and Grabska-Barwińska, Agnieszka and Donner, Christoph and Cubuk, Ekin D. and Schoenholz, Samuel S. and Obika, Akos and Nelson, Andrew W. R. and Back, Tobias and Hassabis, Demis and Kohli, Pushmeet},
  doi = {10.1038/s41567-020-0842-8},
  journal = {Nature Physics},
  pages = {448--454},
  title = {Unveiling the predictive power of graph neural networks for glassy dynamics},
  volume = {16},
  year = {2020}
}

@article{nelson1983order,
  author = {Nelson, David R},
  doi = {10.1103/PhysRevB.28.5515},
  journal = {Physical Review B},
  number = {10},
  pages = {5515},
  title = {Order, frustration, and defects in liquids and glasses},
  volume = {28},
  year = {1983}
}

@article{kelton2003icosahedral,
  author = {Kelton, K. F. and Lee, G. W. and Gangopadhyay, A. K. and Hyers, R. W. and Rathz, T. J. and Rogers, J. R. and Robinson, M. B. and Robinson, D. S.},
  doi = {10.1103/PhysRevLett.90.195504},
  issue = {19},
  journal = {Phys. Rev. Lett.},
  month = {May},
  numpages = {4},
  pages = {195504},
  publisher = {American Physical Society},
  title = {First X-Ray Scattering Studies on Electrostatically Levitated Metallic Liquids: Demonstrated Influence of Local Icosahedral Order on the Nucleation Barrier},
  volume = {90},
  year = {2003}
}

@article{shen2009icosahedral,
  author = {Shen, YT and Kim, TH and Gangopadhyay, AK and Kelton, KF},
  doi = {10.1103/PhysRevLett.102.057801},
  journal = {Physical Review Letters},
  number = {5},
  pages = {057801},
  title = {Icosahedral order, frustration, and the glass transition: Evidence from time-dependent nucleation and supercooled liquid structure studies},
  volume = {102},
  year = {2009}
}

@article{tarjus2005frustration,
  author = {Tarjus, G and Kivelson, S A and Nussinov, Z and Viot, P},
  doi = {10.1088/0953-8984/17/50/R01},
  journal = {Journal of Physics: Condensed Matter},
  month = {dec},
  number = {50},
  pages = {R1143},
  publisher = {},
  title = {The frustration-based approach of supercooled liquids and the glass transition: a review and critical assessment},
  url = {https://doi.org/10.1088/0953-8984/17/50/R01},
  volume = {17},
  year = {2005}
}

@article{tarjus2000fld,
  author = {Tarjus, Gilles and Kivelson, Daniel and Viot, Pascal},
  doi = {10.1088/0953-8984/12/29/321},
  journal = {Journal of Physics: Condensed Matter},
  number = {29},
  pages = {6497--6512},
  title = {The viscous slowing down of supercooled liquids as a temperature-controlled super-Arrhenius activated process: a description in terms of frustration-limited domains},
  volume = {12},
  year = {2000}
}

@article{boattini2021averaging,
  author = {Boattini, Emanuele and Smallenburg, Frank and Filion, Laura},
  doi = {10.1103/PhysRevLett.127.088007},
  issue = {8},
  journal = {Phys. Rev. Lett.},
  month = {Aug},
  numpages = {6},
  pages = {088007},
  publisher = {American Physical Society},
  title = {Averaging Local Structure to Predict the Dynamic Propensity in Supercooled Liquids},
  url = {https://link.aps.org/doi/10.1103/PhysRevLett.127.088007},
  volume = {127},
  year = {2021}
}

@article{tong2018hidden,
  author = {Tong, Hua and Tanaka, Hajime},
  doi = {10.1103/PhysRevX.8.011041},
  issue = {1},
  journal = {Phys. Rev. X},
  month = {Mar},
  numpages = {18},
  pages = {011041},
  publisher = {American Physical Society},
  title = {Revealing Hidden Structural Order Controlling Both Fast and Slow Glassy Dynamics in Supercooled Liquids},
  url = {https://link.aps.org/doi/10.1103/PhysRevX.8.011041},
  volume = {8},
  year = {2018}
}

@article{Boattini2020,
  author = {Emanuele Boattini and Susana Marín-Aguilar and Saheli Mitra and Giuseppe Foffi and Frank Smallenburg and Laura Filion},
  doi = {10.1038/s41467-020-19286-8},
  issn = {2041-1723},
  issue = {1},
  journal = {Nature Communications},
  month = {10},
  pages = {5479},
  title = {Autonomously revealing hidden local structures in supercooled liquids},
  volume = {11},
  year = {2020}
}

@article{alkemade2022mlcomparison,
  author = {Alkemade, Rinske M. and Boattini, Emanuele and Filion, Laura and Smallenburg, Frank},
  doi = {10.1063/5.0088581},
  journal = {The Journal of Chemical Physics},
  number = {20},
  pages = {204503},
  title = {Comparing machine learning techniques for predicting glassy dynamics},
  volume = {156},
  year = {2022}
}

@article{Jung2025,
  author = {Gerhard Jung and Rinske M. Alkemade and Victor Bapst and Daniele Coslovich and Laura Filion and François P. Landes and Andrea J. Liu and Francesco Saverio Pezzicoli and Hayato Shiba and Giovanni Volpe and Francesco Zamponi and Ludovic Berthier and Giulio Biroli},
  doi = {10.1038/s42254-024-00791-4},
  issn = {2522-5820},
  issue = {2},
  journal = {Nature Reviews Physics},
  month = {1},
  pages = {91-104},
  title = {Roadmap on machine learning glassy dynamics},
  volume = {7},
  year = {2025}
}

@article{Tanaka2010b,
  author = {Hajime Tanaka and Takeshi Kawasaki and Hiroshi Shintani and Keiji Watanabe},
  doi = {10.1038/NMAT2634},
  journal = {Nature Materials},
  title = {Critical-like behaviour of glass-forming liquids},
  url = {https://pdfs.semanticscholar.org/4033/568b378c70f580b4ff03bbac2571f56bbff1.pdf},
  volume = {9},
  year = {2010}
}

@article{Russo2015,
  author = {John Russo and Hajime Tanaka},
  doi = {10.1073/pnas.1501911112},
  issn = {0027-8424},
  issue = {22},
  journal = {Proceedings of the National Academy of Sciences},
  month = {6},
  pages = {6920-6924},
  title = {Assessing the role of static length scales behind glassy dynamics in polydisperse hard disks},
  volume = {112},
  year = {2015}
}

@article{GurevichParshinSchober2003,
  author = {Gurevich, Vladimir L. and Parshin, Dmitry A. and Schober, Hans R.},
  doi = {10.1103/PhysRevB.67.094203},
  journal = {Physical Review B},
  pages = {094203},
  title = {Anharmonicity, Instability, and the Boson Peak in Glasses},
  volume = {67},
  year = {2003}
}

@article{ParshinSchoberGurevich2007,
  author = {Parshin, Dmitry A. and Schober, Hans R. and Gurevich, Vladimir L.},
  doi = {10.1103/PhysRevB.76.064206},
  journal = {Physical Review B},
  pages = {064206},
  title = {Vibrational Instabilities, Two-Level Systems, and the Boson Peak in Glasses},
  volume = {76},
  year = {2007}
}

@article{Oligschleger1999Collective,
  author = {Oligschleger, Christina and Schober, H. R.},
  doi = {10.1103/PhysRevB.59.811},
  journal = {Physical Review B},
  number = {2},
  pages = {811--821},
  title = {Collective Jumps in a Soft-Sphere Glass},
  volume = {59},
  year = {1999}
}

@article{Yu2017JGstrings,
  author = {Yu, Hai-Bo and Richert, Ranko and Samwer, Konrad},
  doi = {10.1126/sciadv.1701577},
  journal = {Science Advances},
  number = {10},
  pages = {e1701577},
  title = {Structural Rearrangements Governing Johari--Goldstein Relaxations in Metallic Glasses},
  volume = {3},
  year = {2017}
}

@article{SchoberOligschleger1996,
  author = {Schober, Hans R. and Oligschleger, Christian},
  doi = {10.1103/PhysRevB.53.11469},
  journal = {Physical Review B},
  pages = {11469--11480},
  title = {Low-Frequency Vibrational Modes in a Model Glass},
  volume = {53},
  year = {1996}
}

@article{MalandroLacks1999,
  author = {Malandro, David L. and Lacks, Daniel J.},
  doi = {10.1103/PhysRevE.60.461},
  journal = {Physical Review E},
  pages = {461--467},
  title = {Relationships of Shear-Induced Changes in the Potential Energy Landscape to the Mechanical Properties of Ductile Glasses},
  volume = {60},
  year = {1999}
}

@article{TanguyEtAl2002,
  author = {Tanguy, Anne and Wittmer, Jean-Pierre and Leonforte, Fabio and Barrat, Jean-Louis},
  doi = {10.1103/PhysRevB.66.174205},
  journal = {Physical Review B},
  pages = {174205},
  title = {Continuum Limit of Amorphous Elastic Bodies: A Finite-Size Study of Low-Frequency Vibrational Modes},
  volume = {66},
  year = {2002}
}

@article{khomenko2021relationship,
  author = {Khomenko, Dmytro and Reichman, David R and Zamponi, Francesco},
  journal = {Physical Review Materials},
  number = {5},
  pages = {055602},
  publisher = {APS},
  title = {Relationship between two-level systems and quasilocalized normal modes in glasses},
  volume = {5},
  year = {2021}
}

@article{Coslovich2018,
  author = {Coslovich, Daniele and Ozawa, Misaki and Kob, Walter},
  doi = {10.1140/epje/i2018-11671-2},
  issn = {1292-8941},
  journal = {The European Physical Journal E},
  month = {may},
  number = {5},
  pages = {62},
  title = {{Dynamic and thermodynamic crossover scenarios in the Kob-Andersen mixture: Insights from multi-CPU and multi-GPU simulations}},
  volume = {41},
  year = {2018}
}

@article{bir05,
  author = {Biroli, Giulio and Bouchaud, Jean-Philippe and Tarjus, Gilles},
  journal = {The Journal of chemical physics},
  number = {4},
  publisher = {AIP Publishing},
  title = {Are defect models consistent with the entropy and specific heat of glass formers?},
  volume = {123},
  year = {2005}
}

@article{fra23,
  author = {Fraggedakis, Dimitrios and Hasyim, Muhammad R and Mandadapu, Kranthi K},
  journal = {Proceedings of the National Academy of Sciences},
  number = {14},
  pages = {e2209144120},
  publisher = {National Academy of Sciences},
  title = {Inherent-state melting and the onset of glassy dynamics in two-dimensional supercooled liquids},
  volume = {120},
  year = {2023}
}

@article{fis86,
  author = {Fisher, Daniel S and Huse, David A},
  journal = {Physical review letters},
  number = {15},
  pages = {1601},
  publisher = {APS},
  title = {Ordered phase of short-range Ising spin-glasses},
  volume = {56},
  year = {1986}
}

@article{Struik1997,
  author = {L.C.E. Struik},
  doi = {10.1016/S0032-3861(96)00699-4},
  issn = {00323861},
  issue = {3},
  journal = {Polymer},
  month = {2},
  pages = {733-735},
  title = {The apparent activation energy for mechanical and dielectric relaxation in glass-forming (polymeric) liquids: a misconception?},
  volume = {38},
  year = {1997}
}

@article{Berthier2020,
  author = {L Berthier and M D Ediger},
  doi = {10.1063/5.0015227},
  issue = {4},
  journal = {The Journal of Chemical Physics},
  pages = {44501},
  title = {How to 'measure' a structural relaxation time that is too long to be measured?},
  volume = {153},
  year = {2020}
}

@article{Mehri2022,
  author = {Saeed Mehri and Lorenzo Costigliola and Jeppe C Dyre},
  doi = {10.3390/thermo2030013},
  issn = {2673-7264},
  issue = {3},
  journal = {Thermo},
  pages = {160-170},
  title = {Single-Parameter Aging in the Weakly Nonlinear Limit},
  volume = {2},
  year = {2022}
}

@article{kor25,
  author = {Korchinski, Daniel J and Shohat, Dor and Lahini, Yoav and Wyart, Matthieu},
  journal = {Physical Review X},
  number = {3},
  pages = {031024},
  publisher = {APS},
  title = {Thermal Avalanches Drive Logarithmic Creep in Disordered Media},
  volume = {15},
  year = {2025}
}

@article{cot52,
  author = {Cottrell, AH},
  journal = {Journal of the Mechanics and Physics of Solids},
  number = {1},
  pages = {53--63},
  publisher = {Elsevier},
  title = {The time laws of creep},
  volume = {1},
  year = {1952}
}

@article{deg15b,
  author = {Degiuli, Eric and Lerner, E and Wyart, M},
  journal = {The Journal of chemical physics},
  number = {16},
  publisher = {AIP Publishing},
  title = {Theory of the jamming transition at finite temperature},
  volume = {142},
  year = {2015}
}

@article{cha00,
  author = {Chauve, Pascal and Giamarchi, Thierry and Le Doussal, Pierre},
  journal = {Physical Review B},
  number = {10},
  pages = {6241},
  publisher = {APS},
  title = {Creep and depinning in disordered media},
  volume = {62},
  year = {2000}
}

@article{fer21,
  author = {Ferrero, Ezequiel E and Kolton, Alejandro B and Jagla, Eduardo A},
  journal = {Physical Review Materials},
  number = {11},
  pages = {115602},
  publisher = {APS},
  title = {Yielding of amorphous solids at finite temperatures},
  volume = {5},
  year = {2021}
}

@article{kol06,
  author = {Kolton, Alejandro B and Rosso, Alberto and Giamarchi, Thierry and Krauth, Werner},
  journal = {Physical review letters},
  number = {5},
  pages = {057001},
  publisher = {APS},
  title = {Dynamics below the depinning threshold in disordered elastic systems},
  volume = {97},
  year = {2006}
}

@article{kol09,
  author = {Kolton, Alejandro B and Rosso, Alberto and Giamarchi, Thierry and Krauth, Werner},
  journal = {Physical Review B},
  number = {18},
  pages = {184207},
  publisher = {APS},
  title = {Creep dynamics of elastic manifolds via exact transition pathways},
  volume = {79},
  year = {2009}
}

@article{sho23,
  title={Logarithmic aging via instability cascades in disordered systems},
  author={Shohat, Dor and Friedman, Yaniv and Lahini, Yoav},
  journal={Nature Physics},
  volume={19},
  number={12},
  pages={1890--1895},
  year={2023},
  publisher={Nature Publishing Group UK London}
}

@article{dur23,
  author = {Durin, Gianfranco and Schimmenti, Vincenzo Maria and Baiesi, Marco and Casiraghi, Arianna and Magni, Alessandro and Herrera-Diez, Liza and Ravelosona, Dafin{\'e} and Foini, Laura and Rosso, Alberto},
  journal = {arXiv preprint arXiv:2309.12898},
  title = {Earthquake-like dynamics in ultrathin magnetic film},
  year = {2023}
}

@article{sie12,
  author = {Siebenb{\" u}rger, M. and Ballauff, M. and Voigtmann, T.},
  doi = {10.1103/PhysRevLett.108.255701},
  journal = {Phys. Rev. Lett.},
  number = {25},
  pages = {255701},
  title = {Creep in Colloidal Glasses},
  volume = {108},
  year = {2012}
}

@article{liu21,
  author = {Liu, Chen and Dutta, Suman and Chaudhuri, Pinaki and Martens, Kirsten},
  journal = {Physical Review Letters},
  number = {13},
  pages = {138005},
  publisher = {APS},
  title = {Elastoplastic approach based on microscopic insights for the steady state and transient dynamics of sheared disordered solids},
  volume = {126},
  year = {2021}
}

@article{vas22,
  author = {Vasisht, Vishwas V and Chaudhuri, Pinaki and Martens, Kirsten},
  journal = {Soft Matter},
  number = {34},
  pages = {6426--6436},
  publisher = {Royal Society of Chemistry},
  title = {Residual stress in athermal soft disordered solids: insights from microscopic and mesoscale models},
  volume = {18},
  year = {2022}
}

@article{cab19,
  author = {Cabriolu, R. and Horbach, J. and Chaudhuri, P. and Martens, K.},
  doi = {10.1039/C8SM01432A},
  journal = {Soft Matter},
  pages = {415-423},
  title = {Precursors of fluidisation in the creep response of a soft glass},
  volume = {15},
  year = {2019}
}

@article{div11,
  author = {Divoux, T. and Barentin, C. and Manneville, S.},
  doi = {10.1039/c1sm05607g},
  journal = {Soft Matter},
  number = {18},
  pages = {8409},
  title = {From stress-induced fluidization processes to Herschel-Bulkley behaviour in simple yield stress fluids},
  volume = {7},
  year = {2011}
}

@article{kol05,
  author = {Kolton, Alejandro B and Rosso, Alberto and Giamarchi, Thierry},
  journal = {Physical review letters},
  number = {4},
  pages = {047002},
  publisher = {APS},
  title = {Creep motion of an elastic string in a random potential},
  volume = {94},
  year = {2005}
}

@article{lah17,
  author = {Lahini, Yoav and Gottesman, Omer and Amir, Ariel and Rubinstein, Shmuel M},
  journal = {Physical review letters},
  number = {8},
  pages = {085501},
  publisher = {APS},
  title = {Nonmonotonic aging and memory retention in disordered mechanical systems},
  volume = {118},
  year = {2017}
}

@article{nic00,
  author = {Nicolas, Maxime and Duru, P and Pouliquen, Olivier},
  journal = {The European Physical Journal E},
  pages = {309--314},
  publisher = {Springer},
  title = {Compaction of a granular material under cyclic shear},
  volume = {3},
  year = {2000}
}

@article{ben96,
  author = {Ben-Naim, Eli and Knight, JB and Nowak, ER and Jaeger, HM and Nagel, SR},
  journal = {Physica D: Nonlinear Phenomena},
  number = {1-4},
  pages = {380--385},
  publisher = {Elsevier},
  title = {Slow relaxation in granular compaction},
  volume = {123},
  year = {1998}
}

@article{kni95,
  author = {Knight, James B and Fandrich, Christopher G and Lau, Chun Ning and Jaeger, Heinrich M and Nagel, Sidney R},
  journal = {Physical review E},
  number = {5},
  pages = {3957},
  publisher = {APS},
  title = {Density relaxation in a vibrated granular material},
  volume = {51},
  year = {1995}
}

@article{mat02,
  author = {Matan, Kittiwit and Williams, Rachel B and Witten, Thomas A and Nagel, Sidney R},
  journal = {Physical Review Letters},
  number = {7},
  pages = {076101},
  publisher = {APS},
  title = {Crumpling a thin sheet},
  volume = {88},
  year = {2002}
}

@article{bau06,
  author = {Bauer, T. and Oberdisse, J. and Ramos, L.},
  doi = {10.1103/PhysRevLett.97.258303},
  journal = {Phys. Rev. Lett.},
  number = {25},
  title = {Collective Rearrangement at the Onset of Flow of a Polycrystalline Hexagonal Columnar Phase},
  volume = {97},
  year = {2006}
}

@article{bus07,
  author = {Bustingorry, Sebastian and Kolton, AB and Giamarchi, Thierry},
  journal = {Europhysics Letters},
  number = {2},
  pages = {26005},
  publisher = {IOP Publishing},
  title = {Thermal rounding of the depinning transition},
  volume = {81},
  year = {2007}
}

@article{bou97,
  author = {Boutreux, T and de Gennes, PG},
  journal = {Physica A: Statistical Mechanics and its Applications},
  number = {1-4},
  pages = {59--67},
  publisher = {Elsevier},
  title = {Compaction of granular mixtures: a free volume model},
  volume = {244},
  year = {1997}
}

@article{and62,
  author = {Anderson, Philip W},
  journal = {Physical Review Letters},
  number = {7},
  pages = {309},
  publisher = {APS},
  title = {Theory of flux creep in hard superconductors},
  volume = {9},
  year = {1962}
}

@article{duv78,
  author = {Duval, Paul},
  journal = {Journal of Glaciology},
  number = {85},
  pages = {621--628},
  publisher = {Cambridge University Press},
  title = {Anelastic behaviour of polycrystalline ice},
  volume = {21},
  year = {1978}
}

@article{rub06,
  author = {Rubinstein, Shmuel M and Cohen, Gil and Fineberg, Jay},
  journal = {Physical review letters},
  number = {25},
  pages = {256103},
  publisher = {APS},
  title = {Contact area measurements reveal loading-history dependence of static friction},
  volume = {96},
  year = {2006}
}

@article{ham93,
  author = {Woltjer, Reinout and Hamada, Akemi and Takeda, Eiji},
  journal = {IEEE transactions on electron devices},
  number = {2},
  pages = {392--401},
  publisher = {IEEE},
  title = {Time dependence of p-MOSFET hot-carrier degradation measured and interpreted consistently over ten orders of magnitude},
  volume = {40},
  year = {1993}
}

@article{vak00,
  author = {Vaknin, A. and Ovadyahu, Z. and Pollak, M.},
  doi = {10.1103/PhysRevLett.84.3402},
  issue = {15},
  journal = {Phys. Rev. Lett.},
  month = {Apr},
  numpages = {0},
  pages = {3402--3405},
  publisher = {American Physical Society},
  title = {Aging Effects in an Anderson Insulator},
  url = {https://link.aps.org/doi/10.1103/PhysRevLett.84.3402},
  volume = {84},
  year = {2000}
}

@article{ben10,
  author = {Ben-David, Oded and Rubinstein, Shmuel M and Fineberg, Jay},
  journal = {Nature},
  number = {7277},
  pages = {76--79},
  publisher = {Nature Publishing Group UK London},
  title = {Slip-stick and the evolution of frictional strength},
  volume = {463},
  year = {2010}
}

@article{ami11,
  author = {Amir, Ariel and Borini, Stefano and Oreg, Yuval and Imry, Yoseph},
  journal = {Physical review letters},
  number = {18},
  pages = {186407},
  publisher = {APS},
  title = {Huge (but finite) time scales in slow relaxations: Beyond simple aging},
  volume = {107},
  year = {2011}
}

@article{gre07,
  author = {Grenet, Thierry and Delahaye, Julien and Sabra, Maher and Gay, Fr{\'e}d{\'e}ric},
  journal = {The European Physical Journal B},
  pages = {183--197},
  publisher = {Springer},
  title = {Anomalous electric-field effect and glassy behaviour in granular aluminium thin films: electron glass?},
  volume = {56},
  year = {2007}
}

@article{set01,
  author = {Sethna, James P and Dahmen, Karin A and Myers, Christopher R},
  journal = {Nature},
  number = {6825},
  pages = {242--250},
  publisher = {Nature Publishing Group UK London},
  title = {Crackling noise},
  volume = {410},
  year = {2001}
}

@article{bar13,
  author = {Bar{\'o}, Jordi and Corral, {\'A}lvaro and Illa, Xavier and Planes, Antoni and Salje, Ekhard KH and Schranz, Wilfried and Soto-Parra, Daniel E and Vives, Eduard},
  journal = {Physical review letters},
  number = {8},
  pages = {088702},
  publisher = {APS},
  title = {Statistical similarity between the compression of a porous material and earthquakes},
  volume = {110},
  year = {2013}
}

@article{sal14,
  author = {Salje, Ekhard KH and Dahmen, Karin A},
  journal = {Annu. Rev. Condens. Matter Phys.},
  number = {1},
  pages = {233--254},
  publisher = {Annual Reviews},
  title = {Crackling noise in disordered materials},
  volume = {5},
  year = {2014}
}

@article{cor04,
  author = {Corral, {\'A}lvaro},
  journal = {Physical Review Letters},
  number = {10},
  pages = {108501},
  publisher = {APS},
  title = {Long-term clustering, scaling, and universality in the temporal occurrence of earthquakes},
  volume = {92},
  year = {2004}
}

@book{gut54,
  author = {Gutenberg, B.U. and Richter, C.F.},
  publisher = {Princeton (NJ)},
  title = {{Seismicity of the earth and related phenomena}},
  year = {1954}
}

@article{wes94,
  author = {Wesnousky, S.G.},
  journal = {Bull. Seismol. Soc. Am.},
  number = {6},
  pages = {1940--1959},
  title = {The Gutenberg-Richter or characteristic earthquake distribution, which is it?},
  volume = {84},
  year = {1994}
}

@article{mor11,
  author = {Mora, Thierry and Bialek, William},
  journal = {Journal of Statistical Physics},
  pages = {268--302},
  publisher = {Springer},
  title = {Are biological systems poised at criticality?},
  volume = {144},
  year = {2011}
}

@article{fri12,
  author = {Friedman, Nir and Ito, Shinya and Brinkman, Braden AW and Shimono, Masanori and DeVille, RE Lee and Dahmen, Karin A and Beggs, John M and Butler, Thomas C},
  journal = {Physical review letters},
  number = {20},
  pages = {208102},
  publisher = {APS},
  title = {Universal critical dynamics in high resolution neuronal avalanche data},
  volume = {108},
  year = {2012}
}

@article{beg03,
  author = {Beggs, John M and Plenz, Dietmar},
  journal = {Journal of neuroscience},
  number = {35},
  pages = {11167--11177},
  publisher = {Soc Neuroscience},
  title = {Neuronal avalanches in neocortical circuits},
  volume = {23},
  year = {2003}
}

@article{paz99,
  author = {P{\'a}zm{\'a}ndi, Ferenc and Zar{\'a}nd, Gergely and Zim{\'a}nyi, Gergely T},
  journal = {Physical review letters},
  number = {5},
  pages = {1034},
  publisher = {APS},
  title = {Self-organized criticality in the hysteresis of the Sherrington-Kirkpatrick model},
  volume = {83},
  year = {1999}
}

@article{alt04,
  author = {Altshuler, Ernesto and Johansen, TH},
  journal = {Reviews of Modern Physics},
  number = {2},
  pages = {471},
  publisher = {APS},
  title = {Colloquium: Experiments in vortex avalanches},
  volume = {76},
  year = {2004}
}

@article{kim03,
  author = {Kim, Dong-Hyun and Choe, Sug-Bong and Shin, Sung-Chul},
  journal = {Physical review letters},
  number = {8},
  pages = {087203},
  publisher = {APS},
  title = {Direct observation of Barkhausen avalanche in Co thin films},
  volume = {90},
  year = {2003}
}

@article{ros22,
  author = {Rosso, Alberto and Sethna, James P and Wyart, Matthieu},
  journal = {arXiv preprint arXiv:2208.04090},
  title = {Avalanches and deformation in glasses and disordered systems},
  year = {2022}
}

@article{bud17,
  author = {Budrikis, Z.\ and Castellanos, D.F.\ and Sandfeld, S.\ and Zaiser, M.\ and Zapperi, S.\},
  doi = {10.1038/ncomms15928},
  journal = {Nat. Commun.},
  pages = {15928},
  title = {{{Universal features of amorphous plasticity}}},
  volume = {8},
  year = {2017}
}

@article{mul14,
  author = {M{\" u}ller, M. and Wyart, M.},
  doi = {10.1146/annurev-conmatphys-031214-014614},
  journal = {Annu. Rev. Condens. Matter Phys.},
  number = {1},
  pages = {177–200},
  title = {Marginal Stability in Structural, Spin, and Electron Glasses},
  volume = {6},
  year = {2015}
}

@article{per95,
  author = {Perkovi{\'c}, Olga and Dahmen, Karin and Sethna, James P},
  journal = {Physical review letters},
  number = {24},
  pages = {4528},
  publisher = {APS},
  title = {Avalanches, Barkhausen noise, and plain old criticality},
  volume = {75},
  year = {1995}
}

@article{fry22,
  author = {Fryer, Barnaby and Giorgetti, Carolina and Passel{\`e}gue, Fran{\c{c}}ois and Momeni, Seyyedmaalek and Lecampion, Brice and Violay, Marie},
  journal = {Journal of Geophysical Research: Solid Earth},
  number = {8},
  pages = {e2022JB025113},
  publisher = {Wiley Online Library},
  title = {The influence of roughness on experimental fault mechanical behavior and associated microseismicity},
  volume = {127},
  year = {2022}
}

@article{gav24,
  author = {Gavazzoni, Cristina and Brito, Carolina and Wyart, Matthieu},
  journal = {Physical Review Letters},
  number = {24},
  pages = {248201},
  publisher = {APS},
  title = {Testing theories of the glass transition with the same liquid butmany kinetic rules},
  doi = {10.1103/PhysRevLett.132.248201},
  volume = {132},
  year = {2024}
}

@article{Ji2022,
  author = {Wencheng Ji and Tom W.J. De Geus and Elisabeth Agoritsas and Matthieu Wyart},
  doi = {10.1103/PHYSREVE.105.044601/FIGURES/8/MEDIUM},
  issn = {24700053},
  issue = {4},
  journal = {Physical Review E},
  month = {4},
  pages = {044601},
  pmid = {35590661},
  publisher = {American Physical Society},
  title = {Mean-field description for the architecture of low-energy excitations in glasses},
  volume = {105},
  year = {2022}
}

@article{Glandt84,
  author = {Briano, J.G. and Glandt, E.D.},
  doi = {10.1063/1.447087},
  journal = {J. Chem. Phys.},
  number = {7},
  pages = {3336--3343},
  title = {Statistical thermodynamics of polydisperse fluids},
  volume = {80},
  year = {1984}
}

@article{gop22,
  author = {Gopinath, Gautham and Lee, Chun-Shing and Gao, Xin-Yuan and An, Xiao-Dong and Chan, Chor-Hoi and Yip, Cho-Tung and Deng, Hai-Yao and Lam, Chi-Hang},
  journal = {Physical review letters},
  number = {16},
  pages = {168002},
  publisher = {APS},
  title = {Diffusion-coefficient power laws and defect-driven glassy dynamics in swap acceleration},
  volume = {129},
  year = {2022}
}

@article{wya17,
  author = {Wyart, M. and Cates, M.E.},
  doi = {10.1103/PhysRevLett.119.195501},
  journal = {Phys. Rev. Lett.},
  number = {19},
  pages = {195501},
  title = {{Does a Growing Static Length Scale Control the Glass Transition?}},
  volume = {119},
  year = {2017}
}

@article{jic25,
  author = {Wencheng Ji and Massimo Pica Ciamarra and Matthieu Wyart},
  doi = {10.1073/pnas.2416800122},
  journal = {Proceedings of the National Academy of Sciences},
  number = {11},
  pages = {e2416800122},
  title = {The role of excitations in supercooled liquids: Density, geometry, and relaxation dynamics},
  volume = {122},
  year = {2025}
}

@article{Denny2003,
  author = {R. Aldrin Denny and David R. Reichman and Jean Philippe Bouchaud},
  doi = {10.1103/PhysRevLett.90.025503},
  issn = {10797114},
  issue = {2},
  journal = {Physical Review Letters},
  pages = {4},
  title = {Trap Models and Slow Dynamics in Supercooled Liquids},
  volume = {90},
  year = {2003}
}

@article{bou96,
  author = {Bouchaud, J.-P. and Cugliandolo, L. and Kurchan, J. and M{\'e}zard, M.},
  journal = {Phys. A},
  number = {3-4},
  pages = {243--273},
  title = {Mode-coupling approximations, glass theory and disordered systems},
  volume = {226},
  year = {1996}
}

@article{ike19,
  author = {Ikeda, Harukuni and Urbani, Pierfrancesco and Zamponi, Francesco},
  journal = {Journal of Physics A: Mathematical and Theoretical},
  number = {34},
  pages = {344001},
  publisher = {IOP Publishing},
  title = {Mean field theory of jamming of nonspherical particles},
  volume = {52},
  year = {2019}
}

@article{sza19,
  author = {Szamel, Grzegorz},
  journal = {Journal of Statistical Mechanics: Theory and Experiment},
  number = {10},
  pages = {104016},
  publisher = {IOP Publishing},
  title = {Theory for the single-particle dynamics in glassy mixtures with particle size swaps},
  volume = {2019},
  year = {2019}
}

@article{lan25,
  author = {Lang, Danqi and Scalliet, Camille and Royall, C Patrick},
  journal = {Physical Review E},
  number = {5},
  pages = {055415},
  publisher = {APS},
  title = {Anticorrelation between excitations and locally favored structures in glass-forming systems},
  volume = {111},
  year = {2025}
}

@article{deg14,
  author = {DeGiuli, Eric and Laversanne-Finot, Adrien and D{\"u}ring, Gustavo and Lerner, Edan and Wyart, Matthieu},
  journal = {Soft matter},
  number = {30},
  pages = {5628--5644},
  publisher = {Royal Society of Chemistry},
  title = {Effects of coordination and pressure on sound attenuation, boson peak and elasticity in amorphous solids},
  volume = {10},
  year = {2014}
}

@article{ji20,
  author = {Ji, Wencheng and De Geus, Tom WJ and Popovi{\'c}, Marko and Agoritsas, Elisabeth and Wyart, Matthieu},
  journal = {Physical Review E},
  number = {6},
  pages = {062110},
  publisher = {APS},
  title = {Thermal origin of quasilocalized excitations in glasses},
  volume = {102},
  year = {2020}
}

@article{ler13,
  title={Low-energy non-linear excitations in sphere packings},
  author={Lerner, Edan and D{\"u}ring, Gustavo and Wyart, Matthieu},
  journal={Soft Matter},
  volume={9},
  number={34},
  pages={8252--8263},
  year={2013},
  publisher={Royal Society of Chemistry}
}

@article{ler14b,
  author = {Lerner, Edan and DeGiuli, Eric and D{\"u}ring, Gustavo and Wyart, Matthieu},
  journal = {Soft Matter},
  number = {28},
  pages = {5085--5092},
  publisher = {Royal Society of Chemistry},
  title = {Breakdown of continuum elasticity in amorphous solids},
  volume = {10},
  year = {2014}
}

@article{bir06,
  author = {Biroli, G. and Bouchaud, J.-P. and Miyazaki, K. and Reichman, D.R.},
  doi = {10.1103/PhysRevLett.97.195701},
  journal = {Phys. Rev. Lett.},
  number = {19},
  pages = {195701},
  title = {Inhomogeneous mode-coupling theory and growing dynamic length in supercooled liquids},
  volume = {97},
  year = {2006}
}

@article{got99,
  author = {G{\"o}tze, Wolfgang},
  journal = {Journal of Physics: condensed matter},
  number = {10A},
  pages = {A1},
  publisher = {IOP Publishing},
  title = {Recent tests of the mode-coupling theory for glassy dynamics},
  volume = {11},
  year = {1999}
}

@article{kir89,
  author = {Kirkpatrick, Theodore R and Thirumalai, Devarajan and Wolynes, Peter G},
  journal = {Physical Review A},
  number = {2},
  pages = {1045},
  publisher = {APS},
  title = {Scaling concepts for the dynamics of viscous liquids near an ideal glassy state},
  volume = {40},
  year = {1989}
}

@article{Gar03,
  author = {J. P. Garrahan and D. Chandler},
  doi = {10.1073/pnas.1233719100},
  journal = {Proceedings of the National Academy of Sciences},
  number = {17},
  pages = {9710--9714},
  publisher = {National Academy of Sciences},
  title = {Coarse-grained microscopic model of glass formers},
  volume = {100},
  year = {2003}
}

@incollection{can10,
  address = {Dordrecht},
  author = {N. Cancrini and F. Martinelli and C. Roberto and C. Toninelli},
  booktitle = {Probability and Phase Transition},
  doi = {10.1007/978-90-481-2810-5_47},
  editor = {N. Cerf and R. B. Israel and E. Presutti},
  pages = {451--471},
  publisher = {Springer},
  series = {Springer Proceedings in Mathematics},
  title = {Kinetically Constrained Models},
  year = {2010}
}

@article{Ton04,
  author = {Cristina Toninelli and Giulio Biroli and David S. Fisher},
  doi = {10.1103/PhysRevLett.92.185504},
  journal = {Physical Review Letters},
  number = {18},
  pages = {185504},
  publisher = {American Physical Society},
  title = {Spatial Structures and Dynamics of Kinetically Constrained Models of Glasses},
  volume = {92},
  year = {2004}
}

@article{Jac91,
  author = {J. J{\"a}ckle and S. Eisinger},
  doi = {10.1007/BF01307650},
  journal = {Zeitschrift f{\"u}r Physik B Condensed Matter},
  number = {1},
  pages = {115--124},
  title = {A hierarchically constrained kinetic Ising model},
  volume = {84},
  year = {1991}
}

@article{gar07,
  author = {Garrahan, Juan P and Jack, Robert L and Lecomte, Vivien and Pitard, Estelle and van Duijvendijk, Kristina and van Wijland, Fr{\'e}d{\'e}ric},
  journal = {Physical review letters},
  number = {19},
  pages = {195702},
  publisher = {APS},
  title = {Dynamical first-order phase transition in kinetically constrained models of glasses},
  volume = {98},
  year = {2007}
}

@article{wee00,
  author = {Weeks, Eric R and Crocker, John C and Levitt, Andrew C and Schofield, Andrew and Weitz, David A},
  journal = {Science},
  number = {5453},
  pages = {627--631},
  publisher = {American Association for the Advancement of Science},
  title = {Three-dimensional direct imaging of structural relaxation near the colloidal glass transition},
  volume = {287},
  year = {2000}
}

@article{sho07,
  author = {Shortland, Andrew and Rogers, Nick and Eremin, Katherine},
  journal = {Journal of Archaeological Science},
  number = {5},
  pages = {781--789},
  publisher = {Elsevier},
  title = {Trace element discriminants between Egyptian and Mesopotamian late Bronze Age glasses},
  volume = {34},
  year = {2007}
}

@book{fle99,
  author = {Fleming, Stuart J},
  publisher = {UPenn Museum of Archaeology},
  title = {Roman glass: reflections on cultural change},
  year = {1999}
}

@article{and95,
  author = {P. W. Anderson},
  journal = {Science},
  number = {5204},
  pages = {1618--1618},
  publisher = {American Association for the Advancement of Science},
  title = {Through the glass lightly},
  volume = {267},
  year = {1995}
}

@article{ang95,
  author = {C. A. Angell},
  doi = {10.1126/science.267.5206.19},
  journal = {Science},
  pages = {1924--1935},
  title = {Formation of glasses from liquids and biopolymers},
  volume = {267},
  year = {1995}
}

@article{bri18,
  author = {Brito, C. and Lerner, E. and Wyart, M.},
  doi = {10.1103/PhysRevX.8.031050},
  journal = {Phys. Rev. X},
  number = {3},
  pages = {031050},
  title = {{Theory for Swap Acceleration near the Glass and Jamming Transitions for Continuously Polydisperse Particles}},
  volume = {8},
  year = {2018}
}

@article{deb01,
  author = {P. G. Debenedetti and F. H. Stillinger},
  doi = {10.1038/35065704},
  journal = {Nature},
  pages = {259},
  title = {Supercooled liquids and the glass transition},
  volume = {410},
  year = {2001}
}

@article{franz2000non,
  author = {Franz, S. and Parisi, G.},
  doi = {10.1088/0953-8984/12/29/305},
  journal = {J. Phys.: Condens. Matter},
  number = {29},
  pages = {6335},
  title = {On non-linear susceptibility in supercooled liquids},
  volume = {12},
  year = {2000}
}

@article{franz2011field,
  author = {Franz, S. and Parisi, G. and Ricci-Tersenghi, F. and Rizzo, T.},
  doi = {10.1140/epje/i2011-11102-0},
  journal = {Eur. Phys. J. E},
  number = {9},
  pages = {1--17},
  title = {Field theory of fluctuations in glasses},
  volume = {34},
  year = {2011}
}

@article{gutierrez2015static,
  author = {Guti{\'e}rrez, R. and Karmakar, S. and Pollack, Y.G. and Procaccia, I.},
  doi = {10.1088/0022-3719/21/18/007},
  journal = {Europhys. Lett.},
  number = {5},
  pages = {56009},
  title = {The static lengthscale characterizing the glass transition at lower temperatures},
  volume = {111},
  year = {2015}
}

@article{ton05,
  author = {Toninelli, Cristina and Wyart, Matthieu and Berthier, Ludovic and Biroli, Giulio and Bouchaud, Jean-Philippe},
  journal = {Physical Review E—Statistical, Nonlinear, and Soft Matter Physics},
  number = {4},
  pages = {041505},
  publisher = {APS},
  title = {Dynamical susceptibility of glass formers: Contrasting the predictions of theoretical scenarios},
  volume = {71},
  year = {2005}
}

@article{kau48,
  author = {W. Kauzmann},
  doi = {10.1021/cr60135a002},
  journal = {Chem. Rev.},
  pages = {219},
  title = {The nature of the glassy state and the behavior of liquids at low temperatures},
  volume = {43},
  year = {1948}
}

@article{deg25,
  author = {de Geus, Tom WJ and Rosso, Alberto and Wyart, Matthieu},
  journal = {Physical Review E},
  number = {1},
  pages = {L013503},
  publisher = {APS},
  title = {Dynamical heterogeneities of thermal creep in pinned interfaces},
  volume = {111},
  year = {2025}
}

@article{reh10b,
  author = {A. Rehwald and A. Heuer},
  doi = {10.1103/PhysRevLett.105.117801},
  journal = {Physical Review Letters},
  number = {11},
  pages = {117801},
  title = {From coupled elementary units to the complexity of the glass transition},
  volume = {105},
  year = {2010}
}

@article{Lemaitre14,
  author = {Lema{\^{i}}tre, A.},
  journal = {Phys. Rev. Lett.},
  number = {24},
  pages = {245702},
  title = {{Structural Relaxation is a Scale-Free Process}},
  volume = {113},
  year = {2014}
}

@article{wu15,
  author = {Wu, Bin and Iwashita, Takuya and Egami, Takeshi},
  journal = {Physical Review E},
  number = {3},
  pages = {032301},
  publisher = {APS},
  title = {Anisotropic stress correlations in two-dimensional liquids},
  volume = {91},
  year = {2015}
}

@article{kaw17,
  author = {Kawasaki, Takeshi and Kim, Kang},
  journal = {Science Advances},
  number = {8},
  pages = {e1700399},
  publisher = {American Association for the Advancement of Science},
  title = {Identifying time scales for violation/preservation of Stokes-Einstein relation in supercooled water},
  volume = {3},
  year = {2017}
}

@article{Bul94,
  author = {V. V. Bulatov and A. S. Argon},
  doi = {10.1088/0965-0393/2/2/003},
  journal = {Modeling and Simulation in Materials Science and Engineering},
  number = {2},
  pages = {185--202},
  title = {A stochastic model for continuum elasto-plastic behavior. II. A study of the glass transition and structural relaxation},
  volume = {2},
  year = {1994}
}

@article{Got92,
  author = {Wolfgang G{\"o}tze and Lennart Sj{\"o}gren},
  doi = {10.1088/0034-4885/55/3/001},
  journal = {Reports on Progress in Physics},
  number = {3},
  pages = {241--376},
  title = {Relaxation processes in supercooled liquids},
  volume = {55},
  year = {1992}
}

@article{cha10,
  title={Dynamics on the way to forming glass: Bubbles in space-time},
  author={Chandler, David and Garrahan, Juan P},
  journal={Annual review of physical chemistry},
  volume={61},
  number={1},
  pages={191--217},
  year={2010},
  publisher={Annual Reviews}
}

@article{wya05,
  title={Effects of compression on the vibrational modes of marginally jammed solids},
  author={Wyart, Matthieu and Silbert, Leonardo E and Nagel, Sidney R and Witten, Thomas A},
  journal={Physical Review E—Statistical, Nonlinear, and Soft Matter Physics},
  volume={72},
  number={5},
  pages={051306},
  year={2005},
  publisher={APS}
}

@article{Doliwa2003,
  author = {B. Doliwa and A. Heuer},
  doi = {10.1103/PhysRevE.67.031506},
  issn = {1063651X},
  issue = {3},
  journal = {Physical Review E},
  pages = {16},
  title = {Energy barriers and activated dynamics in a supercooled Lennard-Jones liquid},
  volume = {67},
  year = {2003}
}

@article{neb1,
  author = {Zarkevich, Nikolai A and Johnson, Duane D},
  doi = {10.1063/1.4905209},
  journal = {The Journal of Chemical Physics},
  number = {2},
  pages = {24106},
  title = {{Nudged-elastic band method with two climbing images: Finding transition states in complex energy landscapes}},
  volume = {142},
  year = {2015}
}

@article{pop20,
  author = {Popovi{\' c}, M. and de Geus, T.W.J. and Ji, W. and Wyart, M.},
  doi = {10.48550/arXiv.2009.04963},
  journal = {arXiv preprint 2009.04963},
  title = {Thermally activated flow in models of amorphous solids},
  year = {2020}
}

@article{pur17,
  arxivid = {1704.01489},
  author = {Purrello, V. H. and Iguain, J. L. and Kolton, A. B. and Jagla, E. A.},
  doi = {10.1103/PhysRevE.96.022112},
  journal = {Phys. Rev. E},
  number = {2},
  pages = {022112},
  title = {Creep and Thermal Rounding Close to the Elastic Depinning Threshold},
  volume = {96},
  year = {2017}
}

@article{pac96,
  author = {Paczuski, Maya and Maslov, Sergei and Bak, Per},
  journal = {Physical Review E},
  number = {1},
  pages = {414},
  publisher = {APS},
  title = {Avalanche dynamics in evolution, growth, and depinning models},
  volume = {53},
  year = {1996}
}

@article{bou24,
  author = {Bouchaud, Jean-Philippe},
  journal = {arXiv preprint arXiv:2402.01883},
  title = {Why is the dynamics of glasses super-Arrhenius?},
  year = {2024}
}

@article{lac03,
  author = {N. Lacevic and F. W. Starr and T. B. Schr\"oder and S. C. Glotzer},
  doi = {10.1063/1.1605094},
  journal = {J. Chem. Phys.},
  pages = {7372--7387},
  title = {Spatially heterogeneous dynamics investigated via a time-dependent four-point density correlation function},
  volume = {119},
  year = {2003}
}

@article{Bernini_2017,
  author = {S Bernini and F Puosi and D Leporini},
  doi = {10.1088/1361-648X/aa5a7e},
  journal = {J. Condens. Matter Phys.},
  month = {feb},
  number = {13},
  pages = {135101},
  publisher = {IOP Publishing},
  title = {Thermodynamic scaling of relaxation: insights from anharmonic elasticity},
  url = {https://dx.doi.org/10.1088/1361-648X/aa5a7e},
  volume = {29},
  year = {2017}
}

@article{Puosi_jcp_2012,
  author = {Puosi, F. and Leporini, D.},
  doi = {10.1063/1.3681291},
  issn = {0021-9606},
  journal = {J. Chem. Phys.},
  month = {01},
  number = {4},
  pages = {041104},
  title = {{Communication: Correlation of the instantaneous and the intermediate-time elasticity with the structural relaxation in glassforming systems}},
  url = {https://doi.org/10.1063/1.3681291},
  volume = {136},
  year = {2012}
}

@article{lerner_jcp_2018,
  author = {Lerner,Edan and Bouchbinder,Eran},
  doi = {10.1063/1.5024776},
  journal = {J. Chem. Phys.},
  number = {21},
  pages = {214502},
  title = {A characteristic energy scale in glasses},
  url = {https://doi.org/10.1063/1.5024776},
  volume = {148},
  year = {2018}
}

@article{isoconfigurational_prl_2004,
  author = {Widmer-Cooper, Asaph and Harrowell, Peter and Fynewever, H.},
  doi = {10.1103/PhysRevLett.93.135701},
  issue = {13},
  journal = {Phys. Rev. Lett.},
  month = {Sep},
  numpages = {4},
  pages = {135701},
  publisher = {American Physical Society},
  title = {How Reproducible Are Dynamic Heterogeneities in a Supercooled Liquid?},
  url = {https://link.aps.org/doi/10.1103/PhysRevLett.93.135701},
  volume = {93},
  year = {2004}
}

@article{new_variational_argument_epl_2016,
  author = {Le Yan and Eric DeGiuli and Matthieu Wyart},
  journal = {Europhys. Lett.},
  number = {2},
  pages = {26003},
  title = {On variational arguments for vibrational modes near jamming},
  url = {http://stacks.iop.org/0295-5075/114/i=2/a=26003},
  volume = {114},
  year = {2016}
}

@article{Harrowell_jcp_2009,
  author = {Widmer-Cooper, Asaph and Perry, Heidi and Harrowell, Peter and Reichman, David R.},
  doi = {10.1063/1.3265983},
  issn = {0021-9606},
  journal = {J. Chem. Phys.},
  month = {11},
  number = {19},
  pages = {194508},
  title = {{Localized soft modes and the supercooled liquid's irreversible passage through its configuration space}},
  url = {https://doi.org/10.1063/1.3265983},
  volume = {131},
  year = {2009}
}

@article{dipole_stiffness_jcp_2021,
  author = {Rainone, Corrado and Bouchbinder, Eran and Lerner, Edan},
  doi = {10.1063/5.0005655},
  issn = {0021-9606},
  journal = {J. Chem. Phys.},
  month = {05},
  number = {19},
  pages = {194503},
  title = {{Statistical mechanics of local force dipole responses in computer glasses}},
  url = {https://doi.org/10.1063/5.0005655},
  volume = {152},
  year = {2020}
}

@article{pinching_pnas,
  author = {Rainone, Corrado and Bouchbinder, Eran and Lerner, Edan},
  doi = {10.1073/pnas.1919958117},
  issn = {0027-8424},
  journal = {Proc. Natl. Acad. Sci. U.S.A.},
  number = {10},
  pages = {5228--5234},
  publisher = {National Academy of Sciences},
  title = {Pinching a glass reveals key properties of its soft spots},
  url = {https://www.pnas.org/content/117/10/5228},
  volume = {117},
  year = {2020}
}

@article{Tong_pre_2014,
  author = {Tong, Hua and Xu, Ning},
  doi = {10.1103/PhysRevE.90.010401},
  issue = {1},
  journal = {Phys. Rev. E},
  month = {Jul},
  numpages = {5},
  pages = {010401},
  publisher = {American Physical Society},
  title = {Order parameter for structural heterogeneity in disordered solids},
  url = {https://link.aps.org/doi/10.1103/PhysRevE.90.010401},
  volume = {90},
  year = {2014}
}

@article{nov05pre,
  author = {V. N. Novikov and A. P. Sokolov},
  doi = {10.1103/PhysRevE.71.061501},
  journal = {Physical Review E},
  number = {6},
  pages = {061501},
  title = {Correlation of fragility of supercooled liquids with elastic properties of their glasses},
  volume = {71},
  year = {2005}
}

@article{tan08nm,
  author = {H. Shintani and Hajime Tanaka},
  doi = {10.1038/nmat2293},
  journal = {Nature Materials},
  number = {11},
  pages = {870--877},
  title = {Universal link between the boson peak and transverse phonons in glass},
  volume = {7},
  year = {2008}
}

@article{ngai97jcp,
  author = {K. L. Ngai},
  doi = {10.1063/1.474271},
  journal = {The Journal of Chemical Physics},
  number = {13},
  pages = {5268--5276},
  title = {Correlations between boson-peak strength and characteristics of local segmental relaxation in polymers},
  volume = {107},
  year = {1997}
}

@article{tatsu90prl,
  author = {M. Tatsumisago and B. L. Halfpap and J. L. Green and S. M. Lindsay and C. A. Angell},
  doi = {10.1103/PhysRevLett.64.1549},
  journal = {Physical Review Letters},
  number = {13},
  pages = {1549--1552},
  title = {Fragility of {Ge--As--Se} glass-forming liquids in relation to rigidity percolation, and the {Kauzmann} paradox},
  volume = {64},
  year = {1990}
}

@article{bohmer91prb,
  author = {Roland B{\"o}hmer and C. Austen Angell},
  doi = {10.1103/PhysRevB.45.10091},
  journal = {Physical Review B},
  number = {17},
  pages = {10091--10094},
  title = {Correlations of the nonexponentiality and state dependence of mechanical relaxations with bond connectivity in {Ge--As--Se} supercooled liquids},
  volume = {45},
  year = {1992}
}

@article{kam91prb,
  author = {W. A. Kamitakahara and R. L. Cappelletti and P. Boolchand and B. Halfpap and C. A. Angell and D. J. Bresser},
  doi = {10.1103/PhysRevB.44.94},
  journal = {Physical Review B},
  number = {1},
  pages = {94--100},
  title = {Vibrational densities of states and network rigidity in chalcogenide glasses},
  volume = {44},
  year = {1991}
}

@article{phillips79jncs,
  author = {J. C. Phillips},
  doi = {10.1016/0022-3093(79)90133-4},
  journal = {Journal of Non-Crystalline Solids},
  number = {2},
  pages = {153--181},
  title = {Topology of covalent non-crystalline solids {I}: Short-range order in chalcogenide alloys},
  volume = {34},
  year = {1979}
}

@article{thorpe85ssc,
  author = {M. F. Thorpe},
  doi = {10.1016/0038-1098(85)90381-3},
  journal = {Solid State Communications},
  number = {8},
  pages = {699--702},
  title = {Continuous deformations in random networks},
  volume = {53},
  year = {1985}
}

@article{Buc84,
  author = {U. Buchenau and N. N{\"u}cker and A. J. Dianoux},
  doi = {10.1103/PhysRevLett.53.2316},
  journal = {Physical Review Letters},
  pages = {2316},
  title = {Neutron Scattering Study of the Low-Frequency Vibrations in Vitreous Silica},
  volume = {53},
  year = {1984}
}

@article{Bal10,
  author = {G. Baldi and V. M. Giordano and G. Monaco and B. Ruta},
  doi = {10.1103/PhysRevLett.104.195501},
  journal = {Physical Review Letters},
  number = {19},
  pages = {195501},
  title = {Sound Attenuation at Terahertz Frequencies and the Boson Peak of Vitreous Silica},
  volume = {104},
  year = {2010}
}

\end{document}